\documentclass[11pt, a4paper]{book} %[11pt,openany] or [11pt,oneside]

\usepackage[table, dvipsnames]{xcolor} %Loaded before loading {ku-frontpage}.

\usepackage{amsmath, amssymb, graphicx, mathtools} %geometry
\usepackage{varioref, subcaption, float, booktabs}

\usepackage[sorting=none]{biblatex}
\usepackage{blindtext, sectsty}

\usepackage[utf8]{inputenc}
\usepackage[english]{babel}
\usepackage[T1]{fontenc}

\usepackage[hidelinks]{hyperref} %It seems hyperref needs to be loaded before loading ku-frontpage %Note that position of hyperref and ku-frontpage were changed. Both of them moved together.
\usepackage[english, science, titlepage]{ku-frontpage}

\usepackage{cleveref} %cleveref need to be loaded last time!

\DeclareFieldFormat[article]{citetitle}{#1}
\DeclareFieldFormat[article]{title}{#1}

\makeatletter
\newcommand{\Rmnum}[1]{\expandafter\@slowromancap\romannumeral #1@}
\makeatother

\assignment{Master Thesis}
\title{Dihedral angle prediction using generative adversarial networks }
\subtitle{} % No subtitle
\date{Submitted on: March 5, 2018}

\author{Hyeongki Kim}
\advisor{Supervisor: Thomas Wim Hamelryck\\Co-supervisor: Wouter Krogh Boomsma}
\AtBeginDocument{
  
}

\begin{document}
\allowdisplaybreaks
\pagenumbering{roman}
\maketitle
%\newpage\noindent
%\newline\\[5\baselineskip]
%Name of department:	Department of Biology\\\\
%Author: Hyeongki Kim\\\\
%Title:	Dihedral angle prediction using generative adversarial networks \\\\
%Topic description:	.\\\\
%Supervisor:	Assoc. Prof. Thomas Wim Hamelryck\\\\
%Co-supervisor: Asst. Prof. Wouter Krogh Boomsma\\\\
%Submitted on:	\today \\\\
%Grade:	\\\\
%Number of study units:
%\begin{itemize}
%\item[$\square$] 2  %$\boxtimes$
%\item[$\square$] 3
%\end{itemize}
%~\\
%Number of characters:\\\\

%\newpage
%\chapter*{\centering Abstract}
\chapter*{Abstract}
\addcontentsline{toc}{chapter}{Abstract}
Several dihedral angles prediction methods were developed for protein structure prediction and their other applications. However, distribution of predicted angles would not be similar to that of real angles. To address this we employed generative adversarial networks (GAN) which showed promising results in image generation tasks. Generative adversarial networks are composed of two adversarially trained networks: a discriminator and a generator. A discriminator is trained to distinguish between samples from a dataset and generated samples while a generator is trained to generate realistic samples.\\\\
Although the discriminator of GANs is trained to estimate density, the explicit density of the model is not tractable. On the other hand, noise-contrastive estimation (NCE) was introduced to estimate a normalization constant of an unnormalized statistical model and thus the density function. \\\\
In this thesis, we introduce noise-contrastive estimation generative adversarial networks (NCE-GAN) which enables explicit density estimation of the generative adversarial networks by feeding noise samples from a known distribution, like noise-contrastive estimation, and adding a corresponding class for the discriminator. We analyzed the minibatch discrimination and new loss for the generator is proposed. We also propose residue-wise variants of auxiliary classifier GAN (AC-GAN) and Semi-supervised GAN to handle sequence information in a window. \\\\
In our experiment, the conditional generative adversarial network (C-GAN), AC-GAN and Semi-supervised GAN were compared. And experiments done with improved conditions were invested.\\\\
We identified a phenomenon of AC-GAN that distribution of its predicted angles is composed of unusual clusters. The distribution of the predicted angles of Semi-supervised GAN was most similar to the Ramachandran plot. We found that adding the output of the NCE as an additional input of the discriminator is helpful to stabilize the training of the GANs and to capture the detailed structures. However, using the newly proposed loss in the generator were only helpful in C-GAN and AC-GAN. %\textcolor{red}{In this part, I'm planning to write that we could predict realistic angles using GANs.}\\
Adding regression loss to the object of the generator and using predicted angles by regression loss only model as an additional input of the generator could improve the conditional generation performance of the C-GAN and AC-GAN.% which was found by weighted mean conditional log-likelihood estimated by NCE-GAN.
%Distribution of predicted angles by AC-GAN is composed of some clusters which are not observed in real angle distribution. 
%By estimating weighted mean conditional log-likelihood, we found that adding regression loss to the object of the generator and using predicted angles by another model as an additional input of the generator can improve the conditional generation performance of the C-GAN and AC-GAN.

%\newpage
\tableofcontents

\newpage
\pagenumbering{arabic}
\chapter{Introduction}
\section{Protein structure}
\subsection{Protein and amino acids}
Protein is one of the macromolecules that have a diverse role in living organisms.\\\\
An amino acid, which is a building block (also called monomer) of a protein, contains an asymmetric carbon atom called the alpha carbon at the center of its structure (see figure \ref{fig:amino acid} left). %Alpha carbon in the amino acid is often called C-alpha. 
It has four different partners: an amino group, a carboxyl group, a hydrogen atom, and a variable R group. The R group is also called the side chain of the amino acid \cite{campbell_cain_minorsky_reece_urry_wasserman}.\\\\
The side chain of an amino acid determines characteristics of that amino acid (see figure \ref{fig:amino acid table}). One group of amino acids have non-polar side chains and they are hydrophobic. Another group of amino acids have polar side chains and they are hydrophilic. Acidic amino acids have generally negatively charged side chains. Basic amino acids have generally positively charged side chains. Acidic and basic amino acids are also hydrophilic because they contain electronically charged side chains \cite{campbell_cain_minorsky_reece_urry_wasserman}. \\\\
Covalently bonded chains (also called polymer) of many amino acids are called polypeptide. The covalent bond between amino acids is called a peptide bond (see figure \ref{fig:amino acid} right). The peptide bond is formed when an amino group of a amino acid and a carboxyl group of another amino acid bind together through dehydration reaction \cite{campbell_cain_minorsky_reece_urry_wasserman}. \\\\
\begin{figure}[htbp]
	\centering
	\begin{subfigure}[b]{0.45\textwidth}
        \includegraphics[width=\textwidth]{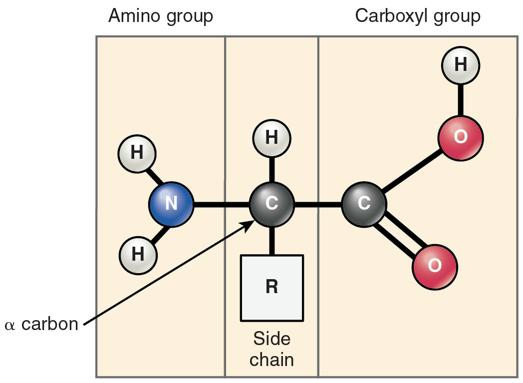}
    \end{subfigure}	\hspace{0.05\textwidth}
    \begin{subfigure}[b]{0.4\textwidth}
        \includegraphics[width=\textwidth]{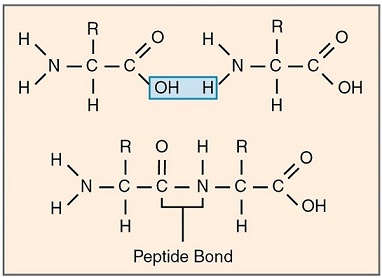}
    \end{subfigure}	
    \caption{Left: Structure of an amino acid from \cite{Essential_to_Human_Functioning}, right: formation of a peptide bond modified from \cite{Essential_to_Human_Functioning}. }
\end{figure}\label{fig:amino acid}

\begin{figure}[htbp]
	\centering
    \begin{subfigure}[b]{0.85\textwidth}
        \includegraphics[width=\textwidth]{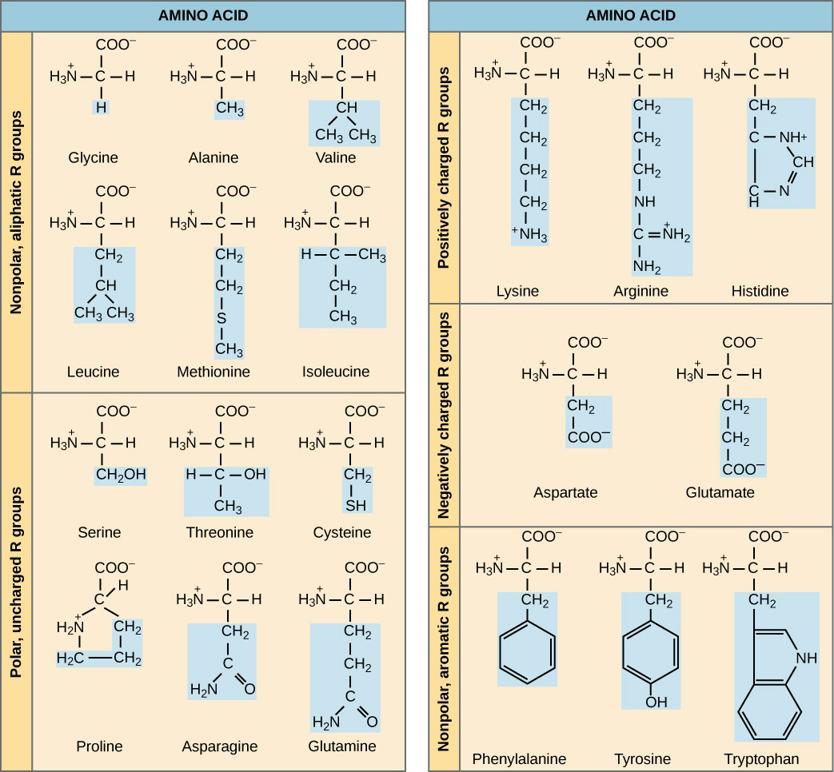}
    \end{subfigure}	
    \caption{Table of 20 common amino acids from \cite{Proteins}. Aliphatic and aromatic amino acids are hydrophoic.}
\end{figure}\label{fig:amino acid table}

\newpage
\subsection{Protein structure}
\begin{figure}[H]
	\centering
    \begin{subfigure}[b]{0.65\textwidth}
        \includegraphics[width=\textwidth]{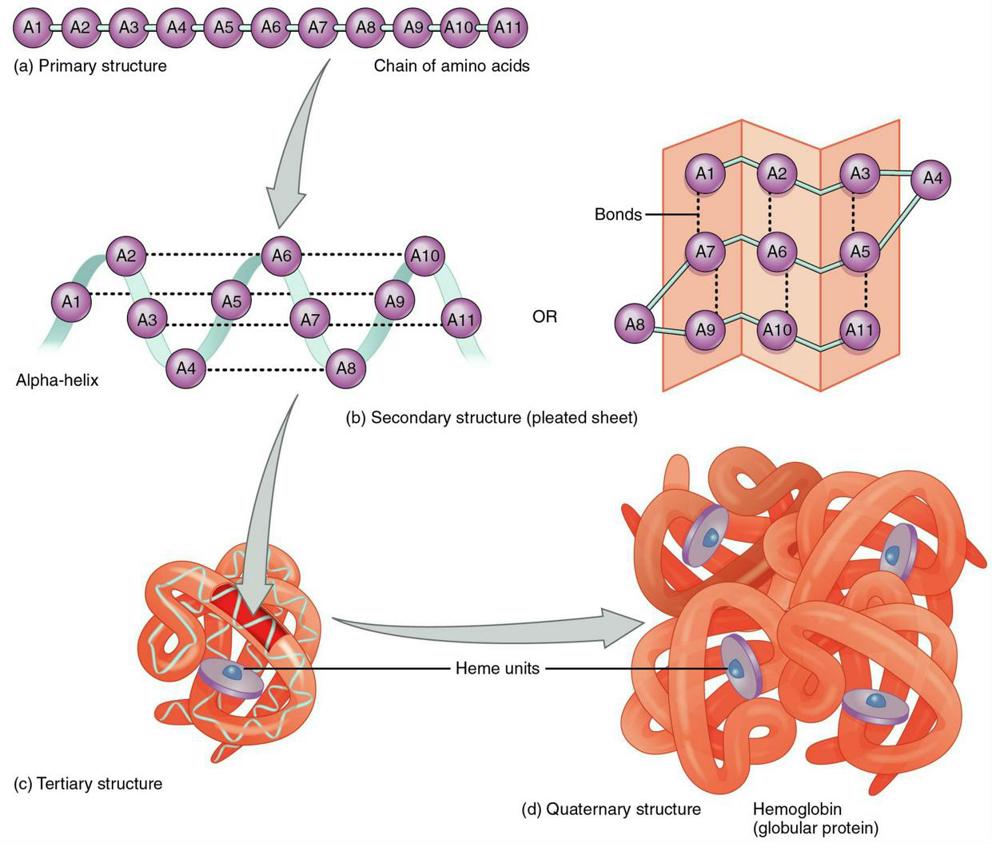}
    \end{subfigure}	
    \caption{4 level hierarchical structure of a protein from \cite{Essential_to_Human_Functioning}. Hemoglobin is shown in this example. }\label{fig:Hierarchical structure}
\end{figure}

\begin{figure}[H]
	\centering
	\begin{subfigure}[b]{0.65\textwidth}
        \includegraphics[width=\textwidth]{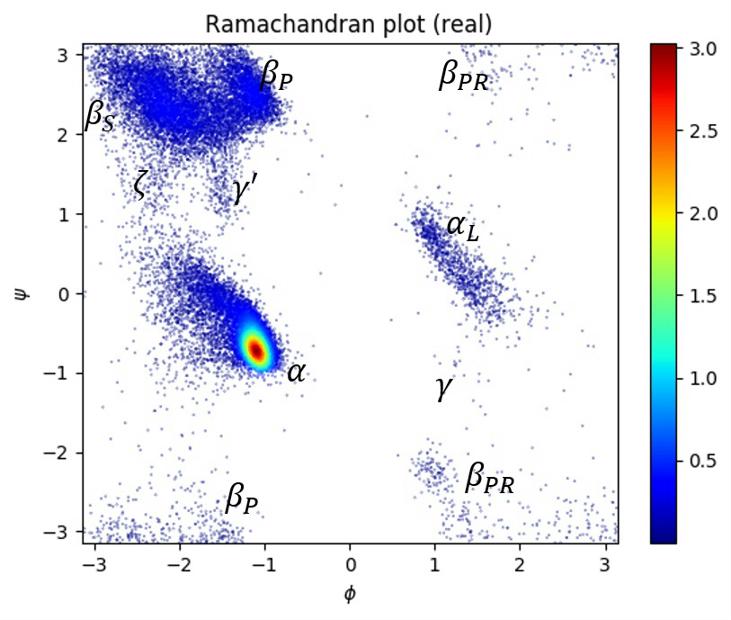}
    \end{subfigure}	
    \caption{Ramachandran plot of the test set \ref{dataset} used in the experiments. Allowed regions of $\phi$, $\psi$ angles are labeled. \\
$\beta_{S}$: extended conformation of residues in $\beta$-sheets \cite{ho2005ramachandran}, $\beta_{P}$: polyproline II conformation region \cite{ho2005ramachandran}, \\
$\zeta$: the region near (-2.3,1.4) = (${-130}^{\circ}$, ${80}^{\circ}$) that is preferentially populated in pre-Proline \cite{lovell2003structure, ho2005ramachandran, karplus1996experimentally}, \\
$\gamma$: $\gamma$-turn conformation \cite{lovell2003structure}, $\gamma '$: mirror image region of $\gamma$ region \cite{lovell2003structure},\\
$\alpha$: right-handed $\alpha$ helical region,
$\alpha_{L}$: left-handed $\alpha$ helical region, $\beta_{PR}$: reflection of $\beta_{P}$ region \cite{ho2005ramachandran}.}\label{fig:Ramachandran plot regions}
\end{figure}

Protein structure can be explained by four level hierarchical structure: primary structure, secondary structure, tertiary structure, and quaternary structure \cite{campbell_cain_minorsky_reece_urry_wasserman}. (see figure \ref{fig:Hierarchical structure})\\\\
The primary structure indicates the amino acids sequence of a protein. The sequence of a protein is determined by genetic information. DNA information is turned into messenger RNA (mRNA) by transcription process. Polypeptide chain is synthesized by ribosome using mRNA as a template. This process is called translation \cite{campbell_cain_minorsky_reece_urry_wasserman}.\\\\
Weak hydrogen bonds between closely positioned polypeptide backbone form certain patterns. These patterns are called secondary structure. There are 3 broad groups of these shapes: $\alpha$-helix, $\beta$-pleated sheet and loops \cite{campbell_cain_minorsky_reece_urry_wasserman}.\\\\
The overall shape of a polypeptide is referred to as its tertiary structure. Tertiary structure is formed by interaction between side chains in the polypeptide. These interactions include hydrophobic interaction, van der Waals interactions, ionic bonds between polar amino acids and disulfide bridge \cite{campbell_cain_minorsky_reece_urry_wasserman}.\\\\
Often more than one polypeptide subunit aggregated together to form one protein. The quaternary structure indicates protein structure that is consist of multiple polypeptides \cite{campbell_cain_minorsky_reece_urry_wasserman}. \\\\
X-ray crystallography and nuclear magnetic resonance (NMR) are the mainly used experimental determination methods for protein structure. Determined structures of proteins can be accessible via Protein Data Bank (PDB) \cite{berman2002protein} which is an archive of biological macromolecular structural data.\\\\
Dihedral angles, or often called torsion angles, are defined by three consecutive bonds among four atoms.
The backbone structure of a protein ignores the side chains of the protein and it is determined by three dihedral angles: $\phi$, $\psi$ and $\omega$. $\phi$ is determined by the atoms $C-N-C_{alpha}-C$ and $\psi$ is determined by the atoms $N-C_{alpha}-C-N$. Likewise, $\omega$ is determined by the atoms $C_{alpha}-C-N-C_{alpha}$. $\omega$ angle is approximately planar (either 0$^{\circ}$ or 180$^{\circ}$) because of its partial double bond character. As trans-configuration of two side chains are favored, most $\omega$ angles are close to 180$^{\circ}$ \cite{Zimmermann2017}. \\\\
Ramachandran et al \cite{ramachandran1963stereochemistry, ramachandran1968conformation} provided analysis on favored and disallowed regions of $\phi$, $\psi$ angles based on steric constraints and concluded that almost 3/4 of the space is unavailable for dihedral angles 
\cite{Zimmermann2017}. Figure \ref{fig:Ramachandran plot regions} shows $\phi$, $\psi$ angles of the test set \ref{dataset} used in the experiments.

%\textcolor{red}{As I' planning to add estimated density plots for each amino acids (or only Ramachandran plots for Glycine and Proline) in the result section, I might need to add some explanation of different Ramachandran plot for each amino acids.}\\\\
%\textcolor{red}{I said I will add some explanation of solvent accessibility because I was planning to use this in the experiment, but I have changed my plan and I might not use solvent accessibility in the experiment and thus I will skip the explanation of it.}\\\\
%Angles and relative solvent accessibility can be easily calculated by Biopython [].
\section{Deep learning}
\subsection{Artificial neural network}
A neuron is a biological information processing unit of a brain. It receives signals from other neurons through its dendrites and electronic signals can be sent along the axon. If collected signals are higher than a certain threshold, the neuron can fire signals and neurotransmitter will send the information to other neurons through the synapse \cite{neural-networks-website}.\\\\
Artificial neuron is a mathematical information processing unit that is inspired by biological neuron. It gets input (column) vector $x$ and this information is collected by multiplying with some weight matrix $W$ and adding some bias vector $b$. Bias term can also be integrated into weight by setting $\tilde{W}=[b \quad W]$  and $\tilde{x}=\left[ \begin{matrix}  1\\x  \end{matrix} \right] $. Activation function $f$ is then applied to the collected information $Wx+b$ and the output of the neuron will be $f(Wx+b)$. Most of the time, nonlinear function is used for activation function \cite{NNslides}. In the case of the perceptron, activation function is the Heaviside step function.\\ \newline
\begin{figure}[H] %htbp
	\centering
	\begin{subfigure}[b]{0.45\textwidth}
        \includegraphics[width=\textwidth]{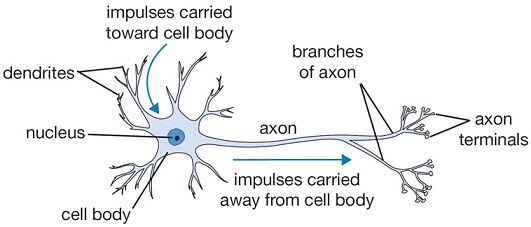}
        %\caption{Biological neuron}
    \end{subfigure}	%\hspace{0.05\textwidth}
    \begin{subfigure}[b]{0.45\textwidth}
        \includegraphics[width=\textwidth]{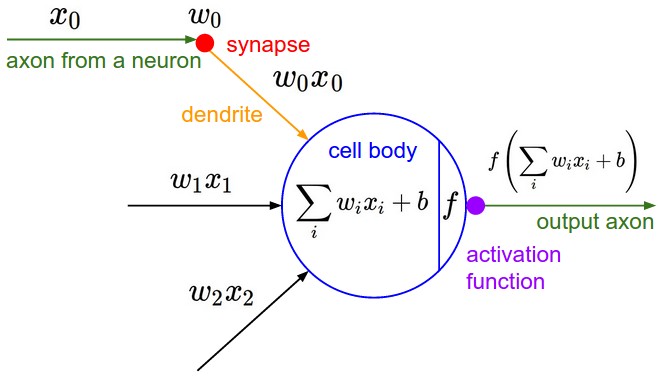}
        %\caption{Mathematical model of neuron}
    \end{subfigure} 
    \caption{Left: A cartoon drawing of a biological neuron,  right: corresponding mathematical model. Images from \cite{neural-networks-website}.}
\end{figure}%\\
A multilayer perceptron (MLP) is a feedforward neural network which contains at least one hidden layer. Feedforward neural network indicates a neural network which has no loop. MLP can be used to optimize classifier or regression problem. Backpropagation is used for training of the MLP. \\ \newline
\begin{figure}[H]
	\centering
	\begin{subfigure}[b]{0.3\textwidth}
        \includegraphics[width=\textwidth]{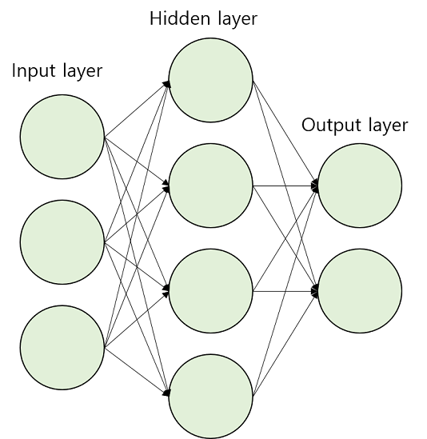}
    \end{subfigure}	
    \caption{architecture of a multilayer perceptron network}
\end{figure}%\\
Backpropagation method repeats forward pass and backward pass to optimize the objective function.\\\\
Forward pass: Input signals and current parameters (weights and biases) are used to output the prediction. Thereafter, loss (object) function is calculated using the outputs and the corresponding labels (label of the autoencoder network is the input itself).\\
Backward pass: Updates parameters to minimize the loss function in reverse order. Variants of gradient descent algorithm such as resilient backpropagation (RProp) \cite{riedmiller1993direct} or adaptive moments (Adam) algorithm \cite{kingma2014adam} can be used for this process \cite{NNslides}.\\\\
Since derivative of activation function is multiplied several times as the network has more layers, vanishing gradient problem happens when the multilayer perceptron is too deep.\\\\
Commonly used traditional activation functions in deep learning include sigmoid function $\frac{1}{(1+e^{-x})}$ and hyperbolic tangent $\tanh(x)=\frac{2}{(1+e^{-2 x})}-1$ (sometimes called bipolar sigmoid).\\\\
Rectified linear unit (ReLU) which can be formulated as: 
\begin{align}
f(x) = \begin{cases} x \text{    if } x\ge 0 \\ 0 \text{  otherwise} \end{cases}
\end{align}
is non-saturated activation function \cite{Nair:2010:RLU:3104322.3104425}. Using non-saturated activation function is advantageous as it helps to prevent vanishing gradient problem, and hence it enables to avoid networks converge to a bad local minimum \cite{maas2013rectifier}.
However, using ReLU could lead to lack of activation in some neurons because their gradient update will always be zero \cite{maas2013rectifier}. To alleviate this problem leaky rectified linear unit (LReLU) was suggested with the following formula for a fixed $a\in (1,\infty )$ \cite{maas2013rectifier}.
\begin{align}
f(x) = \begin{cases} x \text{    if } x\ge 0 \\ \frac{x}{a} \text{  otherwise} \end{cases}
\end{align}
Empirical comparison of rectified linear units showed that LReLU outperforms ReLU when $a=5.5$ \cite{2015arXiv150500853X}. \\\\
One problem of rectified linear units is that they are not bounded and sometimes we require boundedness to solve some regression tasks. We explain another activation function called softsign which can be formulated as \cite{Bergstra+2009}:
\begin{align}
f(x)=\frac{x}{1+|x|}
\end{align} 
Derivative of softsign activation function approaches zero more slowly than tanh function \cite{Bergstra+2009} and experimental results show that softsign function become less saturated than sigmoid and hyperbolic tangent \cite{pmlr-v9-glorot10a}.\\

\section{Deep learning for protein structure prediction}
Protein sequence information can be relatively cheaply obtained through automated procedures, but the determination of protein structure is labor-intensive and costly \cite{faraggi2017accurate}. Thus, computational prediction of protein three-dimensional structure for given sequence information is important.\\\\ %Many sequence to one dimensional structural properties prediction methods have been developed [Prediction of Protein Secondary Structure page v].\\\\
Currently predicted one-dimensional structural properties can be classified into two subclasses: Local structural properties mean structural features that only depend on locally connected residues. Examples of this category are secondary structure and backbone torsion angles. Global structural properties could also depend on structural neighbors that are not necessarily their neighbors in term of their sequence positions. Global structural properties include solvent exposure properties like solvent accessible surface area (ASA), residue coordination (contact) number, residue contact order, and residue depth
\cite{eltit}.

\subsection{Secondary structure prediction methods}
First generation secondary structure prediction methods utilized statistical propensities of amino acid. 
Second generation methods used the sliding window to take into account their neighboring residues. 
The third generation techniques used evolutionary information derived from multiple sequence alignment \cite{yangsixty, JIANG2017379}.\\\\
%State-of-the-art methods for secondary structure prediction employ a sequence profile derived from multiple sequence alignment of homologous sequences. Position specific substitution matrix (PSSM) calculated by PSI-BLAST \cite{altschul1997gapped} is one of the sequence profile \cite{yangsixty}.

\subsection{Dihedral angle prediction methods}
Because of bond distances between neighboring $C_{\alpha}$ atoms are almost fixed and $\omega$ angle is almost fixed, torsion angles $\phi$, $\psi$ can represent the backbone structure of a protein 
\cite{Li2017, Zimmermann2017}.\\\\
%Real value dihedral angle prediction was started by 
%\textcolor{red}{I'm planning to explain some dihedral angle prediction methods, like SPINE and SPIDER. It will include suggested angle periodicity handling methods (shifting and indirect prediction method) and previous effort of generating realistic angles.}\\\\
DESTRUCT is the first publication that predicted real value dihedral angles. It reported Pearson's correlation coefficient (PCC) of $\psi$ angle. Both three-state secondary structure and $\psi$ dihedral angle were predicted using iterative cascade-correlation neural network \cite{Zimmermann2017}.\\\\
Real-SPINE is the first method dedicated to dihedral angle prediction \cite{Zimmermann2017, dor2007real}. Position specific scoring matrix (PSSM) calculated by PSI-BLAST \cite{altschul1997gapped}, one parameter which describes nonexistence of the amino acids near the terminus of the chain, seven representative amino-acid properties (PP) \cite{meiler2001generation} and predicted secondary structures by SPINE \cite{dor2007achieving} were used as inputs \cite{dor2007real}. To minimize the impact of the periodicity of the angles, their improved work applied shifting transformation on both $\phi$, $\psi$ angles so that probabilities of the shifted angles are close to zero around boundary angles \cite{PROT:PROT21940}. \\\\
To avoid predicting angles in the sterically forbidden regions, SPINE-X and SPINE-XI were suggested. The idea of the SPINE-X is that as both $\phi$ and $\psi$ have a bimodal distribution, split the prediction process into two phase. First, each angle is classified into two states (peak I and peak II), and then the real-value prediction is done from the peak. However, predicted angles are too narrowly distributed. In SPINE-XI, they applied conditional random field (CRF) model and it improved both angle distribution and prediction accuracy \cite{Zimmermann2017, FARAGGI20091515}.\\\\
SPIDER predicted $\theta$ and $\tau$ angles rather than 
$\phi$ and $\psi$ angles \cite{Zimmermann2017, JCC:JCC23718}. Such representation is possible because distance between neighbouring $C_{\alpha}$ atoms are almost fixed ($3.8\mathring {A}$). They also suggested another periodicity handling method that predicts sine and cosine transformed value of angles and indirectly calculating the angles by the following formula \eqref{eq:Arctangent} \cite{JCC:JCC23718}.
\begin{align}
x=\tan^{-1} [\frac{\sin(x)}{\cos(x)}]\label{eq:Arctangent}
\end{align}\\
SPIDER2, improved work of SPIDER, could improve the prediction of the secondary structure (SS) and torsion angles ($\theta$, $\tau$, $\phi$, $\psi$) by iteratively using predicted SS, angles and solvent accessible surface area (ASA) results \cite{Zimmermann2017, heffernan2015improving}.\\\\
\begin{figure}[htbp]
	\centering
	\includegraphics[width=0.7\textwidth]{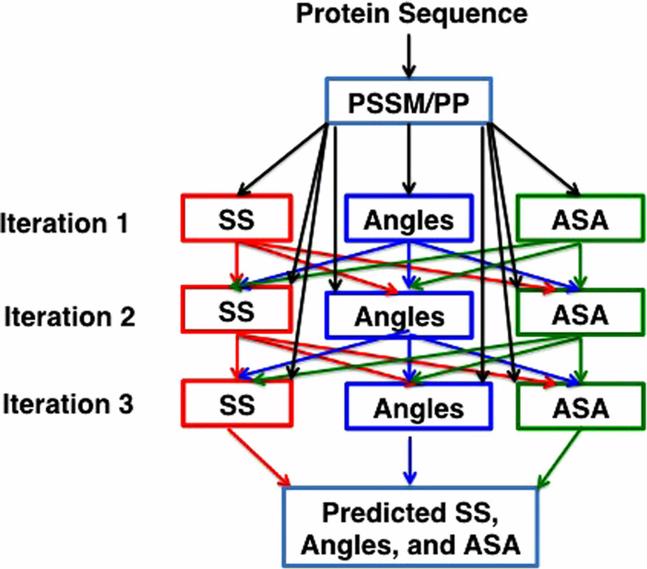}
	\caption{The general architecture of the SPIDER2 method. Image is from \cite{heffernan2015improving}.}
\end{figure}\\
Previously explained methods for angle prediction used the sliding window to make predictions. Recently, SPIDER3 utilized bidirectional recurrent neural networks (BRNNs) with long short-term memory (LSTM) cells to capture non-local interactions \cite{heffernan2017capturing}. Iterative learning is also used as the previous work \cite{heffernan2015improving}. Inputs of the model composed of seven physio-chemical properties (PP) \cite{meiler2001generation}, PSSM from PSI-BLAST \cite{altschul1997gapped}, and 30-dimensional hidden Markov model sequence profiles from HHBlits \cite{remmert2012hhblits}. Their model predicted SS, ASA, backbone angles ($\theta$, $\tau$, $\phi$, $\psi$), Half-Sphere Exposure (HSE) \cite{hamelryck2005amino} and Contact Number (CN). \\

\subsection{Other structure prediction methods}
Recently, RaptorX-Contact \cite{wang2017accurate} ranked 1st in CASP12 contact prediction \cite{PROT:PROT25407}. RaptorX-Contact used deep residual neural network \cite{he2016deep} for contact prediction by concatenating evolutionary coupling and sequence conservation information. This method also scored high quality in contact-assisted protein folding.

\section{Generative adversarial networks}
\subsection{Introduction}
Generative adversarial network (GAN) is a framework for training a generative model \cite{2014arXiv1406.2661G}. It has been successfully applied to image generation tasks like text to image generation \cite{reed2016learning, reed2016generative, 2016arXiv161203242Z, 2017arXiv170306412D}, image-to-image translation \cite{2016arXiv161107004I} which includes domain transfer \cite{kim2017learning, choi2017stargan}, image super-resolution \cite{ledig2016photo}, image inpainting \cite{pathak2016context}. It has also been used to improve the semi-supervised learning performance \cite{2016arXiv160601583O, 2017arXiv170509783D, 2016arXiv160603498S}. From these successful applications of GAN framework, we investigated applying this approach in protein dihedral angle prediction.
\\\\
A generative adversarial network is composed of two adversarial networks: a generator $G$ tries to generate samples which look realistic using random noise $z$ as input and a discriminator $D$ tries to estimate the probability that samples are coming from the data set rather than the generative model $G$, i.e., $p(S=real|x)$ where $S$ refers to source and $x$ refers to input \cite{2014arXiv1406.2661G}.\\\\ 
Input noise $p_{z} (z)$ is mapped to data space $G(z;\theta_g)$ by the generative model $G$ whose parameters are $\theta_g$. And the output of the discriminator $D(x;\theta_d)$ is the probability that $x$ is coming from the data rather than the generator's distribution $p_g$ \cite{2014arXiv1406.2661G}.
\\\\
During the training of the GAN, both $G$ and $D$ are trained simultaneously \cite{2014arXiv1406.2661G}. \\\\
The training process can be considered as playing a two-player minimax game with the following objective function $V(D,G)$ \cite{2014arXiv1406.2661G}:
\begin{align}
\underset{G}{\min} \,\underset{D}{\max}\, V(D,G) = \mathbb{E}_{x \sim p_{data}(x)} [\log{D(x)}] + \mathbb{E}_{z \sim p_{z}(z)}[\log{(1-D(G(z)))  }]
\end{align}
\begin{figure}[htbp]
	\centering
	\includegraphics[width=0.3\textwidth]{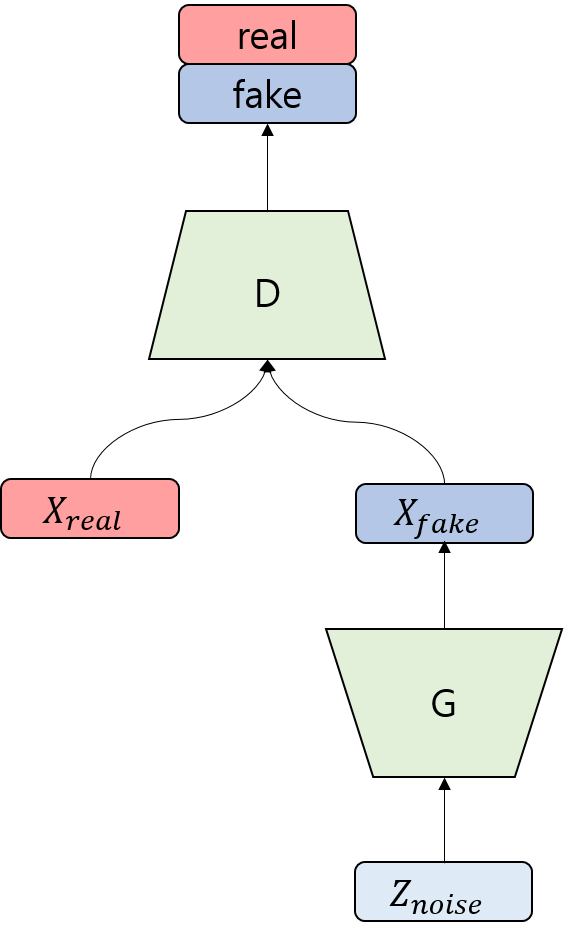}
	\caption{Architecture of the generative adversarial network}
\end{figure}\\
Any differentiable function can model the networks \cite{2014arXiv1406.2661G}. %Deep convolutional GANs (DCGAN) used convolutional network to generate images \cite{}.
 
\subsection{Problems and suggested solutions for training of generative adversarial networks}
%Training of GANs can be considered as finding a Nash equilibrium \footnote{•} of two-player non-cooperative game \cite{2016arXiv160603498S}. \textcolor{red}{Here, I'm planning to briefly explain about Nash equlibruim.} Unfortunately, finding Nash equilibrium of such game is difficult. And 
As explained in \cite{2014arXiv1412.6515G}, 
the generator is not guaranteed to converge on its optimal solution when the object function of it is non-convex.\\\\
One problem is instability occurring during the training of GANs. And another problem is mode collapse which means the generator only generates highly probable similar samples to fool the discriminator. Once mode collapse occurs, the discriminator is trained to classify such a mode as coming from the generator. However, it cannot easily distinguish samples from real data if such a mode is actually highly probable in the data distribution \cite{2016arXiv160603498S}.\\\\
Feature matching is introduced to resolve the instability of GAN training by changing the object of the generator. The generator tries to match the statistics of features on an intermediate layer. The expected value of the features on an intermediate layer is used in the suggested paper, and thus the new object for the generator is $\left\| \mathbb{E}_{x \sim p_{data}(x)} f(x) - \mathbb{E}_{z \sim p_{z}(z)} f(G(z))\right\| $ (where $f(x)$ denotes the activation on an intermediate layer of a discriminator) \cite{2016arXiv160603498S}. On the other hand, McGAN matches mean and covariance feature statistics \cite{2017arXiv170208398M}.\\\\
Minibatch discrimination is introduced to circumvent mode collapse. Minibatch discrimination uses a discriminator that also takes into account the combination of samples rather than only discriminating each sample separately. In their implementation of minibatch discrimination, the activation of an intermediate layer of a discriminator $f(x_i) \in \mathbb{R}^{A}$ is multiplied by a tensor $T\in \mathbb{R}^{A \times B \times C}$ which results in a matrix $M_i \in \mathbb{R}^{B \times C}$. Then the negative exponential of the $L_1$ distance between the rows of the matrix across samples $i \in \left\{  1, \dots ,n \right\} $ is computed which results $c_b (x_i, x_j ) =exp (-\left\| M_{i,b}-M_{j,b}\right\|_{L1}) \in \mathbb{R}$. The output $o(x_i)$ of the minibatch layer for sample $x_i$ is then calculated as follows \cite{2016arXiv160603498S}.
\begin{align*}
o(x_i)_b &= \sum_{j=1}^{n} {c_b (x_i, x_j)} \in \mathbb{R}\\
o(x_i) &= \left[o(x_i)_1, \dots , o(x_i)_B \right] \in \mathbb{R}^B, \, o(X) \in \mathbb{R}^{n \times B}
\end{align*}
Then intermediate activation $f(x_i)$ and minibatch output $o(x_i)$ are concatenated to be fed to the next layer of the discriminator. The discriminator is then outputting whether a sample cames from the data set or from the generator like the usual GANs except it can use $o(x_i)$ as side information. They could generate visually appealing images by this process \cite{2016arXiv160603498S}. \\\\
 
\subsection{Metrics for generative adversarial networks}
Several evaluation measures were used for evaluation of generative adversarial networks. We will explain some metrics that do not require human annotators.\\\\
Parzen window-based log-likelihood estimation is used to evaluate the performance of a model since the arrival of the generative adversarial networks \cite{2014arXiv1406.2661G}. The density of the data is estimated by Parzen window density estimation using generated samples only or by also adding some training samples, and then log-likelihood is calculated for test data. Parzen window estimation cannot capture true log-likelihood of the model when the dimension of the data is high even if a large data set is fed \cite{2015arXiv151101844T}. Hence, Parzen window estimation based measure may not be the best measure for high dimensional data.\\\\
Inception score $exp(\mathbb{E}_{\boldsymbol{x}} KL(p(y|\boldsymbol{x})  \parallel p(y)))$ is a measure of GANs when corresponding label $y$ exists for $\boldsymbol{x}$. Inception model \cite{2015arXiv151200567S} is trained to get the conditional label distribution $p(y|\boldsymbol{x})$. It assumes samples that contain meaningful information should have a conditional label distribution $p(y|\boldsymbol{x})$ with low entropy and marginal distribution $\int{ p(y|\boldsymbol{x} = G(z)) dz}$ should have high entropy as we expect good generator to generate diverse samples \cite{2016arXiv160603498S}. Higher inception score indicates better performance of the generator. However, as noted in \cite{2018arXiv180101973B} one can make a generator that achieves nearly optimal Inception score by outputting adversarial examples \footnote{Adversarial examples are perturbated images after applying small perturbation on the input images to change model's prediction \cite{2013arXiv1312.6199S}}, and thus it should only be used as a rough guide to evaluate the performance of GANs \cite{2016arXiv160603498S, 2018arXiv180101973B}.\\\\
Diversity of the generated images can be measured by mean multi-scale structural similarity (MS-SSIM) \cite{2016arXiv161009585O}. MS-SSIM is a variant of single scale structural similarity \cite{1292216}. %\textcolor{red}{Here, I might need to briefly explain the meaning of structural similarity.} 
Higher MS-SSIM values mean images are perceptually similar. Samples from classes that are high in diversity are expected to have low mean MS-SSIM scores \cite{2016arXiv161009585O}.\\\\
Generative adversarial metric (GAM) is introduced to measure the comparative strength of GANs. Let that there are two Generative adversarial networks $M_1 = \left\{ (G_1, D_1) \right\}, \, M_2 = \left\{ (G_2, D_2) \right\}$ where $G_1$ and $D_1$ are adversarially trained and same for the $G_2$ and $D_2$. At test time, the following two ratios are calculated to compare their strength \cite{2016arXiv160205110J}. 
\begin{align}
r_{test} = \frac{\epsilon (D_1 (x_{test}))}{\epsilon (D_2 (x_{test}))} \text{ and  } r_{sample} = \frac{\epsilon (D_1 (G_2(z)))}{\epsilon (D_2 (G_1(z)))}
\end{align}
 where $\epsilon(.)$ is the classification error rate, and $x_{test}$ is the test set. They proposed to judge the winning model considering sample ratio $r_{sample}$ and test ratio $r_{test}$ as follows: 
\begin{align}
\text{Winner} = \begin{cases}  M_1 \text{  if } r_{sample}<1 \text{ and } r_{test} \simeq 1\\ M_2 \text{  if } r_{sample}>1 \text{ and } r_{test} \simeq 1  \\ \text{Tie  otherwise} \end{cases}
\end{align}
The reason for also using the $r_{test}$ is due to the possibility of biased sample ratio $r_{sample}$. For example, one of the discriminators could be overfitted on the training data \cite{2016arXiv160205110J}.\\
\subsection{Conditional generative adversarial networks (C-GAN)}
Conditional generative adversarial network (C-GAN) generates samples by putting additional information $y$ into both generator and discriminator of a GAN \cite{2014arXiv1411.1784M}. $y$ could be class labels \cite{2014arXiv1411.1784M} or text embedding \cite{2016arXiv161203242Z} or images \cite{2016arXiv161107004I}.\\\\
Training process can be considered as playing two-player minimax game with the following objective function $V(D,G)$ \cite{2014arXiv1411.1784M}:
\begin{align}
\underset{G}{\min} \,\underset{D}{\max}\, V(D,G) = \mathbb{E}_{x \sim p_{data}(x)} [\log{D(x|y)}] + \mathbb{E}_{z \sim p_{z}(z)}[\log{(1-D(G(z|y)))  }]
\end{align}
\\Architecture of the conditional generative adversarial network is depicted in figure \ref{fig:CGAN}.
\subsection{Auxiliary classifier generative adversarial networks (AC-GAN)} \label{AC-GAN explanation}
\begin{figure}[htbp]
	\centering
	\begin{subfigure}[b]{0.275\textwidth}
        \includegraphics[width=\textwidth]{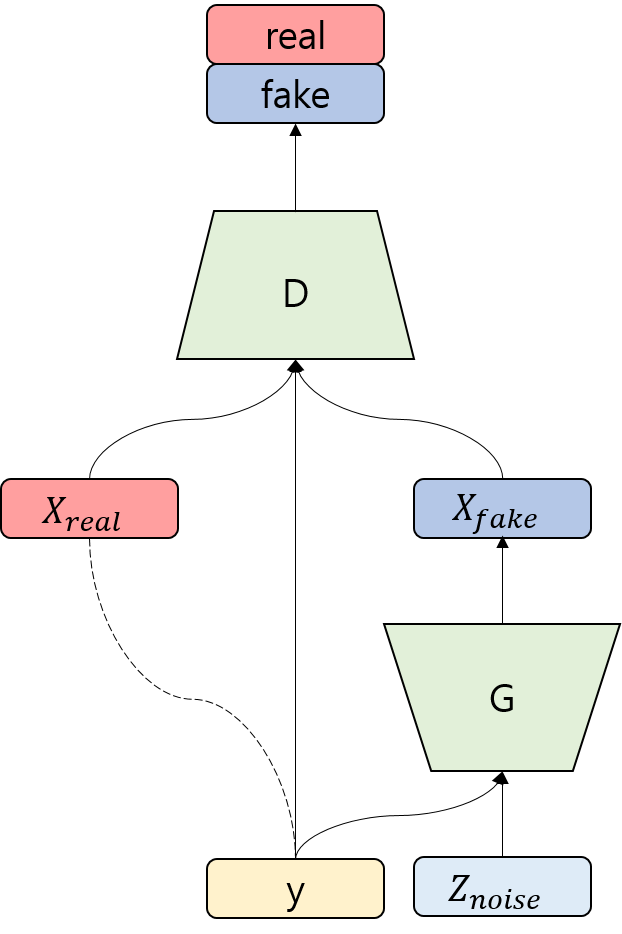}
        \caption{Conditional generative adversarial network (C-GAN)}
        \label{fig:CGAN}
    \end{subfigure}	\hspace{0.05\textwidth}
    \begin{subfigure}[b]{0.275\textwidth}
        \includegraphics[width=\textwidth]{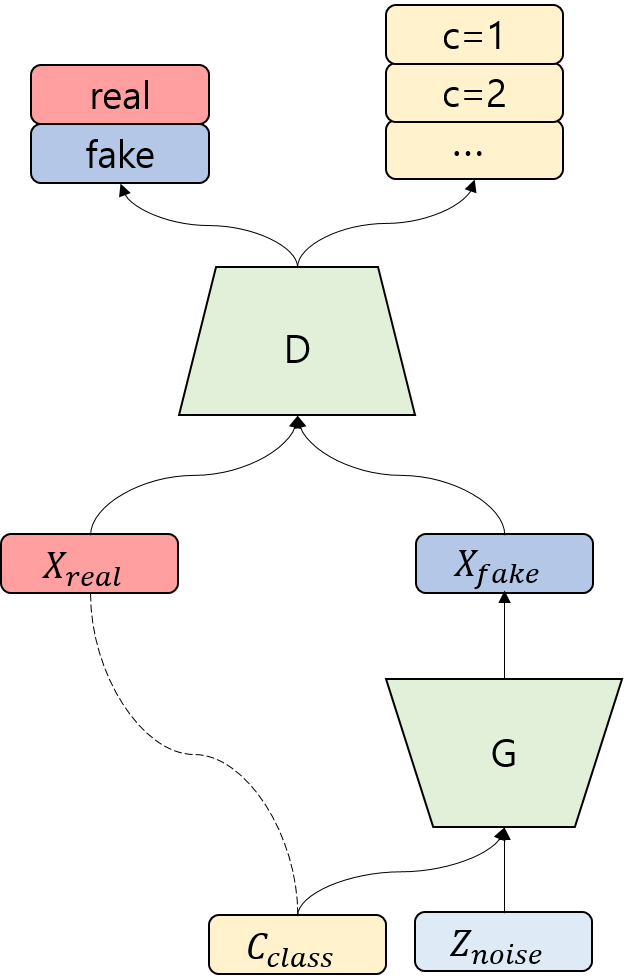}
        \caption{Auxiliary classifier generative adversarial network (AC-GAN)}
        \label{fig:ACGAN}
    \end{subfigure} \hspace{0.05\textwidth}
    \begin{subfigure}[b]{0.275\textwidth}
        \includegraphics[width=\textwidth]{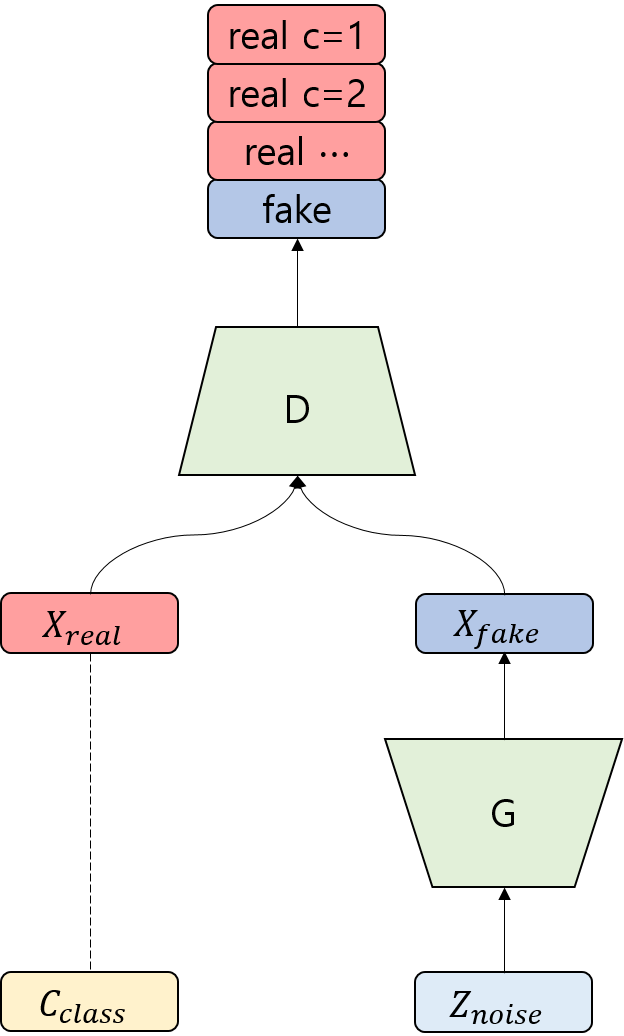}
        \caption{Semi-supervised generative adversarial network (Semi-supervised GAN)}
        \label{fig:SemisupervisedGAN}
    \end{subfigure}	
    \caption{Comparsion of generative adversarial network architectures. Dashed lines indicate correspondence relation between data and its label. Some samples would not have corresponding labels in the Semi-supervised GAN}\label{fig:Compare GANs}
\end{figure}

In auxiliary classifier generative adversarial networks (AC-GAN), architecture of the generator will be same as the C-GAN except the fact that additional information $y$ should be a categorial class label $c$. Unlike conditional generative adversarial network, the discriminator of AC-GAN only gets samples and estimates both a probability distribution over sources $p(S | X)$ and a probability distribution over the class labels $ p(C | X)$ \cite{2016arXiv161009585O}.\\\\
The object function is composed of two parts: the log-likelihood of the
correct source, $L_S$, and the log-likelihood of the correct class, $L_C$. Note that $X_{fake} = G(c, z)$ \cite{2016arXiv161009585O}.
\begin{align}
&L_S = \mathbb{E}[\log p(S = real \,| \,X_{real})] + \mathbb{E}[\log p(S = fake \,| \,X_{fake})]\\
&L_C = \mathbb{E}[\log p(C = c \,| \,X_{real})] + \mathbb{E}[\log p(C = c \,| \,X_{fake})]
\end{align}
The discriminator $D$ maximizes $L_S + L_C$ and the generator $G$ maximizes $L_C - L_S$ during the training \cite{2016arXiv161009585O}. \\\\%i. e. $G$ maximizes $ \mathbb{E}[\log P(C = c \,| \,X_{fake})]-\mathbb{E}[\log P(S = fake \,| \,X_{fake})] = ? $. 
The discriminator maximize $\mathbb{E}[\log p(S = real \,| \,X_{real})] +\mathbb{E}[\log p(C = c \,| \,X_{real})]=\mathbb{E}[\log (p(S = real \,| \,X_{real})p(C = c \,| \,X_{real}))]$ for $X_{real}$ and $\mathbb{E}[\log p(S = fake \,| \,X_{fake})]+\mathbb{E}[\log p(C = c \,| \,X_{fake})]=\mathbb{E}[\log (p(S = fake \,| \,X_{fake})p(C = c \,| \,X_{fake}))]$ for $X_{fake}$. This can be interpreted as the discriminator assuming predicted classes and sources are conditionally independent on the inputs.\\\\
Architecture of the auxiliary classifier generative adversarial network is depicted in figure \ref{fig:ACGAN}.
\subsection{Semi-supervised generative adversarial networks (Semi-supervised GAN)}
We focus on the semi-supervised generative adversarial network whose discriminator $D$ outputs $N+1$ classes when there are $N$ possible classes for data \cite{2016arXiv160603498S, 2016arXiv160601583O}. Output of the discriminator $p_{model}(c|\boldsymbol{x})$ indicates the probability that $\boldsymbol{x}$ is coming from the corresponding class $[Class-1,\, Class-2,\, ...,\, Class-N, \,Fake]$ \cite{2016arXiv160601583O}.\\\\
Training loss of the discriminator is consist of two parts: 
the classification loss for labeled data, 
$L_{supervised}$, and unsupervised loss $L_{unsupervised}$ which is equivalent to standard GAN loss \cite{2016arXiv160603498S}.
\begin{align}
L_{supervised} = - &\mathbb{E}_{\boldsymbol{x},c \sim p_{data} (\boldsymbol{x},c)} [\log p_{model} (c | \boldsymbol{x},c<N+1)] \\
L_{unsupervised} =-& \left\{ \mathbb{E}_{\boldsymbol{x} \sim p_{data} (x)}[ \log (1-p_{model} (c=N+1 | \boldsymbol{x}))] \right. \nonumber\\
 +& \left.  \mathbb{E}_{z \sim p_{z} (z)}[ \log p_{model} (c=N+1|G(z))] \right\}
\end{align}
The discriminator $D$ minimizes $L_{supervised} + L_{unsupervised}$ and the generator $G$ minimizes $\mathbb{E}_{z \sim p_{z} (z)}[ \log p_{model} (c=N+1|G(z))]$ during training \cite{2016arXiv160603498S}.\\\\
Analysis of semi-supervised generative adversarial network showed that theoretically we need a complement generator which generates complement samples in the feature space to improve the performance of the classifier \cite{2017arXiv170509783D}. \\\\
When we compare the discriminators of
AC-GAN with Semi-supervised GAN, we notice that unlike the discriminator of the former model, Semi-supervised GAN has a discriminator that is combined with the classifier. Because of this discriminator and the classifier interact each other and as a result, they would help to improve their counterpart \cite{2016arXiv160601583O}. From this reasoning, we hypothesized that we might be able to improve the generating performance of the AC-GAN by combining the classifier and the discriminator. %In section $\Rmnum{2} .2$, we developed corresponding model and result was compared with the residue-wise variant of the AC-GAN.\\

\section{Noise-contrastive estimation}
Noise-contrastive estimation (NCE) is introduced to estimate the probability distribution of data by estimating parameters of unnormalized models \cite{pmlr-v9-gutmann10a}.\\\\
Unknown probability density function of data $p_{data} (u)$ can be modeled by a model $p_{model} (u;\theta)$ where $\theta=\left\{\alpha, c\right\} $ denotes the vector of parameters and $c$ is a negative logarithm of the normalization parameter, i.e., $c=-\ln(\int{p_{model} (u;\alpha) du ) }$ and thus $p_{model} (u;\theta) = e^{c} \, p_{model} (u;\alpha)$ for an unnormalized model $p_{model} (u;\alpha)$ \cite{pmlr-v9-gutmann10a}.\\\\
Let's say $X=\left\{x_1,\dots, x_T\right\}$ is the set of the observed data and $Y=\left\{y_1,\dots, y_T\right\}$ is the set of the samples drawn from the known noise distribution $p_{noise}$. Noise-contrastive estimation optimizes the parameter vector $\theta$ to maximize the following object function:\\
\begin{align}
J_T (\theta) = \frac{1}{2T} \sum_{t} {\ln [h(x_t ; \theta)] + \ln [1 - h(y_t ; \theta)]}
\end{align}
 where $h(u;\theta)=\frac{1}{1+exp(-G(u;\theta))}$ and $G(u;\theta) = \ln p_{model} (u;\theta) - \ln p_{noise} (u)$ \cite{pmlr-v9-gutmann10a}.\\\\
This method is similar to GAN in that it is trained to discriminate samples from two distributions. However, the generative model itself is trained to discriminate and NCE uses fixed noise distribution comparision while GANs use a generator which keeps changing \cite{2014arXiv1406.2661G}. Another important difference is that NCE can output the estimated density of the data as we use known noise distribution to compare while this is not possible for the generative adversarial networks due to lack of known distribution.\\\\

\chapter{Materials and methods}
\section{Dataset}\label{dataset}
We obtained our dataset via a similar process to that described in \cite{JCC:JCC23718}, however the dataset is different. Proteins with less than $25\%$ sequence similarity, high resolution ($<2.0 \mathring{A}$), sequence length between 40 and 700 and R-value better (smaller) than 0.25 were retained by PISCES \cite{wang2003pisces}. Biopython library \cite{cock2009biopython} was used to parse sequences and DSSP \cite{kabsch1983dictionary} was used for dihedral angle calculation. Only 20 standard amino acids were considered and window size 17 like SPIDER2 \cite{heffernan2015improving} was used to get sequence fragments. 4590 protein chains and 1190 protein chains were obtained for training and test set, respectively. 845521 fragments and 298878 fragments were obtained for training and test set, respectively. To reduce computation time, only a randomly selected sample of 25000 fragments were used for most of the testing except for predicting angles for each central amino acid and calculation of the conditional log-likelihoods. \\
\section{Utilized models in the experiments}
The regression model, which predicts $\phi$, $\psi$ angles, using only regression loss which is shown in equation \ref{eq:Regression loss} \cite{JCC:JCC23718}, was used as a baseline model for the angle prediction.\\
\begin{align}
{L}_{\text{Regression}}&=\mathbb{E}[\left(\cos{({\phi}_{real})}-\cos{({\phi}_{predicted})}\right)^{2}+\left(\sin{({\phi}_{real})}-\sin{({\phi}_{predicted})}\right)^2 \nonumber\\
&+\left(\cos{({\psi}_{real})}-\cos{({\psi}_{predicted})}\right)^{2}+\left(\sin{({\psi}_{real})}-\sin{({\psi}_{predicted})}\right)^2] \label{eq:Regression loss}
\end{align}\\
Noise-contrastive estimation (NCE) model was used as a baseline model for density estimation. As we don't need to generate samples from NCE model, we used usual deep learning classifier for density estimation. Noise samples for density estimation were drawn from a uniform distribution.\\\\
Three generative adversarial network models; conditional generative adversarial network (C-GAN), auxiliary classifier generative adversarial network (AC-GAN) and semi-supervised generative adversarial network (Semi-supervised GAN), were used in order to compare their characters and to see the changes result from possible improvements. NCE-GANs, which will be introduced in section \ref{NCE-GAN explanation}, were used in the generative models to track the estimated densities. Residue-wise AC-GAN and Residue-wise Semi-supervised GAN, which will be introduced in section \ref{Residue-wise variants explanation}, were used in AC-GAN and Semi-supervised GAN to utilize all sequence information in a window.\\\\
\section{Architecture of neural networks}
We used the same architectures for regression model and the generators. The common architecture was a multi-layered perceptron composed of 3 hidden layers and 150 neurons for each layer as the architecture of SPIDER \cite{JCC:JCC23718}. We made predictions using one-hot encoded sequence information as input and no input noise was added in the generators as we are doing deterministic angle prediction, but predicted angles were added in some of the experiments. Leaky rectifier linear unit (LReLu) with a=5 \cite{maas2013rectifier} was used as activation function except for the output activation function. Ranges for predicted $\phi$, $\psi$ angles were $(-\pi, \pi)$. $\pi$ was multiplied to fit the range after softsign activation function was applied to the output layer. No normalization was used for both input and output of the generators except when predicted angles by regression model were added in some of the experiments. Scaling was applied when predicted angles are fed into the generators. And only $\phi, \psi$ angles in the central residue of the window were predicted for window size 17. Shifted angle method \cite{PROT:PROT21940} was used to handle the periodicity of angles. $\phi$ was shifted $\pi$ ($90^{\circ}$) and $\psi$ was shifted around $- 1.40$ ($-80.2^{\circ}$) by calculating proper shift from training data. \\\\
Like the regression model and the generators, multi-layered perceptron composed of 3 hidden layers and 150 neurons for each layer were used in the discriminators which also include NCE model. This implies weights of the auxiliary classifier and the discriminators were shared for AC-GAN, and also the weights of the classifier and the discriminator of the Semi-supervised GAN were shared. Shifted real $\phi$, $\psi$ angles and noise samples were fed into the noise-contrastive estimation (NCE) model. Sequence information and scaled angle information were concatenated, and this combined information was fed into the discriminator of the C-GAN. Only angle information was fed into the discriminators of AC-GAN and Semi-supervised GAN except some experiments. NCE-GAN structure, which will be described in section \ref{NCE-GAN explanation}, was used in the discriminators of GANs for evaluation. Note that, in some of the experiments, the output of the NCE model was also fed into GAN-discriminators. NCE model predicts whether it got the input from the real angles or the noise samples and it also outputs predicted sequence information like Residue-wise Semi-supervised GAN, which will be introduced in section \ref{Residue-wise variants explanation}.\\\\
We used Tensorflow library to train our models \cite{abadi2016tensorflow}. Batch size was 64 and Adam weight optimization method \cite{kingma2014adam} was used.
%\textcolor{red}{I think it would be better to relocate the following paragraphs, but I don't know how to do this.}\\
%For convenience, we will say a model which was trained using regression loss as a regression model and a model which was trained with adversarial loss will be called a generative model or GAN model. Note that training loss of a GAN model could also contain regression loss.\\
\section{Comparison of conditional generative adversarial netowrk methods and suggested improvements} \label{Compare GAN}
To find a model whose Ramachandran plot of a predicted angles looks realistic, three generative adversarial networks models (conditional GAN, AC-GAN and Semi-supervised GAN) were implemented and the results were compared. Note that NCE-GAN in \cref{NCE-GAN explanation} were used for all models to compare their performances. \\\\
In the first experiment, to check the problems of original setting and to compare the results with experiments done with improved condition, we did not add regression loss in the generators and noise class ignoring method, which will be explained in section \ref{NCE-GAN explanation}, was used for training loss of the generators.\\\\
As can be seen in the result of the first experiment \ref{Without regression loss}, training of the discriminators was not stable and estimated $p_{model}(x|real)$ of Semi-supervised GAN did not show detailed shape of the Ramachandran plot. In the second experiment, in an effort to stabilize the training of the discriminator, we added predicted density estimation by NCE model as an additional input of the discriminators. This idea is similar to \cite{heffernan2015improving, 2016arXiv161203242Z} in a way that we are using predicted information as an additional input of the generators. NCE model was pre-trained for 50000 iterations before training of the generative models started. One thing to be considered when using the predicted information as an input is that in order to prevent the generator generating NAN values. Hence, we penalized generating angles in regions near the prediction boundary even though we still encountered NaN values sometimes. The other settings were same with the first experiment.\\\\
To see the effects of minibatch-wise generation, which will be explained in \cref{Analysis MB}, we used minibatch-wise loss in the generators of the third experiment. Loss of C-GAN and AC-GAN generators were replaced by minibatch-wise loss. As the discriminator and the classifier was combined in Semi-supervised GAN, minibatch-wise loss was added to the loss of Semi-supervised generator. The other settings were same with the second experiment.\\\\
In all of the three experiments, MSE and MAE values of the generative models were much higher than regression model. However, reduction of MAE values are necessary for accurate structure prediction. Hence, in the fourth experiment, we added regression loss \eqref{eq:Regression loss} in the loss of the generators in an attempt to reduce the MSEs and MAEs. The other settings were same with the third experiment.\\\\%We also tried different loss like l1 loss of cosine and sine values of angles and l2 loss of angles, but the results were bad in terms of MSE, Parzen window based log-likelihood and Ramachandran plot. (Results were not shown.) \\
%Even though we could see some improvement in the Ramachandran plot we can not get nice Ramachandran plot and comparable MAE values with the regression model at the same time. 
To further reduce the MSE and MAE values, in the last experiment, we used predicted results by regression model as an additional inputs of the generative model like the idea of improving performance by stacking results \cite{heffernan2015improving, 2016arXiv161203242Z}. Regression model was pre-trained for 50000 iterations before training of the generative models started. The other settings were same with the fourth experiment.\\\\

\section{Estimating density of GAN models by combing with NCE} \label{NCE-GAN explanation}
Data distribution $p_{data}$ and distribution of the generated samples $p_{G}$ are intractable in generative adversarial nets as we don't know the density function of either. That's why methods like Parzen window score and Inception score \cite{2016arXiv160603498S} are used to measure the performance of the generative adversarial models. However, we can estimate their density function simply by adding samples from known noise distribution and adding noise distribution class to the discriminator. (Note that noise samples to be discriminated from real samples and fake samples are different from noise samples which will be fed into the generator to generate fake samples.) If an additional class is added, the output of the discriminator will be $[real,\,fake, \, noise]$. The idea is used with the noise-contrastive estimation \cite{pmlr-v9-gutmann10a}. We call this model as noise-contrastive estimation generative adversarial networks or simply NCE-GAN. NCE-GAN can also be applied to AC-GAN and Semi-supervised GAN. Architecture of the noise-contrastive estimation generative adversarial network is depicted in figure \ref{fig:NCE-GAN}.\\\\
The discriminator of the NCE-GAN maximizes the following:
\begin{align}
\mathbb{E} [\log p_{model} (S=real |x_{real})]
+\mathbb{E} [\log p_{model} (S=fake |x_{fake})]
+\mathbb{E} [\log p_{model} (S=noise |x_{noise})] \label{eq:NCE-GAN loss}
\end{align} 
where $x_{real}$ is a sample from the data set, $x_{fake}$ is a sample generated by the generator and $x_{noise}$ is a sample drawn from noise distribution.\\\\
The generator is trained to maximize the $\mathbb{E} [\log p_{model} (S=real |x_{fake})]$.\\
\begin{figure}[htbp]
	\centering
	\includegraphics[width=0.3\textwidth]{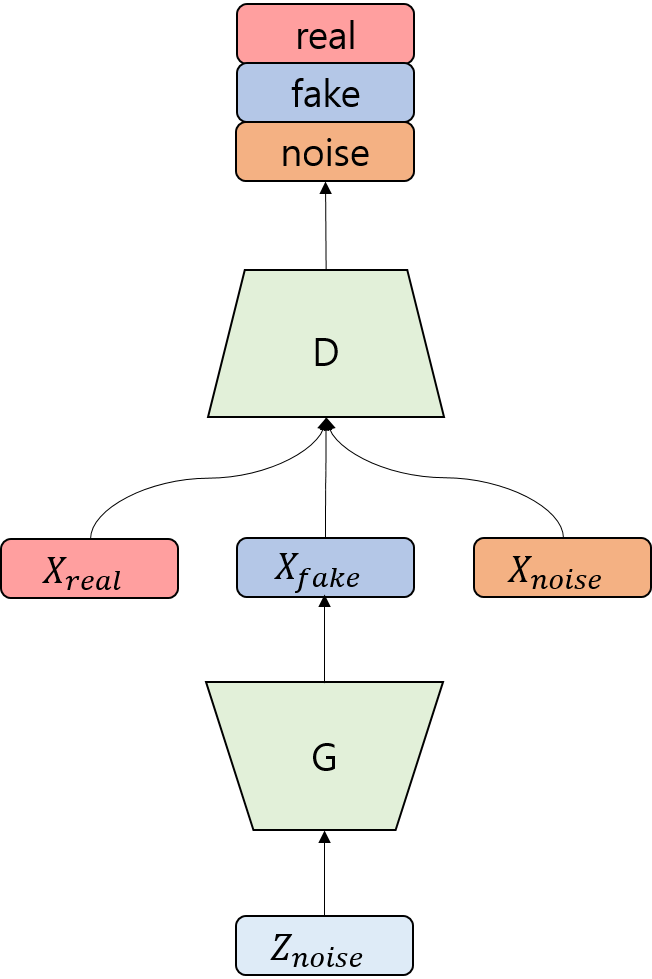}
	\caption{Architecture of the noise-contrastive estimation generative adversarial network. Noise samples to be discriminated $X_{noise}$ should have same shape with the data samples $X_{real}$ while this is not necessary for noise samples which will be fed into the generator $Z_{noise}$.}
\end{figure}\label{fig:NCE-GAN}\\
This model has some advantages over other GANs.
We can generate samples when target distribution is given without requiring another model \cite{2017arXiv170509783D} (see appendix \ref{Appendix generate samples} for details). As we can estimate both densities we can easily measure the log-likelihood of the models and numerically estimated Kullback-Leibler (KL) divergence between data and samples of generator distribution. This would tell us how well the generator mimics estimated real data distribution. And feeding fixed noise samples to the discriminator could possibly stabilize the training of the discriminator.\\\\
%One thing to be careful is we don't want the generator to be affected by noise distribution. i.e, loss of the generator should ignore the noise class of the discriminator.
However as the discriminator also distinguish samples from noise distribution, the generator will be effected by this. For example, once mode collapse occurs in the generator and if routes to go to other modes are surrounded by noise samples, then it will not easy for the the generator to get out of the mode collapse. Hence, other methods need be applied to resolve this problem. One method involves defining the object function of the generator as if there is no the noise class in the discriminator. Let $[l_{real} (x),\,l_{fake} (x), \, l_{noise} (x)]$ be an output of the discriminator before applying exponentiation and normalization (It is sometimes called a logit of a softmax classifier.). In this case, usual object function for the generator will be $\mathbb{E} [\log \frac { e^{l_{real}(x_{fake})} }{e^{l_{real}(x_{fake})}+e^{l_{fake}(x_{fake})}+e^{l_{noise}(x_{fake})}  } ]$. If we ignore the noise class, the object function of the generator will be $\mathbb{E} [\log \frac { e^{l_{real}(x_{fake})} }{e^{l_{real}(x_{fake})}+e^{l_{fake}(x_{fake})} } ]$. Another way to handle this problem is generating samples by taking into account the samples in a batch. This method is explained in \cref{Analysis MB}.\\\\ 
%(Occasionally labelling noise samples as data samples might helps)
We now describe density estimation for each $c$ when training of NCE-GAN is finished. We think about a case where the discriminator classifies its input into some classes whose distribution is unknown and where there is also one noise class. The classes of unknown distributions can include classes of real data and a class for samples generated by an adversarial generator. We explain how can we estimate density function for a class $c$ when we know the noise distribution $p(x|C=noise)$. $p(x|C=noise)$ should be nonzero whenever $p(x|C=c')$ is nonzero for all class $c' \neq noise$ \cite{pmlr-v9-gutmann10a}.
 We assume we know all prior probabilities for each class $c$, i.e., $p(C=c)$ is known and class probabilities are represented by softmax function, i.e., $p_{model}(C=c|x) = \frac{\exp(l_c (x))}{\sum_{c'} {\exp(l_{c'} (x))}}$. And we assume the discriminator can classify samples well even though the generator was not fixed during training process. In other words, $p_{model}(C=c|x)\approx p(C=c|x)$. We also asssume $p_{model}(C=c)\approx p(C=c)$.\\\\
By Bayes' theorem, $p(x|C=c)=\frac{p(C=c|x) p(x)}{p(C=c)}$ holds. \\\\
%\textcolor{red}{Please note that I need to change $=$ in some parts of the following equations to $\approx$. Or, it might be better to rewrite.}
By reformulating the formula and focusing on the noise class of the model we get:\\\\
\begin{align}
p_{model}(x) = \frac{p_{model}(x|C=noise) p_{model}(C=noise)}{p_{model}(C=noise|x)} \label{eq:model noise class}
\end{align} \\\\
 Note that we can estimate $p_{model}(x)$ if we substitute $p_{model}(C=noise)$ with $p(C=noise)$ and $p_{model}(C=noise|x)$ with $p(C=noise|x)$ in the right hand side of the equation \eqref{eq:model noise class}.\\\\
 Then for each class $c$:
\begin{align}
p_{model}(x|C=c)=& \frac{p_{model}(C=c|x) p_{model}(x)}{p_{model}(C=c)}\\
=& \frac{p_{model}(C=c|x) p_{model}(x|C=noise) p_{model}(C=noise)}{p_{model}(C=c)p_{model}(C=noise|x)} \, ( \because \text{equation \ref{eq:model noise class}})\\
=& \frac{\frac{\exp(l_c (x))}{\sum_{c'} {\exp(l_{c'}(x))}} p_{model}(x|C=noise) p_{model}(C=noise)}{p_{model}(C=c)\frac{\exp(l_{noise}(x))}{\sum_{c'} {\exp(l_{c'}(x))}}} \, ( \because \text{softmax representation} )\\
=& p_{model}(x|C=noise) \frac{\exp(l_c (x))}{\exp(l_{noise} (x))} \frac{ p_{model}(C=noise)}{ p_{model}(C=c)}
\end{align}
We can set $\exp(l_{noise} (x))=p(x|C=noise) $ and then we get more simplified formula:
\begin{align}
p_{model}(x|C=c) &= \exp(l_c (x)) \frac{p_{model}(x|C=noise)}{p(x|C=noise)} \frac{ p_{model}(C=noise)}{ p_{model}(C=c)} \\
& \approx \exp(l_c (x)) \frac{ p_{model}(C=noise)}{ p_{model}(C=c)} \, ( \because p_{model}(C=noise|x)\approx p(C=noise|x) )\\
& \approx \exp(l_c (x)) \frac{ p(C=noise)}{ p(C=c)} \, ( \because p_{model}(C=c)\approx p(C=c)) \label{eq:density by NCE}
\end{align}\\
 As we know all terms in the right hand side of equation \eqref{eq:density by NCE}, we can estimate density function for each class $c$.\\\\

\section{Analysis on minibatch discrimination} \label{Analysis MB}
%\textcolor{red}{It might be better to add hyperlinks of some equations in this section.}\\
Even though suggested minibatch discrimination method enabled the generator to successfully generate visually appealing images \cite{2016arXiv160603498S}, this method will encounter two problems when density estimation is needed.\\\\
First, although their discriminator outputs a single number for each sample \cite{2016arXiv160603498S}, the outputs of a batch will be dependent on the other samples in the batch and thus such result can not be used for density estimation. Applying minibatch discrimination to the noise-contrastive estimation \cite{pmlr-v9-gutmann10a} shows this problem (see appendix \ref{Appendix minibatch discrimination} for details). In terms of density estimation, adding minibatch information into the discriminator would be a cheating behaviour. To prevent this problem in density estimation the discriminator should not get batch information. \\\\ 
%We can handle minibatch information by making another minibatch discriminator to give feedback to the generator.\\\\
Second, as the discriminator models closeness \cite{2016arXiv160603498S} it requires $O(n^2)$ computation time and memory. And this is unwanted when evaluation of large number of test samples is needed. 
The following analysis explains mini-batch discrimination might not help the discriminator to correctly classify a minibatch samples, and thus $O(n^2)$ computation complexity would be unnecessary for the discriminator. It is important to note that this doesn't mean mini-batch discrimination can not help the generator.\\\\
If we think about vanilla GANs \cite{2014arXiv1406.2661G}:
\begin{align} 
p_{model}(S=real|x) &\approx p(S=real|x)\\
&=\frac{p(x|real)p(S=real)}{p(x)} \, (\because \text{Bayes' theorem}) \label{eq:real class Bayes's theorm}\\
&=\frac{p(x|real)}{p(x|real)+p(x|fake)} \, (\because p(S=real)=p(S=fake)) \\
&=\frac{1}{1+\frac{p(x|fake)}{p(x|real)}}\\
&=\frac{1}{1+\exp(-\ln(\frac{p(x|real)}{p(x|fake)}))}\\
&=\text{sigmoid}(\ln (\frac{p(x|S=real)}{p(x|S=fake)})) \label{eq:sigmoid ln}
\end{align}\\
By applying logit function which is the inverse of the sigmoid function: 
\begin{align}
\ln (\frac{p(x|S=real)}{p(x|S=fake)})\approx \text{logit}(p_{model}(S=real|x)) \label{eq:ln logit}
\end{align}\\
If we think about samples of mini-batch $x_1, \dots x_n$ which are assumed to be satisfy i.i.d. (Independently identically distributed) condition:\\
\begin{align}
p_{model}(S=real|x_1, \dots x_n ) &\approx p(S=real|x_1, \dots x_n )\\
&=\text{sigmoid}(\ln (\frac{p(x_1, \dots x_n|S=real)}{p(x_1, \dots x_n|S=fake)}))\\ 
( &\because \text{By similar process done in equation \ref{eq:real class Bayes's theorm}-\ref{eq:sigmoid ln}}) \nonumber\\
&=\text{sigmoid}(\ln (\frac{\prod_{i=1}^n {p(x_i|S=real)}}{\prod_{i=1}^n {p(x_i|S=fake)}})) \\ 
\,(&\because \text{ i.i.d. condition} ) \nonumber\\
&=\text{sigmoid}( \sum _{i=1}^n \ln (\frac{ p(x_i|S=real)}{p(x_i|S=fake)}))\\
&\approx\text{sigmoid}( \sum _{i=1}^n \text{logit}(p_{model}(S=real|x_i))) \label{eq:sum logit}\\
( &\because \text{By equation \ref{eq:ln logit}}) \nonumber
\end{align}
The equation \eqref{eq:sum logit} indicates that if the discriminator learned to correctly classify the real source of each sample, then the discriminator can also correctly classify a mini-batch of samples. And thus, we doubt the minibatch discrimination help the discriminator to correctly classify a minibatch samples. However, this doesn't necessary mean this can not help the generator to generate realistic samples. \\\\%We explain why considering minibatch samples can help the generator while this isn't applied for the discriminator.\\\\
The success of minibatch discrimination \cite{2016arXiv160603498S} in generating visually appealing samples could be explained by the fact their minibatch discriminator does point-wise discrimination also using minibatch information. Their discriminator could learn point-wise information and the corresponding generator could get minibatch-wise feedback.\\\\
Stabilizing property of feature matching \cite{2016arXiv160603498S} can also be explained similarly. In feature matching, the discriminator is trained by point-wise discrimination while the generator is trained to follow statistics of the data samples which can be considered as one form of minibatch information.\\\\
We propose a new loss of a generator which takes into account minibatch samples.\\\\
Let $[l_{real} (x),\,l_{fake} (x)]$ be a output of the discriminator before applying exponentiation and normalization when there are only two classes $[real,\,fake]$.\\\\
%We think about the loss of a generator using $\mathbb{E} [l_{real} (G(z))]$ and $\mathbb{E}[l_{fake} (G(z))]$. \\
We maximize the following formula \ref{eq:minibatch loss} rather than maximizing $\mathbb{E}[\ln(p_{model}(S=real|G(z)))]$ \newline 
$=\mathbb{E}[\ln( \frac{\exp(l_{real} (G(z)))}{\exp(l_{real} (G(z)))+\exp(l_{fake} (G(z)))})]$ in suggested object of a generator in \cite{2014arXiv1406.2661G}.\\
\begin{align}
\ln(\frac{\exp(\mathbb{E} [l_{real} (G(z))])}{\exp(\mathbb{E} [l_{real} (G(z))])+\exp(\mathbb{E} [l_{fake} (G(z))])})\label{eq:minibatch loss}
\end{align}\\
We explain why maximizing \ref{eq:minibatch loss} in the generator make the generator to take into account minibatch samples.\\\\
As $p_{model}(S=s|x)=\frac{\exp(l_{s} (x))}{\exp(l_{real} (x))+\exp(l_{fake} (x))}$, we get:
\begin{align}
\exp(l_{s} (x)) = p_{model}(S=s|x) (\exp(l_{real} (x))+\exp(l_{fake} (x)) )\label{exp logit}
\end{align}\\
If we think about generated samples of mini-batch $G(z_1), \dots G(z_n)$ which satisfy i.i.d. condition:\\
%\frac{\exp(\mathbb{E} [l_{real} (x)])}{\exp(\mathbb{E} [l_{real} (x)])+\exp(\mathbb{E} [l_{fake} (x)])}\\
\begin{align} 
& \frac{\exp(\frac{1}{n} \sum_{i=1}^n (l_{real} (G(z_i))))}{\exp(\frac{1}{n} \sum_{i=1}^n  (l_{real} (G(z_i))))+\exp(\frac{1}{n} \sum_{i=1}^n  (l_{fake} (G(z_i))))}\nonumber\\
=&\frac{\prod_{i=1}^n \exp(l_{real} (G(z_i)))^{\frac{1}{n}}}{\prod_{i=1}^n \exp(l_{real} (G(z_i)))^{\frac{1}{n}}+\prod_{i=1}^n \exp(l_{fake} (G(z_i)))^{\frac{1}{n}}}\\
=&\frac{\prod_{i=1}^n [p_{model}(S=real|G(z_i)) (*) ]^{\frac{1}{n}}}{\prod_{i=1}^n [p_{model}(S=real|G(z_i)) (*)]^{\frac{1}{n}}+\prod_{i=1}^n [p_{model}(S=fake|G(z_i)) (*)]^{\frac{1}{n}}}\nonumber\\
& (\text{where } (*)=\exp(l_{real} (G(z_i)))+\exp(l_{fake} (G(z_i))) \,)\nonumber\\
\,(&\because \text{ By equation } \ref{exp logit})\\
=&\frac{\prod_{i=1}^n [p_{model}(S=real|G(z_i)) ]^{\frac{1}{n}}}{\prod_{i=1}^n [p_{model}(S=real|G(z_i))]^{\frac{1}{n}}+\prod_{i=1}^n [p_{model}(S=fake|G(z_i))]^{\frac{1}{n}}}\\
=&\frac{\prod_{i=1}^n [\frac{p_{model}(G(z_i)|real)p_{model}(S=real) }{p_{model}(G(z_i))} ]^{\frac{1}{n}}}{\prod_{i=1}^n [\frac{p_{model}(G(z_i)|real)p_{model}(S=real) }{p_{model}(G(z_i))}]^{\frac{1}{n}}+\prod_{i=1}^n [\frac{p_{model}(G(z_i)|fake)p_{model}(S=fake) }{p_{model}(G(z_i))}]^{\frac{1}{n}}}\nonumber\\
\,(&\because \text{ By Bayes's theorem }) \\
=&\frac{p_{model}(S=real) \prod_{i=1}^n [p_{model}(G(z_i)|real)  ]^{\frac{1}{n}}}{p_{model}(S=real) \prod_{i=1}^n [ p_{model}(G(z_i)|real) ]^{\frac{1}{n}}+p_{model}(S=fake) \prod_{i=1}^n [p_{model}(G(z_i)|fake) ]^{\frac{1}{n}}}\\
=&\frac{p_{model}(S=real) (p_{model}(G(z_1), \dots G(z_n)|real) ^{\frac{1}{n}})}{ \splitfrac{ \left\{p_{model}(S=real)  ( p_{model}(G(z_1), \dots G(z_n)|real) ^{\frac{1}{n}}) \right. }{ \left. \qquad\qquad \qquad+ p_{model}(S=fake) (p_{model}(G(z_1), \dots G(z_n)|fake) ^{\frac{1}{n}}) \right\} } }\nonumber\\
\,(&\because \text{ i.i.d. condition} )\\
=&\frac{  (**) }{(**)+1} \, (\text{where } (**)= \frac{p_{model}(S=real) (p_{model}(G(z_1), \dots G(z_n)|real) ^{\frac{1}{n}})}{p_{model}(S=fake) (p_{model}(G(z_1), \dots G(z_n)|fake) ^{\frac{1}{n}})}\,) \label{eq:(**)}
\end{align}\\
Maximizing \ref{eq:minibatch loss} is equivalent to maximizing (**) in \ref{eq:(**)}. Maximizing (**) can be interpreted as generating samples considering minibatch samples. We call \ref{eq:minibatch loss} as a minibatch-wise loss of the generator. We speculated that using minibatch-wise loss in the generator might help the generator to improve its performance and experiment \ref{D Without regression loss minibatch} was done to check this hypothesis. We used the following loss in the experiment as we are using NCE-GAN structure.
\begin{align}
\ln(\frac{\exp(\mathbb{E} [l_{real} (G(z))])}{\exp(\mathbb{E} [l_{real} (G(z))])+\exp(\mathbb{E} [l_{fake} (G(z))])+ \exp(\mathbb{E} [l_{noise} (G(z))])})\label{eq:minibatch loss NCE-GAN}
\end{align}\\

\section{Residue-wise variants of AC-GAN and Semi-supervised GAN} \label{Residue-wise variants explanation}
\begin{figure}[htbp]
	\centering
	\begin{subfigure}[b]{0.375\textwidth}
        \includegraphics[width=\textwidth]{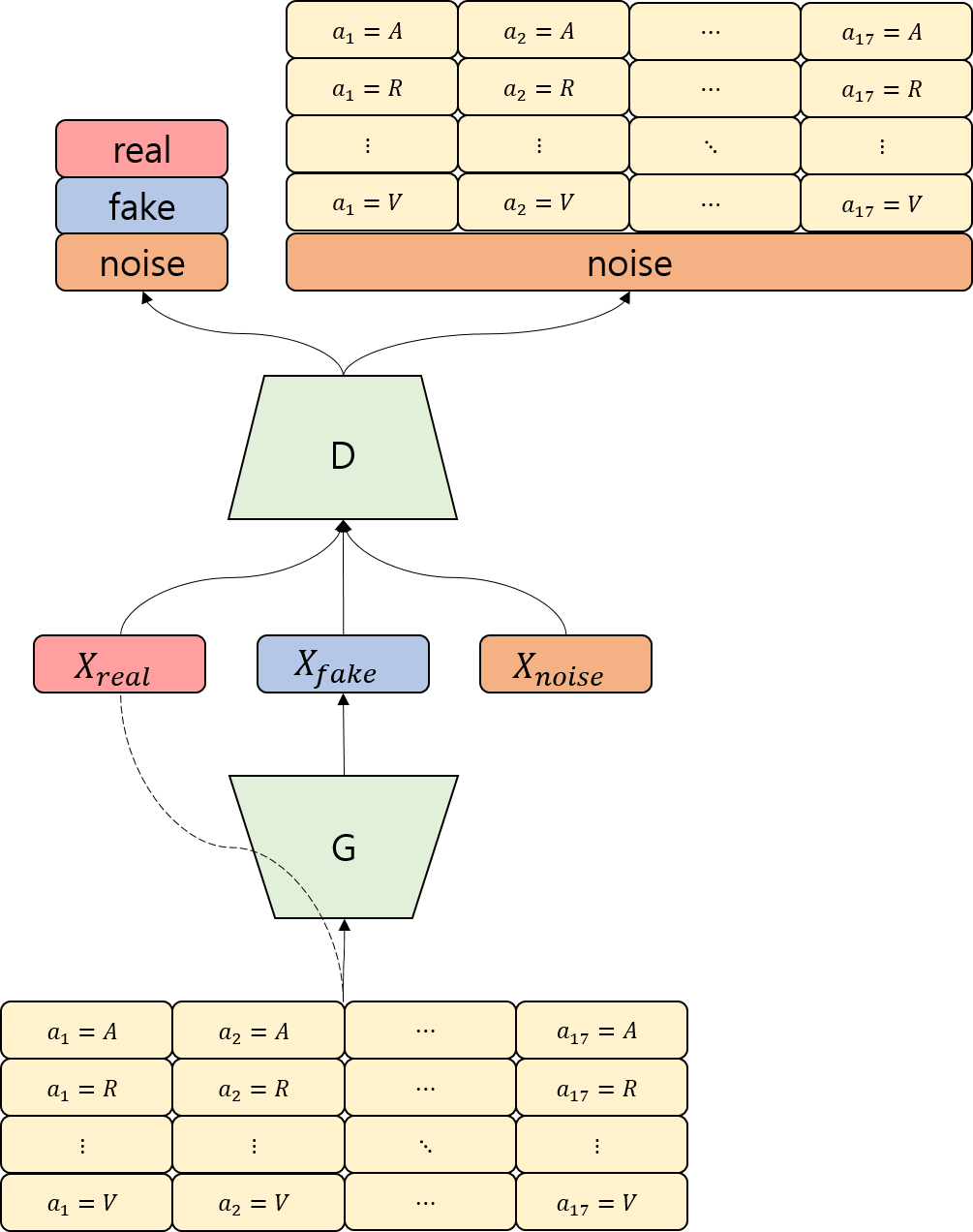}
        \caption{Residue-wise ACGAN} 
        \label{fig:Residue_ACGAN}
    \end{subfigure}	\hspace{0.05\textwidth}
    \begin{subfigure}[b]{0.275\textwidth}
        \includegraphics[width=\textwidth]{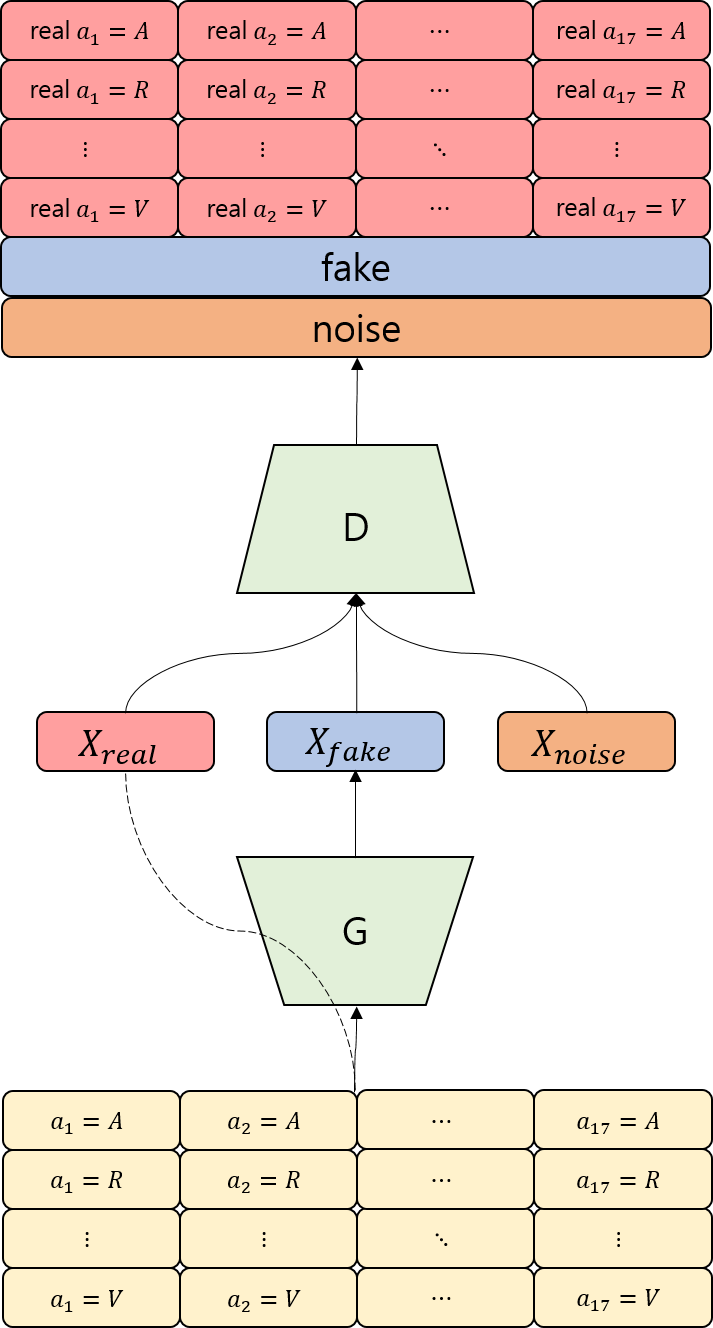}
        \caption{Residue-wise Semi-supervised GAN}
        \label{fig:Residue_Semi_GAN}
    \end{subfigure} 
    \caption{Residue-wise variants of ACGAN and Semi-supervised GAN. Note that these figures show the situations where NCE-GANs were used. Dashed lines indicate correspondence relation between data and its label.}\label{fig:Residue GANs}
\end{figure}

As we want to utilize all sequence information in a window, the discriminator for AC-GAN and Semi-supervised GAN should be able to handle sequence information. To do that we modified AC-GAN and Semi-supervised GAN so that they output corresponding sequence information for the dihedral angles. We call the modified versions as Residue-wise AC-GAN and Residue-wise Semi-supervised GAN, respectively.\\\\
In the Residue-wise AC-GAN, the classifier outputs corresponding sequence information for given input angles residue-wise. Similarly, in Residue-wise Semi-supervised GAN, the discriminator outputs sequence information for real angles residue-wise and fake labels for generated angles by the generator. Note that even though we named Residue-wise Semi-supervised GAN due to its origin of structure, we don't feed unlabelled angles (samples) in our experiments as it's unlikely to have angle information without sequence information (labels).\\\\
One problem of making the Semi-supervised discriminator to handle residue-wise information is that we want to have only one output about $p_{model} (C=fake|x)$ for each input $x$. (The same problem occurs when we also use NCE-GAN and add a class for noise samples as shown in the figure  \ref{fig:Residue GANs}. The class for noise samples can also be handled by a similar process.). We can make the discriminator do that with the following process. \\\\
First, we residue-wise normalize the sequence prediction. And the normalized result is multiplied by $p_{model} (C=real|x)$ which is the estimated probability that input $x$ is real. On the other hand, $p_{model} (C=fake|x)$, the estimated probability that input $x$ is fake, is replicated $window \,size$ times. And then both results are concatenated. This process can also be applied to handle multi labelled data and sequence information whose window size is not fixed, for example, a discriminator which use recurrent neural network.

\section{Evaluation}
To check the regression performance of the models MSE (mean square error) of cosine and sine values of angles and MAE (mean absolute error) of $\phi$ and $\psi$ angles, which can be calculated by equation \eqref{eq:Regression loss}, were reported. Note that $\min(d,2\pi-d)$ was used for calculation of MAE to handle periodicity, where $d$ is the absolute difference.\\\\
To measure the performance of the discriminator's density estimation, mean log-likelihood (LL) were reported for both AC-GAN and Semi-supervised GAN using equation \eqref{eq:density by NCE}. Note that C-GAN can not calculate likelihood, so mean logarithm of $p_{\text{\tiny C-GAN}}(x,y|S=real)$ was reported. As right hand side of the \eqref{eq:density by NCE} is an approximation, its integration might not be 1 and this could affect the estimated mean log-likelihood. To resolve this problem, $\int \exp(l_{real} (x)) \frac{ p(C=noise)}{ p(C=real)} dx$ was estimated by Monte Carlo integration \cite{weinzierl2000introduction} and inverse of this was multiplied by the previously estimated density.\\\\
To measure the generation performance of the generators, mean log-likelihood (LL) for generated class were reported for both AC-GAN and Semi-supervised GAN using equation \eqref{eq:density by NCE}. Note that C-GAN can not calculate likelihood, so mean logarithm of $p_{\text{\tiny C-GAN}}(x,y|S=fake)$ was reported. The same approach was applied to correct the integration of the estimated density.\\\\
To measure the conditional density estimation performance of the discriminators, weighted mean conditional log-likelihood (CLL) were reported for both AC-GAN and Semi-supervised GAN along central amino acids. $p(Center=c)$ was estimated from training data like $\hat{p}(Center=c)=\frac{\text{number of fragments whose central amino acid is c}}{\text{total number of fragments}}$. Conditional likelihood $p_{model}(x|Center=c,S=real)$ of the Semi-supervised GAN can be estimated using equation \eqref{eq:density by NCE}. However, we need different formula for AC-GAN. Remember that the discriminator of the AC-GAN assumes predicted classes and sources are conditionally independent on the inputs as explained in \ref{AC-GAN explanation}. $p_{model}(Center=c,S=s)=p_{model}(Center=c)p_{model}(S=s)$ is also assumed as $p(Center=c,S=s)=p(Center=c)p(S=s)$.
\begin{align}
p_{model}(x|{}& Center=c,S= real)=\frac{p_{model}(x,Center=c,S=real)}{p_{model}(Center=c,S=real)}\\
=&\frac{p_{model}(x,S=real|Center=c)p_{model}(Center=c)}{p_{model}(Center=c,S=real)}\\
=&\frac{p_{model}(S=real|x,Center=c)p_{model}(x|Center=c)p_{model}(Center=c)}{p_{model}(Center=c,S=real)}\\
=&\frac{p_{model}(S=real|x)p_{model}(x|Center=c)p_{model}(Center=c)}{p_{model}(Center=c)p_{model}(S=real)}\nonumber\\
\,(&\because \text{ conditional independence and assumption on prior} )\\
=&\frac{p_{model}(S=real|x)p_{model}(x|Center=c)}{p_{model}(S=real)} \label{AC-GAN density}
\end{align}\\
We estimated $p_{model}(x|Center=c,S=real)$ in AC-GAN by equation \eqref{AC-GAN density}. $p_{model}(S=real)=\frac{1}{3}$ is assumed and the previously explained approach is applied for estimation of $p_{model}(x|Center=c)$.\\\\
To measure the conditional generation performance of the generator, weighted mean conditional log-likelihood (CLL) were reported only for AC-GAN by same method mentioned in the previous paragraph. As Semi-supervised GAN does not have a conditional model for generated samples, we could not report this measure. One can use structure proposed in the future work section of \cite{2016arXiv160601583O} to measure this performance.\\\\
To measure the distribution mimicking performance of the generators, Kullback-Leibler (KL) divergence were reported for AC-GAN and Semi-supervised GAN. $p_{model}(x|real)$ and $p_{model}(x|fake)$ are estimated by \eqref{eq:density by NCE} and integration correcting method were applied for both. Monte Carlo integration \cite{weinzierl2000introduction} is applied for KL divergence estimation.\\\\
MSE, MAE values, LL and KL divergences were checked every 200th iteration and CLL values were checked every 2000th iteration of training in order to create training plots which show trends of the training.

\chapter{Results}
\section{Comparison of various GAN methods for dihedral angle prediction}
\subsection{Without regression loss} \label{Without regression loss}
\begin{figure}[H]
	\centering
	\begin{subfigure}[t]{0.47\textwidth}
        \includegraphics[width=\textwidth]{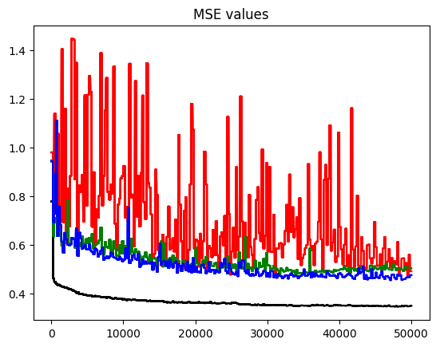}
        \caption{Test mean square errors of cosine and sine values of angles} 
    \end{subfigure}	%\hspace{0.05\textwidth}
    \begin{subfigure}[t]{0.47\textwidth}
        \includegraphics[width=\textwidth]{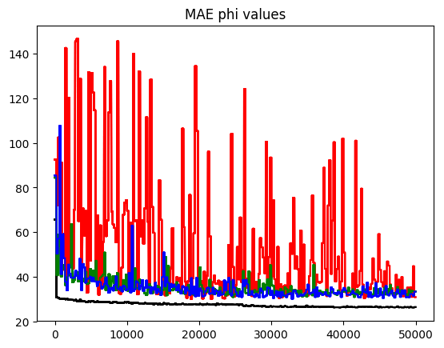}
        \caption{Test mean absolute errors of phi angle}
    \end{subfigure} 
	\begin{subfigure}[t]{0.47\textwidth}
        \includegraphics[width=\textwidth]{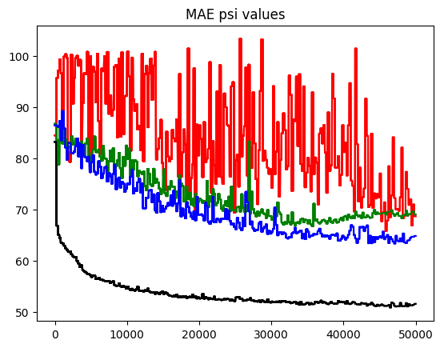}
        \caption{Test mean absolute errors of psi angle}
    \end{subfigure}     
    \caption{The regression errors of the first experiment.\\
Black line: Regression model, Red line: Conditional GANs, Green line: AC-GANs, Blue line: Semi-supervised GANs. The x-axis indicates the number of iterations. }\label{fig:errors without regression loss}
\end{figure}

\begin{figure}[H]
	\centering
	\begin{subfigure}[t]{0.47\textwidth}
        \includegraphics[width=\textwidth]{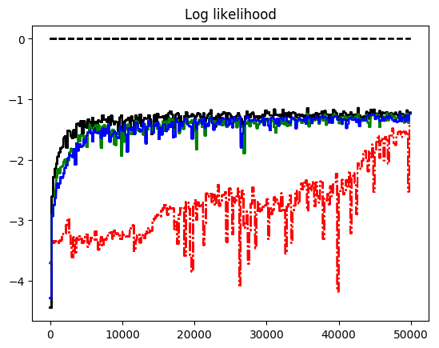}\hfil 
        \caption{Mean log-likelihood} 
    \end{subfigure}	%\hspace{0.05\textwidth}
    \begin{subfigure}[t]{0.47\textwidth}
        \includegraphics[width=\textwidth]{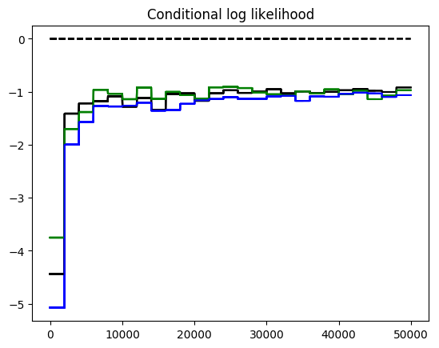}
        \caption{Weighted mean conditional log-likelihood along the middle amino acids}
    \end{subfigure}\par\medskip
	\begin{subfigure}[t]{0.47\textwidth}
        \includegraphics[width=\textwidth]{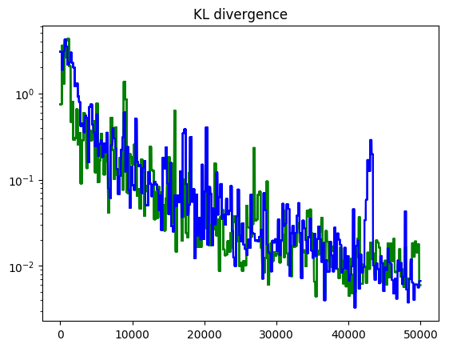}\par
        \caption{Kullback-Leibler divergence between\\ $p_{model}(x|S=real)$ and $p_{model}(x|S=fake)$}
    \end{subfigure}\par\medskip     
    \begin{subfigure}[t]{0.47\textwidth}
        \includegraphics[width=\textwidth]{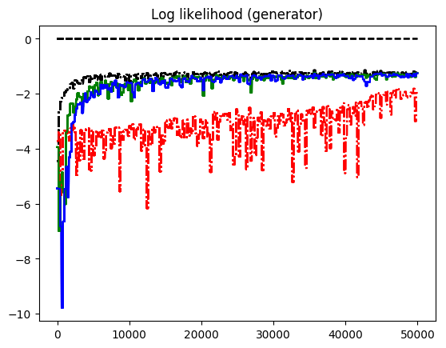}
        \caption{Mean log-likelihood for generated samples} 
    \end{subfigure}\medskip %\hspace{0.05\textwidth}
    \begin{subfigure}[t]{0.47\textwidth}
        \includegraphics[width=\textwidth]{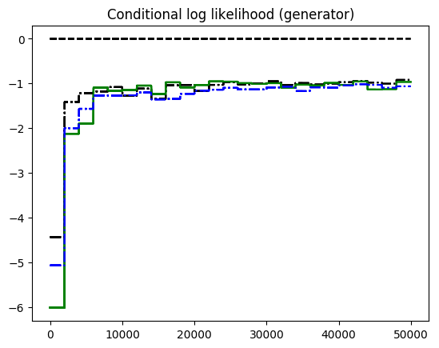}\par
        \caption{Weighted mean conditional log-likelihood along the middle amino acids for generated samples}
    \end{subfigure} 
    \caption{The generation metrics of the first experiment.\\
Black line: NCE model, Red line: Conditional GANs, Green line: AC-GANs, Blue line: Semi-supervised GANs. The x-axis indicates the number of iterations. \\
$\ast$ Red line in (a): $p_{model}(x,y|S=real)$, red line in (d): $p_{model}(x,y|S=fake)$, \\
black and blue line in (e): Weighted mean conditional log-likelihood for real data (not the generated data). They are depicted only for comparison.}\label{fig:estimations without regression loss} %\textcolor{red}{I'm planning to change the KL divergence plot so that y axis indicates log KL divergence with base 10.}
\end{figure}

\begin{figure}[H]
\centering
    \includegraphics[width=0.47\linewidth]{Ramachandran_plot__real_.jpg}\par
    \includegraphics[width=0.47\linewidth]{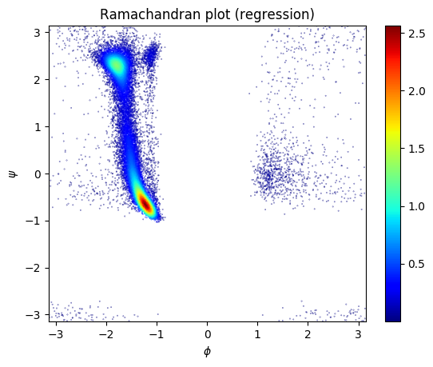}\hfil
    \includegraphics[width=0.47\linewidth]{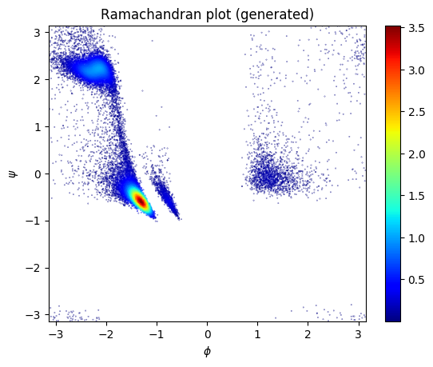}\par\medskip
    \includegraphics[width=0.47\linewidth]{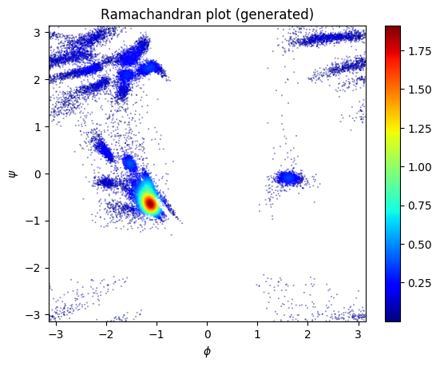}\hfil
    \includegraphics[width=0.47\linewidth]{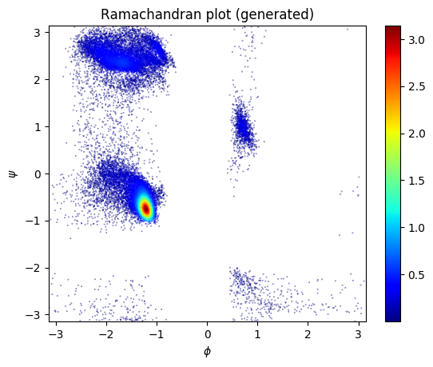}
\caption{Ramachandran plot of test angles and plots using predicted angles for the first experiment.\\
Top: real angles, \\
middle left: predicted angles by using regression model, middle right: predicted angles by using C-GAN,\\
bottom left: predicted angles by using AC-GAN, bottom right: predicted angles by using Semi-supervised GAN }\label{fig:Ramachandran plot without regression loss}
\end{figure}

\begin{figure}[H]
\centering
    \includegraphics[width=0.47\linewidth]{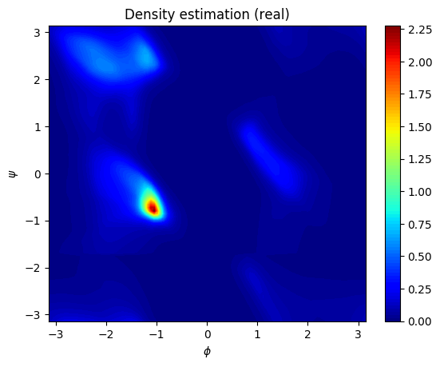}\par
    \includegraphics[width=0.47\linewidth]{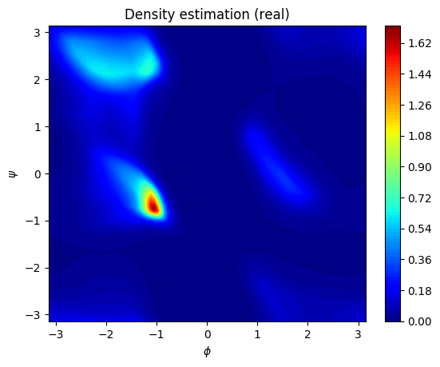}\hfil
    \includegraphics[width=0.47\linewidth]{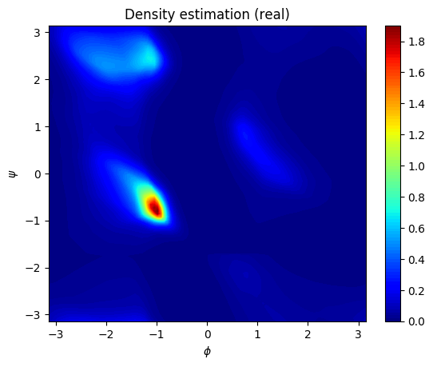}\par\medskip
    \includegraphics[width=0.47\linewidth]{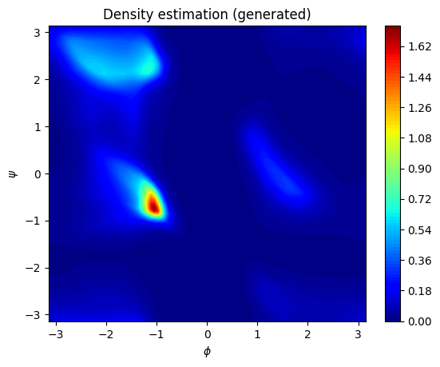}\hfil 
    \includegraphics[width=0.47\linewidth]{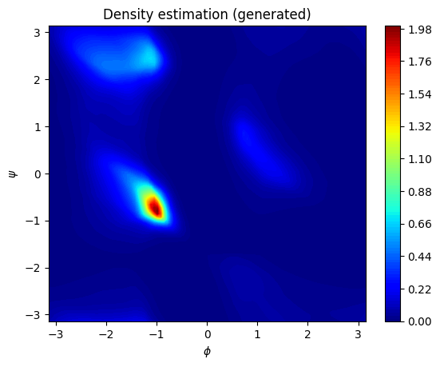}\par\medskip
\caption{Estimated density plots for the first experiment. Note that square root were applied on the estimated densities of angles for better visualization.\\ 
Top: $p_{\text{\tiny NCE}}(x|S=real)$, \\
middle left: $p_{\text{\tiny AC-GAN}}(x|S=real)$, 
middle right: $p_{\text{\tiny Semi-supervised GAN}}(x|S=real)$,\\ 
bottom left: $p_{\text{\tiny AC-GAN}}(x|S=fake)$, bottom right: $p_{\text{\tiny Semi-supervised GAN}}(x|S=fake)$}\label{fig:Density plot without regression loss}
\end{figure}

As mentioned in \cref{Compare GAN}, in this experiment, we did not add regression loss in the generators and noise class ignoring method was used for training loss of the generators.\\\\
When we see MSE (mean square errors) of cosine and sine values of angles and MAE (mean absolute errors) of $\phi$ and $\psi$ angles in figure \ref{fig:errors without regression loss}, we can see training of C-GAN is unstable compared to other models. Figure \ref{fig:estimations without regression loss} (a), (d) also show this trend. MSEs and MAEs of generative models were much higher than regression model. Also see table \ref{table:without regression loss}.\\\\
In table \ref{table:without regression loss} and figure \ref{fig:estimations without regression loss} (a), noise-contrastive estimator got the best performance in terms of mean log-likelihood and weighted mean conditional log-likelihood. This might indicate that adversarially trained discriminators (density estimators) get confused when they learn density of the real data distribution because of instability originating from the adversarial training. However, despite the instability, adversarial training can help the density estimation of the real data if the discriminator has to estimate high-dimensional data such as image data.\\\\
In table \ref{table:without regression loss}, although it could just be a result of the random behaviour, it seems AC-GAN performs better than Semi-supervised GAN in terms of KL (Kullback-Leibler) divergence and mean log-likelihood (LL) for the  generated samples. However, caution is needed when comparing these results because loss, as defined for AC-GAN, is defined as the summation of the classification and discrimination losses and this assumes predicted classes and sources are conditionally independent on the inputs as mentioned in \ref{AC-GAN explanation}. If such assumptions do not hold, the discriminator may not capture the actual density of the generated samples, especially when some weights of the discriminator and the classifier are shared, as in our case. This is explained in the next paragraph.\\\\
When we compared Ramachandran plots based on predicted angles in figure \ref{fig:Ramachandran plot without regression loss}, the distribution of generated samples of C-GAN was similar to those of the regression model. The distribution of generated samples of AC-GAN and Semi-supervised GAN were roughly similar to that of real angles. However, the distribution of predicted angles in AC-GAN is composed of some clusters, which do not exist in the real angle distribution, and was less smooth. %This phenomenon occurred when discrete labels like one-hot labels were used for training of AC-GAN. 
%\textcolor{red}{I was planning to add result when soft label (PSSM information) is used as an input, but I got weird result when I put PSSM information as input and It might better to not adding such result.} 
Such clusters did not show in the estimated density plot of generated samples \cref{fig:Density plot without regression loss} (bottom left) in AC-GAN. This is possibly because it did not have a fixed generator, but it could also be because the independence assumption mentioned in the previous paragraph does not hold. If we see the estimated density plots for each amino acid, even though we have to be careful in analysing estimated densities of AC-GAN, we can observe that AC-GAN's conditional densities for each amino acids are highly peaked compare to real angle distribution. This is shown in table of plots \ref{fig:Amino acids plots without regression loss}. These peaks might correspond to the clusters which occur in figure \ref{fig:Ramachandran plot without regression loss}. These clusters can also be explained by analysis of AC-GAN \cite{shuac}. They showed that if joint distribution of data and label contains points in decision boundary of the classifier, AC-GAN will down-samples those points. From their analysis we hypothesis that the regions the generator fails to generate are the places where decision boundaries are located.\\\\
In figure \ref{fig:Ramachandran plot without regression loss} (bottom right), distributions of predicted angles in Semi-supervised GAN did not show the two separated regions $\zeta$ and $\gamma'$. Figure \ref{fig:Density plot without regression loss} (middle right) shows this was because the discriminator could not capture the two separated regions. The distributions of predicted angles of Semi-supervised GAN also shows distribution of predicted angles in $\alpha_{L}$ region is wrongly positioned (distribution is parallel to $\psi$ axis and too narrowly distributed), but this is not shown in estimated density plot of generated samples \ref{fig:Density plot without regression loss} (bottom right). This can be explained by the instability originating from adversarial training. This can be checked by training more iterations in the discriminator after finishing the training of the generator. Another possible reason is over-fitting of the generator. To check over-fitting of the generator, one can plot the predicted angles using training sequence information. However, we did not invest either of these possible reasons.\\\\
When we compared Ramachandran plots based on predicted angles for each central amino acids in the table of figures \ref{fig:Amino acids plots without regression loss}, the generative models only generated samples in $\alpha_L$ among  $\alpha$ and $\alpha_L$ regions for central amino acid Glycine, while the regression model could generate samples in both regions. AC-GAN generated samples in narrower regions than C-GAN and Semi-supervised GAN for some amino acids. This will make the distribution of the generated angles of AC-GAN composed of unusual clusters.\\\\
%\textcolor{red}{I'm planning to add explanation of the results of mean log-likelihood, weighted mean conditional log-likelihood and  KL divergence. I might say AC-GAN is slightly better than Semi-supervised GAN in terms of these metrics.}\\\\
%\textcolor{red}{I'm also planning to add explanation of the estimated density plots. I will say $p_{model}(x|S=real)$ by using Semi-supervised GAN could not captured two separated populations of angles area around $(\phi, \psi) = (-2,1)$ and that's why its corresponding generator could not generated detailed Ramachandran plot.}\\\\
%\textcolor{red}{I'm also planning to add explanation of the estimated density for each amino acids or only Glycine and proline, but I need to read some papers about them.}
\begin{table}[H]
	\begin{tabular}{ l | l | l | l | l | l }
    	\hline
    	\rowcolor{lightgray}
		 {}  &Regression &NCE &C-GAN &AC-GAN &Semi-supervised GAN\\ \hline
		MSE & $\boldsymbol{0.3508 }$ & - &0.4925&0.5043&0.4770\\ \hline
		Phi MAE & $\boldsymbol{26.3873}$ & - &30.836&33.3909&33.3042\\ \hline
		Psi MAE & $\boldsymbol{51.6076}$ & - &68.6832&69.0326&64.8486\\ \hline
		LL & - & -$\boldsymbol{1.2174}$ & -1.4372${}^\ast$ &-1.2392&-1.2362\\ \hline
		CLL &- & -$\boldsymbol{0.9184}$ & - &-0.9702&-1.0637\\ \hline
		KL divergence & -& - & -& $\boldsymbol{0.005981}$ &0.006797\\ \hline
		LL (generated)&-& -&-1.9650${}^\ast$& -$\boldsymbol{1.2433}$ &-1.2522\\ \hline
		CLL (generated) &- & - & - &-0.9656&-\\ \hline
	\end{tabular} 
\caption{Comparision of MSEs, MAEs, mean log-likelihood, weighted mean conditional log-likelihood along the middle amino acids and Kullback-Leibler divergence between $p_{model}(x|S=real)$ and $p_{model}(x|S=fake)$ where $x$ indicates dihedral angles. The result of the first experiment after 50000 iterations.\\
$\ast$ Note that LL for C-GAN is not actually a mean log-likelihood, but $p_{model}(x,y|S=real)$ where $y$ indicates sequence information. Likewise, LL (generated) for C-GAN indicates $p_{model}(x,y|S=fake)$. }\label{table:without regression loss}%\textcolor{red}{It might be better to use multi-class NCE as a base line model for density estimation to also estimate baseline CLL.} 
\end{table}
\newpage
\newcolumntype{M}[1]{>{\centering\arraybackslash}m{#1}}
\begin{table}[H]
	\begin{tabular}{M{1.8cm}M{2.4cm}M{2.4cm}M{2.4cm}M{2.4cm}M{2.4cm}}
		{}&\footnotesize Aspartate (D)&\footnotesize Cysteine (C)&\footnotesize Glycine (G)&\footnotesize Histidine (H)&\footnotesize Serine (S)\\
		\footnotesize Real angles&\includegraphics[width=1.15\linewidth]{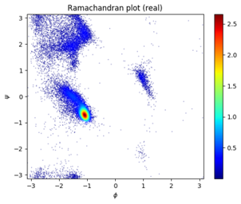}&\includegraphics[width=1.15\linewidth]{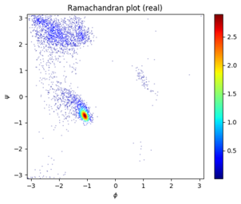}&\includegraphics[width=1.15\linewidth]{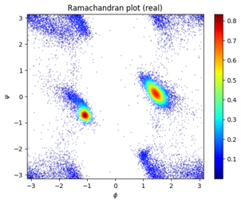}&\includegraphics[width=1.15\linewidth]{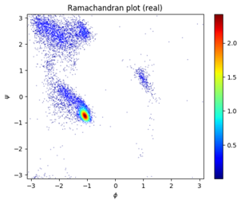}&\includegraphics[width=1.15\linewidth]{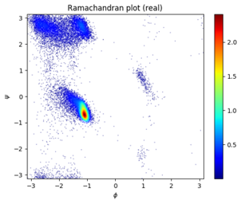}\\ 
		\footnotesize Density estimation by NCE&\includegraphics[width=1.15\linewidth]{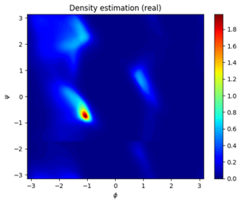}&\includegraphics[width=1.15\linewidth]{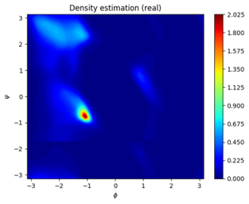}&\includegraphics[width=1.15\linewidth]{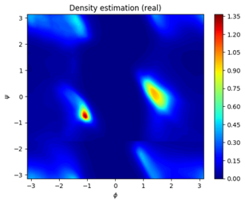}&\includegraphics[width=1.15\linewidth]{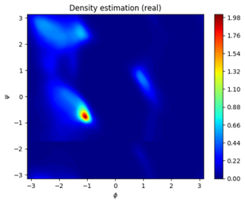}&\includegraphics[width=1.15\linewidth]{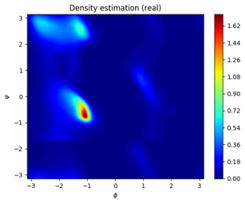}\\
		\footnotesize Density estimation by AC-GAN&\includegraphics[width=1.15\linewidth]{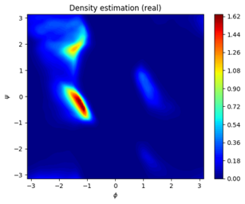}&\includegraphics[width=1.15\linewidth]{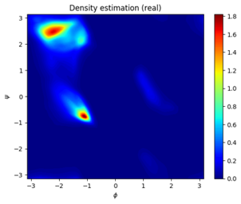}&\includegraphics[width=1.15\linewidth]{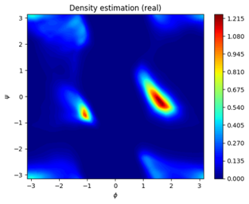}&\includegraphics[width=1.15\linewidth]{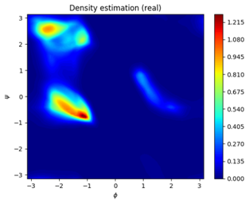}&\includegraphics[width=1.15\linewidth]{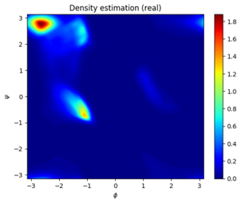}\\
		\footnotesize Density estimation by Semi-supervised GAN&\includegraphics[width=1.15\linewidth]{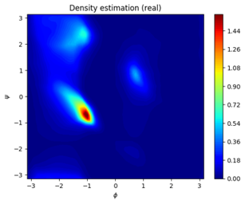}&\includegraphics[width=1.15\linewidth]{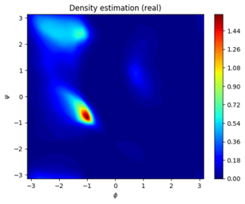}&\includegraphics[width=1.15\linewidth]{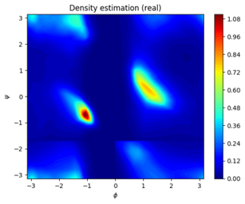}&\includegraphics[width=1.15\linewidth]{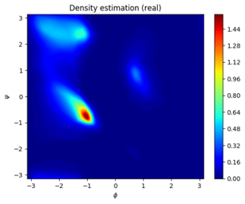}&\includegraphics[width=1.15\linewidth]{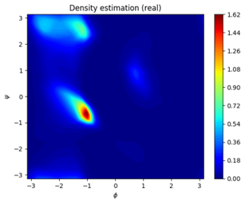}\\
		\footnotesize Predicted angles by regression model&\includegraphics[width=1.15\linewidth]{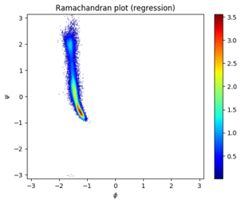}&\includegraphics[width=1.15\linewidth]{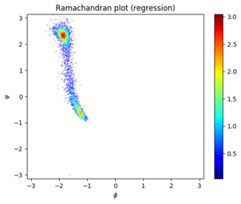}&\includegraphics[width=1.15\linewidth]{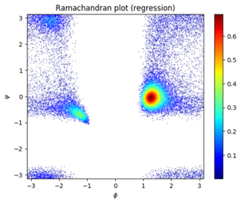}&\includegraphics[width=1.15\linewidth]{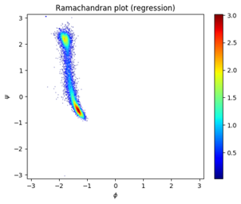}&\includegraphics[width=1.15\linewidth]{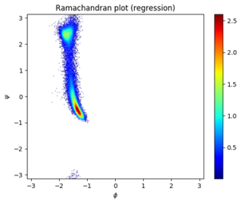}\\
		\footnotesize Predicted angles by C-GAN&\includegraphics[width=1.15\linewidth]{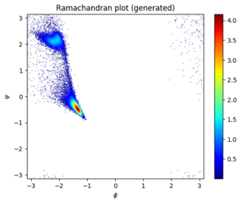}&\includegraphics[width=1.15\linewidth]{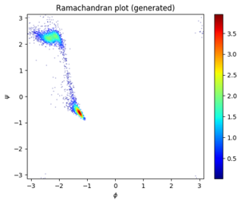}&\includegraphics[width=1.15\linewidth]{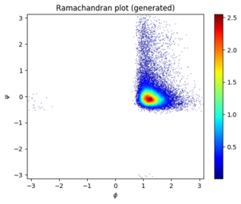}&\includegraphics[width=1.15\linewidth]{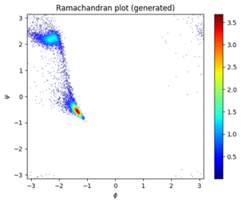}&\includegraphics[width=1.15\linewidth]{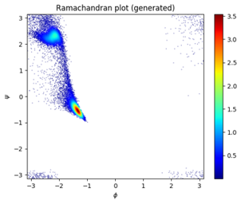}\\
		\footnotesize Predicted angles by AC-GAN&\includegraphics[width=1.15\linewidth]{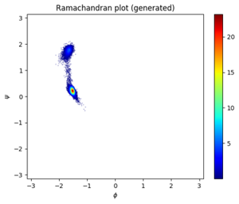}&\includegraphics[width=1.15\linewidth]{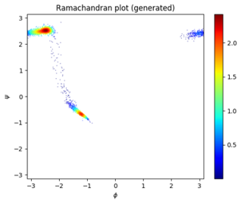}&\includegraphics[width=1.15\linewidth]{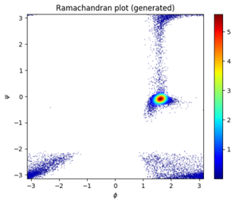}&\includegraphics[width=1.15\linewidth]{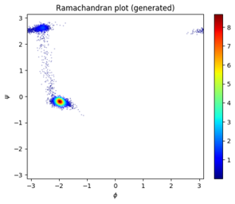}&\includegraphics[width=1.15\linewidth]{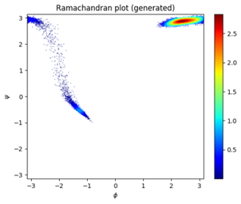}\\
		\footnotesize Predicted angles by Semi-supervised GAN&\includegraphics[width=1.15\linewidth]{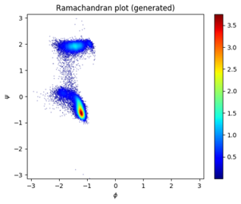}&\includegraphics[width=1.15\linewidth]{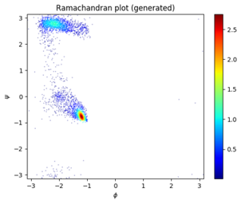}&\includegraphics[width=1.15\linewidth]{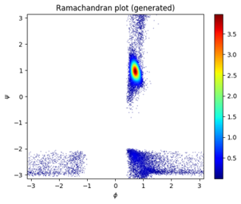}&\includegraphics[width=1.15\linewidth]{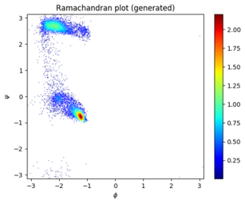}&\includegraphics[width=1.15\linewidth]{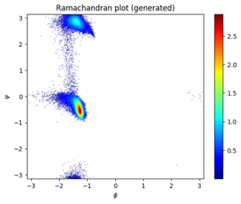}
	\end{tabular}
	\caption{Table of real angles, estimated densities $p_{model}(x|Center=c,S=real)$ and predicted angles plots for some central amino acids for the first experiment. Plotted real angles are from the test set and predicted angles are predicted by sequences in the test set. Plots of amino acid whose estimated density plot of AC-GAN is highly peaked compared to real angles are included. Plots of Glycine are also included due to its unique distribution \cite{ramachandran1963stereochemistry, ramachandran1968conformation, Zimmermann2017} and the difference of the predicted angles among the 4 prediction models.} \label{fig:Amino acids plots without regression loss}
\end{table}
\subsection{Without regression loss and using predicted density by NCE} \label{D Without regression loss}
\begin{figure}[H]
	\centering
	\begin{subfigure}[t]{0.47\textwidth}
        \includegraphics[width=\textwidth]{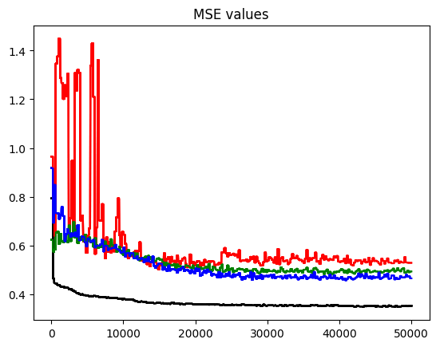}
        \caption{Test mean square errors of cosine and sine values of angles} 
    \end{subfigure}	%\hspace{0.05\textwidth}
    \begin{subfigure}[t]{0.47\textwidth}
        \includegraphics[width=\textwidth]{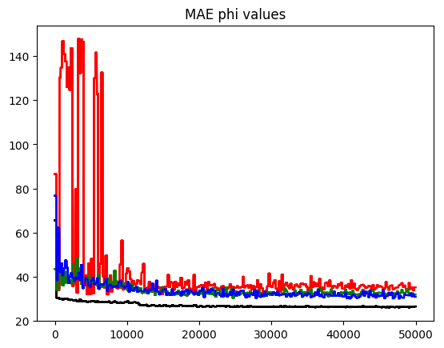}
        \caption{Test mean absolute errors of phi angle}
    \end{subfigure} 
	\begin{subfigure}[t]{0.47\textwidth}
        \includegraphics[width=\textwidth]{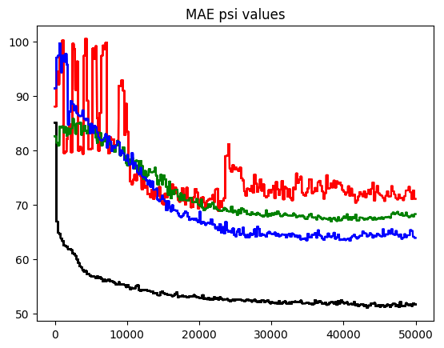}
        \caption{Test mean absolute errors of psi angle}
    \end{subfigure}     
    \caption{The regression errors of the second experiment.\\
Black line: Regression model, Red line: Conditional GANs, Green line: AC-GANs, Blue line: Semi-supervised GANs. The x-axis indicates the number of iterations. }\label{fig:errors D without regression loss}
\end{figure}

\begin{figure}[H]
	\centering
	\begin{subfigure}[t]{0.47\textwidth}
        \includegraphics[width=\textwidth]{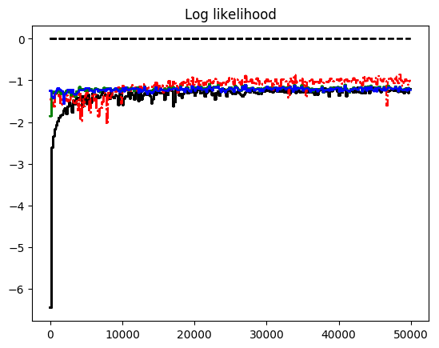}\hfil 
        \caption{Mean log-likelihood} 
    \end{subfigure}	%\hspace{0.05\textwidth}
    \begin{subfigure}[t]{0.47\textwidth}
        \includegraphics[width=\textwidth]{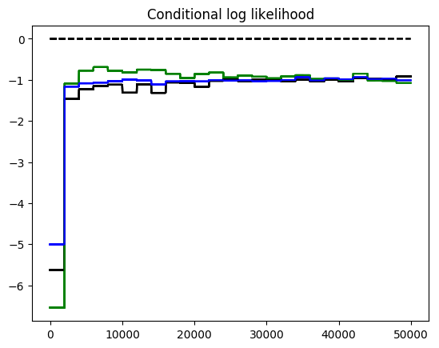}
        \caption{Weighted mean conditional log-likelihood along the middle amino acids}
    \end{subfigure}\par\medskip
	\begin{subfigure}[t]{0.47\textwidth}
        \includegraphics[width=\textwidth]{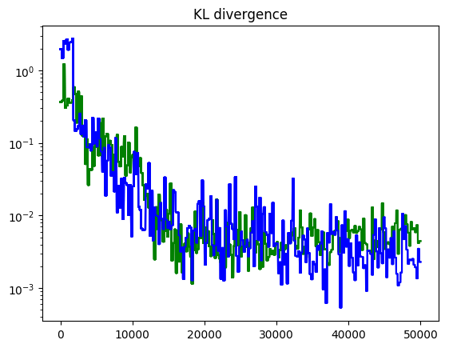}\par
        \caption{Kullback-Leibler divergence between\\ $p_{model}(x|S=real)$ and $p_{model}(x|S=fake)$}
    \end{subfigure}\par\medskip     
    \begin{subfigure}[t]{0.47\textwidth}
        \includegraphics[width=\textwidth]{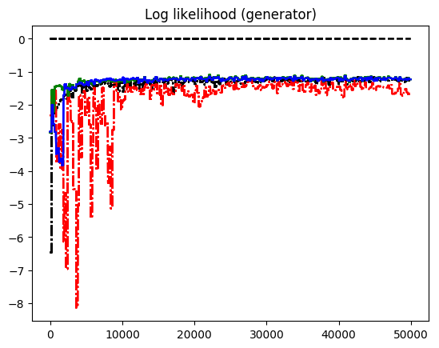}
        \caption{Mean log-likelihood for generated samples} 
    \end{subfigure}\medskip %\hspace{0.05\textwidth}
    \begin{subfigure}[t]{0.47\textwidth}
        \includegraphics[width=\textwidth]{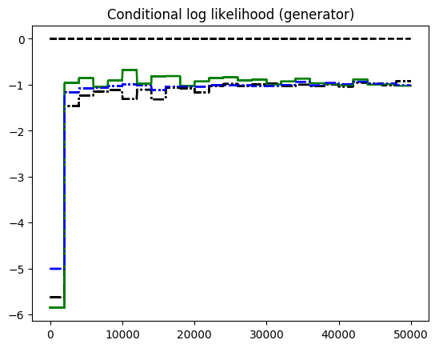}\par
        \caption{Weighted mean conditional log-likelihood along the middle amino acids for generated samples}
    \end{subfigure} 
    \caption{The generation metrics of the second experiment.\\
Black line: NCE model, Red line: Conditional GANs, Green line: AC-GANs, Blue line: Semi-supervised GANs. The x-axis indicates the number of iterations. \\
$\ast$ Red line in (a): $p_{model}(x,y|S=real)$, red line in (d): $p_{model}(x,y|S=fake)$, \\
black and blue line in (e): Weighted mean conditional log-likelihood for real data (not the generated data). They are depicted only for comparison.}\label{fig:estimations D without regression loss} %\textcolor{red}{I'm planning to change the KL divergence plot so that y axis indicates log KL divergence with base 10.}
\end{figure}

\begin{figure}[H]
\centering
    \includegraphics[width=0.47\linewidth]{Ramachandran_plot__real_.jpg}\par
    \includegraphics[width=0.47\linewidth]{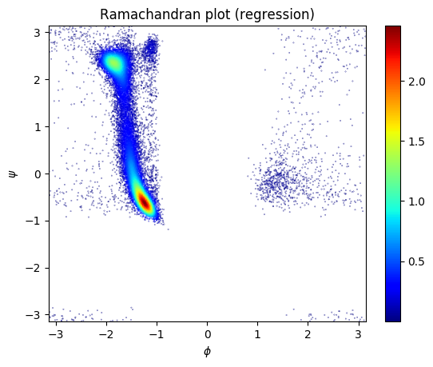}\hfil
    \includegraphics[width=0.47\linewidth]{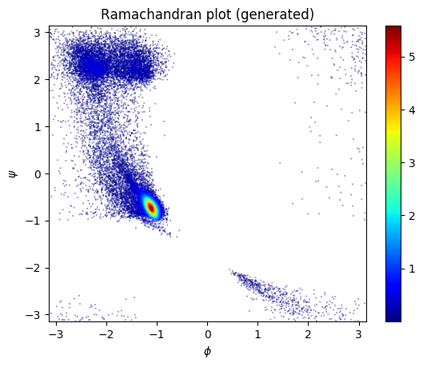}\par\medskip
    \includegraphics[width=0.47\linewidth]{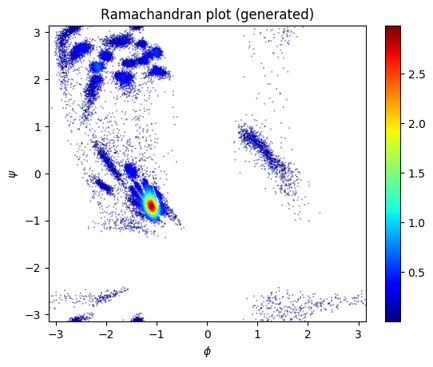}\hfil
    \includegraphics[width=0.47\linewidth]{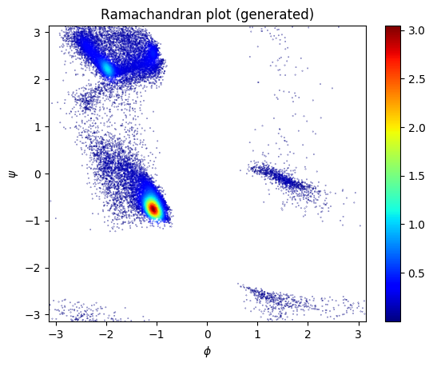}
\caption{Ramachandran plot of test angles and plots using predicted angles for the second experiment.\\
Top: real angles,\\
middle left: predicted angles by using regression model, middle right: predicted angles by using C-GAN, \\
bottom left: predicted angles by using AC-GAN, bottom right: predicted angles by using Semi-supervised GAN}\label{fig:Ramachandran plot D without regression loss}
\end{figure}

\begin{figure}[H]
\centering
    \includegraphics[width=0.47\linewidth]{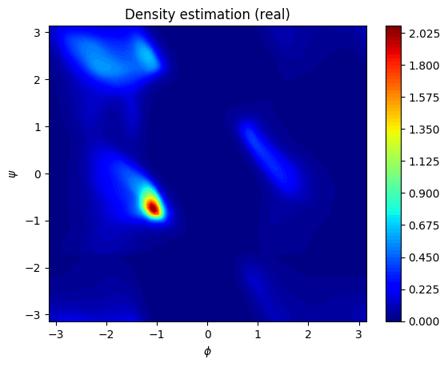}\par
    \includegraphics[width=0.47\linewidth]{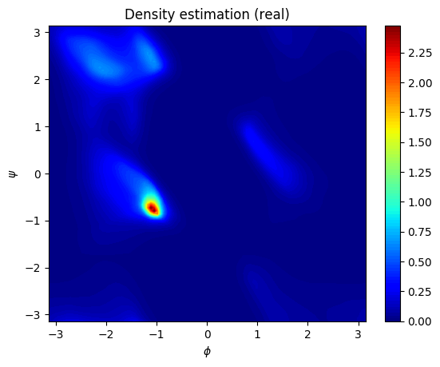}\hfil
    \includegraphics[width=0.47\linewidth]{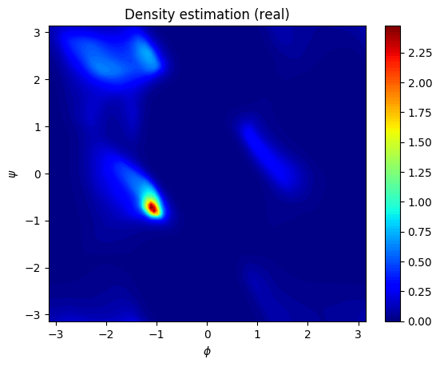}\par\medskip
    \includegraphics[width=0.47\linewidth]{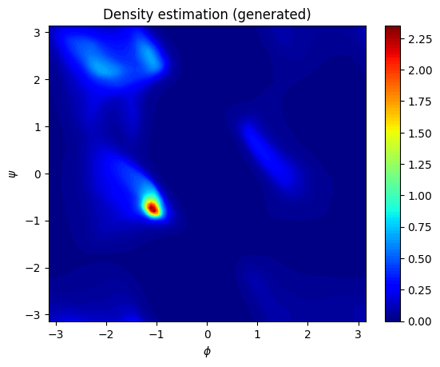}\hfil 
    \includegraphics[width=0.47\linewidth]{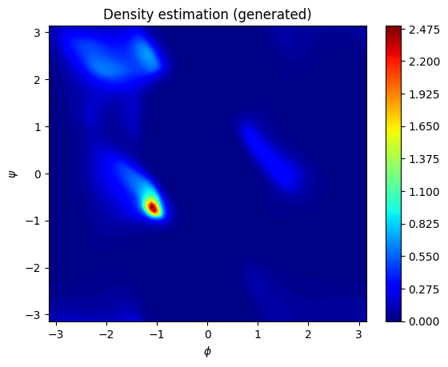}\par\medskip
\caption{Estimated density plots for the second experiment. Note that square root were applied on the estimated densities of angles for better visualization.\\ 
Top: $p_{\text{\tiny NCE}}(x|S=real)$, \\
middle left: $p_{\text{\tiny AC-GAN}}(x|S=real)$, 
middle right: $p_{\text{\tiny Semi-supervised GAN}}(x|S=real)$,\\ 
bottom left: $p_{\text{\tiny AC-GAN}}(x|S=fake)$, bottom right: $p_{\text{\tiny Semi-supervised GAN}}(x|S=fake)$}\label{fig:Density plot D without regression loss}
\end{figure}

Training of the discriminators was not stable and estimated $p_{model}(x|real)$ of Semi-supervised GAN did not capture the detailed shape of the Ramachandran plot in the previous experiment \ref{Without regression loss}. As mentioned in section \ref{Compare GAN}, we added predicted density estimation by NCE model as an additional input of the discriminators in this experiment in an effort to resolve mentioned issues.\\\\
When we see MSE of cosine and sine values of angles and MAE  of $\phi$ and $\psi$ angles in figure \ref{fig:errors D without regression loss}, we can see training of C-GAN become more stable compared to not putting predicted density information \ref{fig:errors without regression loss}. Figure \ref{fig:estimations D without regression loss} (a) shows training of $p_{model}(x,y|S=real)$ was not only stabilized, but also trained faster. Similar trend was observed in figure \ref{fig:estimations D without regression loss} (d) except for early 10000 iterations of weight updates.\\\\
Table \ref{table:D without regression loss} shows both $p_{model}(x,y|S=real)$ %(-1.0072) 
and $p_{model}(x,y|S=fake)$ %(-1.6494) 
for C-GAN were improved. % (corresponding values were -1.4372 and -1.9650 in the previous experiment). 
MSE and MAE values were slightly reduced for both AC-GAN and Semi-supervised GAN. Mean log-likelihood (LL) for both real and generated data were increased in AC-GAN and Semi-supervised GAN. KL-divergence between estimated real data and generated sample distribution were reduced for both AC-GAN and Semi-supervised GAN. This is also shown in figure \ref{fig:estimations D without regression loss} (c).\\\\
Figures in \ref{fig:Ramachandran plot D without regression loss} show more realistic angle distributions except that the C-GAN failed to generate angles in the $\alpha_{L}$ region. For example C-GAN seems to capture two different region $\beta_S$ and $\beta_P$ which was not observed in \ref{fig:Ramachandran plot without regression loss} (middle right). Although distribution of predicted angles in $\alpha_{L}$ region for Semi-supervised GAN seems to be somewhat vertically squeezed, both AC-GAN and Semi-supervised GAN could generate more believable angles in the $\alpha_{L}$ region than the results of the previous section and the  Semi-supervised GAN seems to capture $\zeta$ and $\gamma'$ regions. In general, the discriminator of the Semi-supervised GAN could capture these two regions shown in figure \ref{fig:Density plot D without regression loss} (middle right). It should be noted generated samples from the AC-GAN contain some unusual clusters and figure \ref{fig:Density plot D without regression loss} (bottom left) could not capture this phenomenon. \\\\
When we compared Ramachandran plots based on predicted angles for each central amino acids in the table of figures \ref{fig:Amino acids plots D without regression loss}, AC-GAN and Semi-supervised GAN only generated samples in $\alpha_L$ and C-GAN only generated samples in $\alpha$ among $\alpha$ and $\alpha_L$ regions for central amino acid Glycine. AC-GAN generated samples in narrow regions like the first experiment. Contours of distributions of generated angles of Semi-supervised GAN were well matched with these of real angles.\\\\
\begin{table}[H]
	\begin{tabular}{ l | l | l | l | l | l }
    	\hline
    	\rowcolor{lightgray}
		 {}  &Regression &NCE &C-GAN &AC-GAN &Semi-supervised GAN\\ \hline
		MSE & $\boldsymbol{0.3517 }$ & - &0.5282&0.4929&0.4644\\ \hline
		Phi MAE & $\boldsymbol{26.5431}$ & - &35.2343&31.8829&30.8688\\ \hline
		Psi MAE & $\boldsymbol{51.6854}$ & - &71.0663&68.2798&63.9015\\ \hline
		LL & - & -$\boldsymbol{1.2074}$ & -1.0072${}^\ast$ &-1.2209&-1.2290\\ \hline
		CLL &- & -$\boldsymbol{0.9188}$ & - &-1.0754&-1.0100\\ \hline
		KL divergence & -& - & -& 0.004449 &$\boldsymbol{0.002294}$\\ \hline
		LL (generated)&-& -&-1.6494${}^\ast$& -1.2276 &-$\boldsymbol{1.2273}$\\ \hline
		CLL (generated) &- & - & - &-1.0221&-\\ \hline
	\end{tabular} 
\caption{Comparision of MSEs, MAEs, mean log-likelihood, weighted mean conditional log-likelihood along the middle amino acids and Kullback-Leibler divergence between $p_{model}(x|S=real)$ and $p_{model}(x|S=fake)$ where $x$ indicates dihedral angles. The result of the second experiment after 50000 iterations. \\
$\ast$ Note that LL for C-GAN is not actually a mean log-likelihood, but $p_{model}(x,y|S=real)$. Likewise, LL (generated) for C-GAN indicates $p_{model}(x,y|S=fake)$. }\label{table:D without regression loss}%\textcolor{red}{It might be better to use multi-class NCE as a base line model for density estimation to also estimate baseline CLL.} 
\end{table}
\newpage
\begin{table}[H]
	\begin{tabular}{M{1.8cm}M{2.4cm}M{2.4cm}M{2.4cm}M{2.4cm}M{2.4cm}}
		{}&\footnotesize Aspartate (D)&\footnotesize Cysteine (C)&\footnotesize Glycine (G)&\footnotesize Histidine (H)&\footnotesize Serine (S)\\
		\footnotesize Real angles&\includegraphics[width=1.15\linewidth]{X_without_MSE/Ramachandran_plot__real__Aspartate__D_.png}&\includegraphics[width=1.15\linewidth]{X_without_MSE/Ramachandran_plot__real__Cysteine__C_.png}&\includegraphics[width=1.15\linewidth]{X_without_MSE/Ramachandran_plot__real__Glycine__G_.png}&\includegraphics[width=1.15\linewidth]{X_without_MSE/Ramachandran_plot__real__Histidine__H_.png}&\includegraphics[width=1.15\linewidth]{X_without_MSE/Ramachandran_plot__real__Serine__S_.png}\\ 
		\footnotesize Density estimation by NCE&\includegraphics[width=1.15\linewidth]{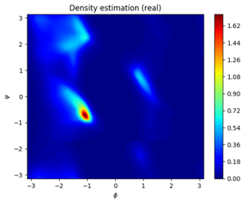}&\includegraphics[width=1.15\linewidth]{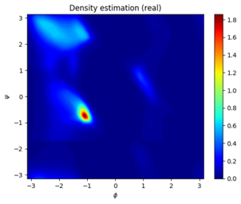}&\includegraphics[width=1.15\linewidth]{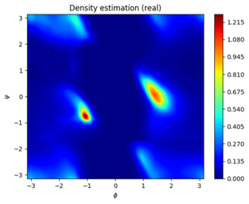}&\includegraphics[width=1.15\linewidth]{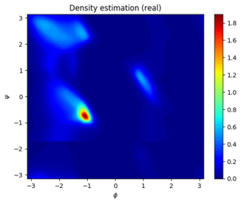}&\includegraphics[width=1.15\linewidth]{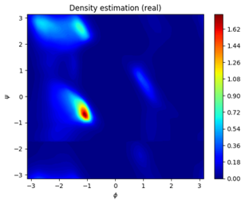}\\
		\footnotesize Density estimation by AC-GAN&\includegraphics[width=1.15\linewidth]{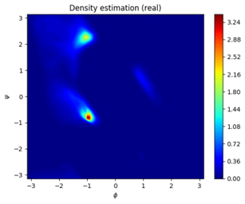}&\includegraphics[width=1.15\linewidth]{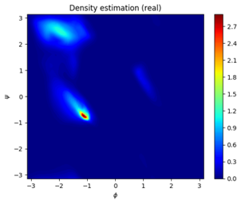}&\includegraphics[width=1.15\linewidth]{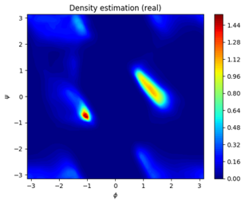}&\includegraphics[width=1.15\linewidth]{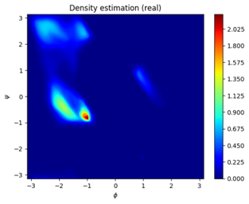}&\includegraphics[width=1.15\linewidth]{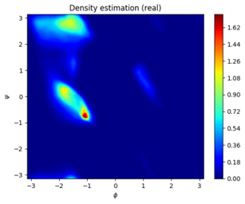}\\
		\footnotesize Density estimation by Semi-supervised GAN&\includegraphics[width=1.15\linewidth]{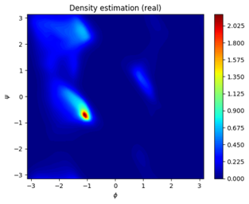}&\includegraphics[width=1.15\linewidth]{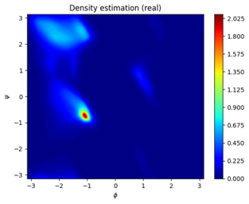}&\includegraphics[width=1.15\linewidth]{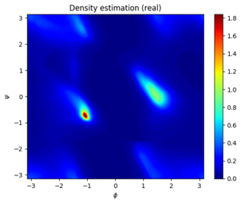}&\includegraphics[width=1.15\linewidth]{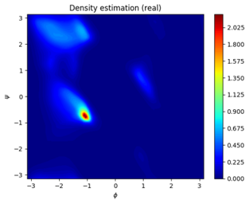}&\includegraphics[width=1.15\linewidth]{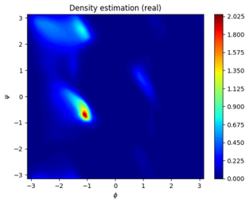}\\
		\footnotesize Predicted angles by regression model&\includegraphics[width=1.15\linewidth]{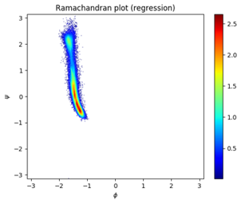}&\includegraphics[width=1.15\linewidth]{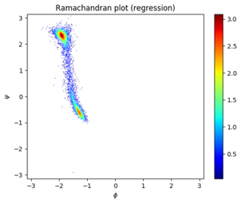}&\includegraphics[width=1.15\linewidth]{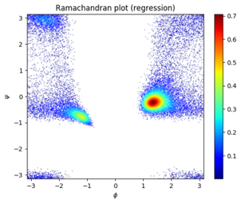}&\includegraphics[width=1.15\linewidth]{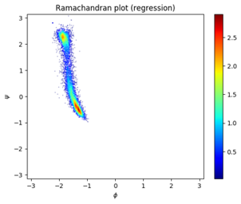}&\includegraphics[width=1.15\linewidth]{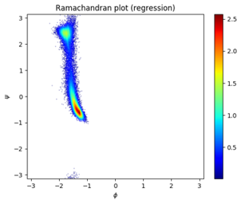}\\
		\footnotesize Predicted angles by C-GAN&\includegraphics[width=1.15\linewidth]{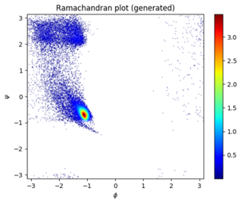}&\includegraphics[width=1.15\linewidth]{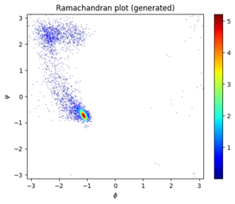}&\includegraphics[width=1.15\linewidth]{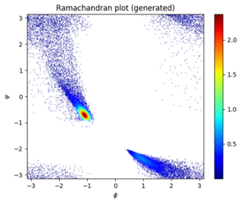}&\includegraphics[width=1.15\linewidth]{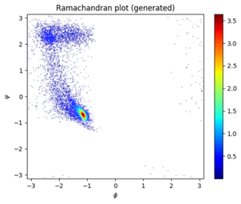}&\includegraphics[width=1.15\linewidth]{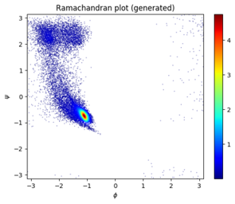}\\
		\footnotesize Predicted angles by AC-GAN&\includegraphics[width=1.15\linewidth]{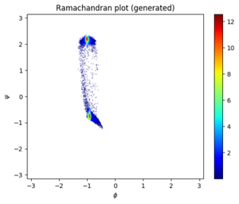}&\includegraphics[width=1.15\linewidth]{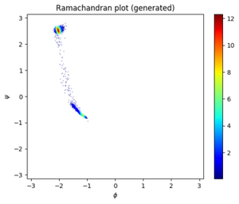}&\includegraphics[width=1.15\linewidth]{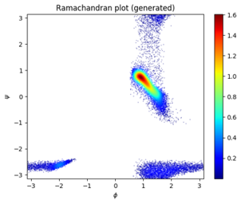}&\includegraphics[width=1.15\linewidth]{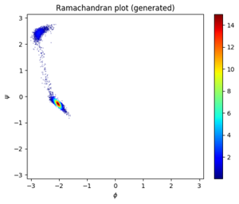}&\includegraphics[width=1.15\linewidth]{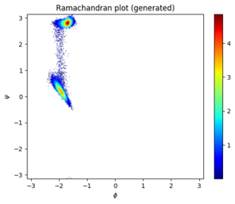}\\
		\footnotesize Predicted angles by Semi-supervised GAN&\includegraphics[width=1.15\linewidth]{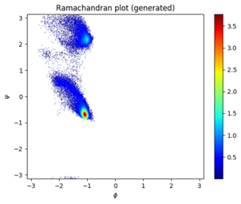}&\includegraphics[width=1.15\linewidth]{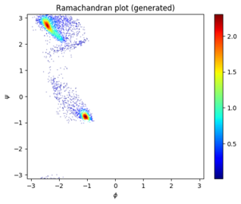}&\includegraphics[width=1.15\linewidth]{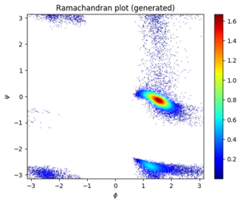}&\includegraphics[width=1.15\linewidth]{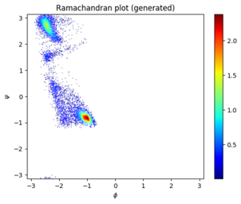}&\includegraphics[width=1.15\linewidth]{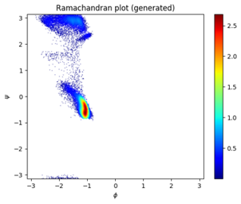}
	\end{tabular}
	\caption{Table of real angles, estimated densities $p_{model}(x|Center=c,S=real)$ and predicted angles plots for some central amino acids for the second experiment. Plotted real angles are from the test set and predicted angles are predicted by sequences in the test set. Same 5 chosen amino acids are depicted as in figure \ref{fig:Amino acids plots without regression loss}.} \label{fig:Amino acids plots D without regression loss}
\end{table}

\subsection{Without regression loss, using predicted density by NCE and minibatch-wise loss in generators} \label{D Without regression loss minibatch}
\begin{figure}[H]
	\centering
	\begin{subfigure}[t]{0.47\textwidth}
        \includegraphics[width=\textwidth]{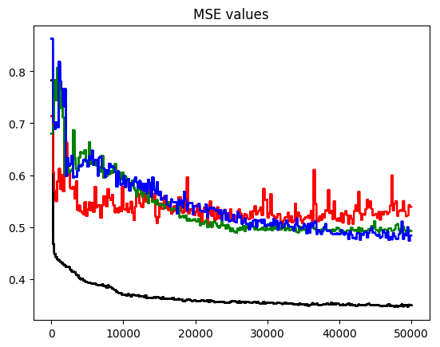}
        \caption{Test mean square errors of cosine and sine values of angles} 
    \end{subfigure}	%\hspace{0.05\textwidth}
    \begin{subfigure}[t]{0.47\textwidth}
        \includegraphics[width=\textwidth]{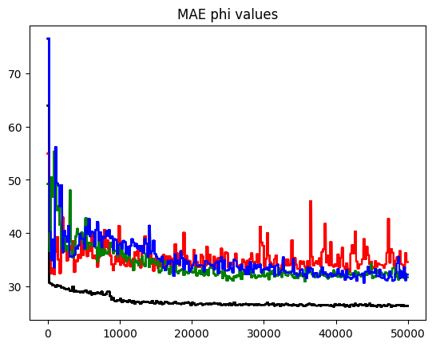}
        \caption{Test mean absolute errors of phi angle}
    \end{subfigure} 
	\begin{subfigure}[t]{0.47\textwidth}
        \includegraphics[width=\textwidth]{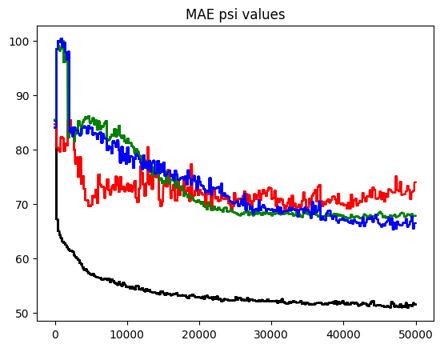}
        \caption{Test mean absolute errors of psi angle}
    \end{subfigure}     
    \caption{The regression errors of the third experiment.\\
Black line: Regression model, Red line: Conditional GANs, Green line: AC-GANs, Blue line: Semi-supervised GANs. The x-axis indicates the number of iterations. }\label{fig:errors D without regression loss logit2}
\end{figure}

\begin{figure}[H]
	\centering
	\begin{subfigure}[t]{0.47\textwidth}
        \includegraphics[width=\textwidth]{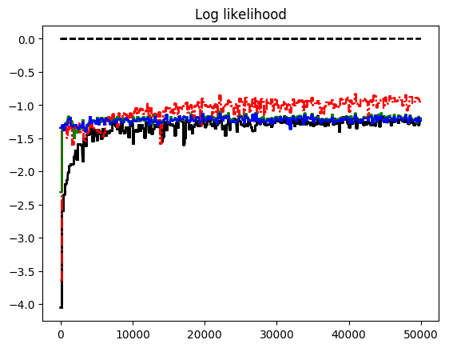}\hfil 
        \caption{Mean log-likelihood} 
    \end{subfigure}	%\hspace{0.05\textwidth}
    \begin{subfigure}[t]{0.47\textwidth}
        \includegraphics[width=\textwidth]{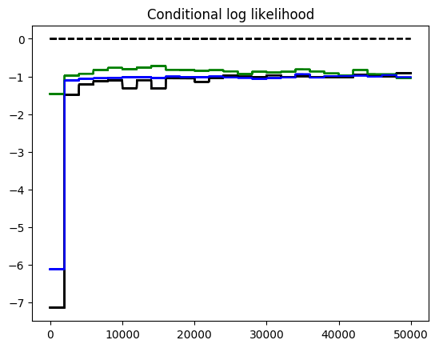}
        \caption{Weighted mean conditional log-likelihood along the middle amino acids}
    \end{subfigure}\par\medskip
	\begin{subfigure}[t]{0.47\textwidth}
        \includegraphics[width=\textwidth]{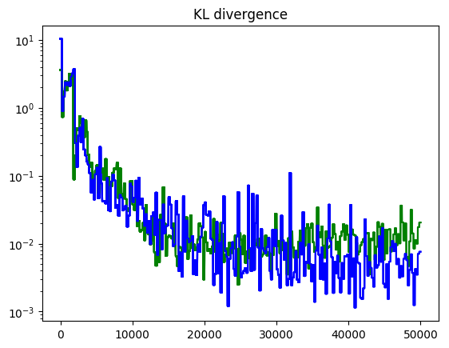}\par
        \caption{Kullback-Leibler divergence between\\ $p_{model}(x|S=real)$ and $p_{model}(x|S=fake)$}
    \end{subfigure}\par\medskip     
    \begin{subfigure}[t]{0.47\textwidth}
        \includegraphics[width=\textwidth]{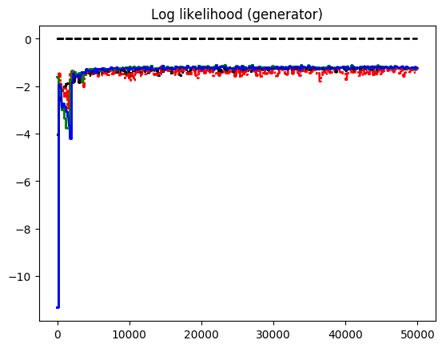}
        \caption{Mean log-likelihood for generated samples} 
    \end{subfigure}\medskip %\hspace{0.05\textwidth}
    \begin{subfigure}[t]{0.47\textwidth}
        \includegraphics[width=\textwidth]{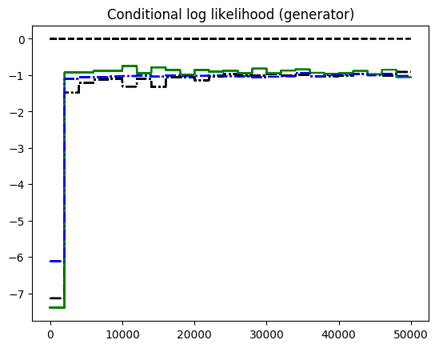}\par
        \caption{Weighted mean conditional log-likelihood along the middle amino acids for generated samples}
    \end{subfigure} 
    \caption{The generation metrics of the third experiment.\\
Black line: NCE model, Red line: Conditional GANs, Green line: AC-GANs, Blue line: Semi-supervised GANs. The x-axis indicates the number of iterations. \\
$\ast$ Red line in (a): $p_{model}(x,y|S=real)$, red line in (d): $p_{model}(x,y|S=fake)$, \\
black and blue line in (e): Weighted mean conditional log-likelihood for real data (not the generated data). They are depicted only for comparison.}\label{fig:estimations D without regression loss logit2} %\textcolor{red}{I'm planning to change the KL divergence plot so that y axis indicates log KL divergence with base 10.}
\end{figure}

\begin{figure}[H]
\centering
    \includegraphics[width=0.47\linewidth]{Ramachandran_plot__real_.jpg}\par
    \includegraphics[width=0.47\linewidth]{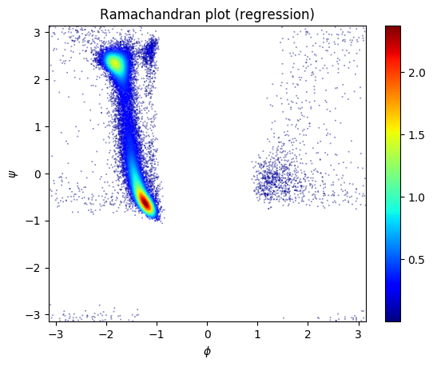}\hfil
    \includegraphics[width=0.47\linewidth]{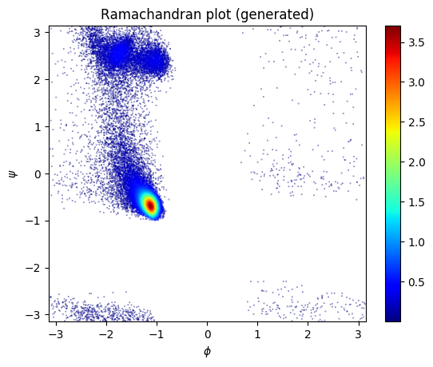}\par\medskip
    \includegraphics[width=0.47\linewidth]{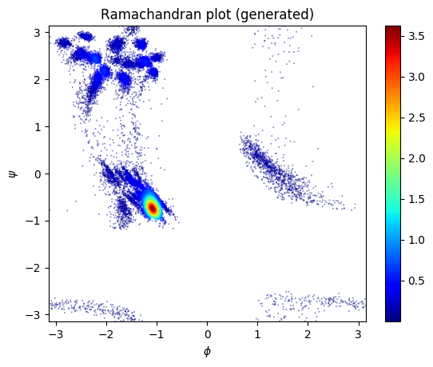}\hfil
    \includegraphics[width=0.47\linewidth]{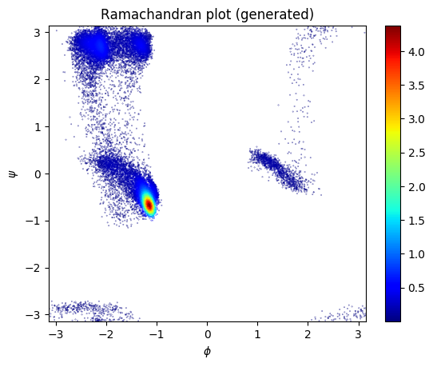}
\caption{Ramachandran plot of test angles and plots using predicted angles for the third experiment.\\
Top: real angles,\\
middle left: predicted angles by using regression model, middle right: predicted angles by using C-GAN,\\
bottom left: predicted angles by using AC-GAN, bottom right: predicted angles by using Semi-supervised GAN}\label{fig:Ramachandran plot D without regression loss logit2}
\end{figure}

As mentioned in section \ref{Compare GAN}, we used minibatch-wise loss in the generators of this experiment to see the effects of minibatch-wise generation loss. \\\\
When we see MSE value and MAE of $\phi$ and $\psi$ angles in figure \ref{fig:errors D without regression loss logit2}, we can see training of C-GAN becoming more stable in the early 10000 iterations of weight updates. This is also confirmed in figure \ref{fig:estimations D without regression loss logit2}. This indicates using minibatch-wise loss stabilize training of the C-GAN generator.\\\\
Table \ref{table:D without regression loss logit2} shows both $p_{model}(x,y|S=real)$ %(-0.9483) 
and $p_{model}(x,y|S=fake)$ %(-1.3046) 
for C-GAN were improved. %(corresponding values were -1.0072 and -1.6494 in the previous experiment). 
This would be because the generation performance of the generator $p_{model}(x,y|S=fake)$ is improved and $p_{model}(x,y|S=real)$ might also be improved because of this. LL and CLL for both real and generated data were slightly increased in AC-GAN except for
CLL of the generated data. However, KL divergence increased in AC-GAN as also shown in figure \ref{fig:estimations D without regression loss logit2}. Semi-supervised GAN got worse result in terms of all reported evaluation metrics.\\\\
Figure \ref{fig:Ramachandran plot D without regression loss logit2} show C-GAN could generate more samples in the $\alpha_L$ region and the AC-GAN seems to be get better at generating angles in regions $\zeta$ and $\gamma'$. Semi-supervised GAN got worse shape in the $\beta_S$ and $\beta_P$ as shown in figure \ref{fig:Ramachandran plot D without regression loss logit2} (bottom right). This indicates using minibatch-wise loss in the loss of the Semi-supervised generator does not improve its generation performance. Hence, different approach to the Semi-supervised generator would be needed, or it might be better to use the non-minibatch-wise loss in its generator.\\\\
When we compared Ramachandran plots based on predicted angles for each central amino acids in the table of figures \ref{fig:Amino acids plots D without regression loss minibatch}, C-GAN generated samples in both $\alpha$ and $\alpha_L$ regions for central amino acid Glycine. AC-GAN generated samples in narrow regions like the first experiment. Contours of distributions of generated angles of Semi-supervised GAN got worse at matching with these of real angles.\\\\

\begin{table}[H]
	\begin{tabular}{ l | l | l | l | l | l }
    	\hline
    	\rowcolor{lightgray}
		 {}  &Regression &NCE &C-GAN &AC-GAN &Semi-supervised GAN\\ \hline
		MSE & $\boldsymbol{0.3487 }$ & - &0.5388&0.4922&0.4836\\ \hline
		Phi MAE & $\boldsymbol{26.2408}$ & - &34.5151&31.6593&32.1611\\ \hline
		Psi MAE & $\boldsymbol{51.4794}$ & - &73.9974&67.8315&66.4762\\ \hline
		LL & - & -$\boldsymbol{1.2088}$ & -0.9483${}^\ast$ &-1.2179&-1.2376\\ \hline
		CLL &- & -$\boldsymbol{0.9131}$ & - &-1.0467&-1.0197\\ \hline
		KL divergence & -& - & -& 0.02048 &$\boldsymbol{0.007591}$\\ \hline
		LL (generated)&-& -&-1.3046${}^\ast$& --$\boldsymbol{1.2214}$ &-1.2453\\ \hline
		CLL (generated) &- & - & - &-1.0554&-\\ \hline
	\end{tabular} 
\caption{Comparision of MSEs, MAEs, mean log-likelihood, weighted mean conditional log-likelihood along the middle amino acids and Kullback-Leibler divergence between $p_{model}(x|S=real)$ and $p_{model}(x|S=fake)$ where $x$ indicates dihedral angles. The result of the third experiment after 50000 iterations. \\
$\ast$ Note that LL for C-GAN is not actually a mean log-likelihood, but $p_{model}(x,y|S=real)$. Likewise, LL (generated) for C-GAN indicates $p_{model}(x,y|S=fake)$. }\label{table:D without regression loss logit2}%\textcolor{red}{It might be better to use multi-class NCE as a base line model for density estimation to also estimate baseline CLL.} 
\end{table}
\newpage
\begin{figure}[H]
\centering
    \includegraphics[width=0.47\linewidth]{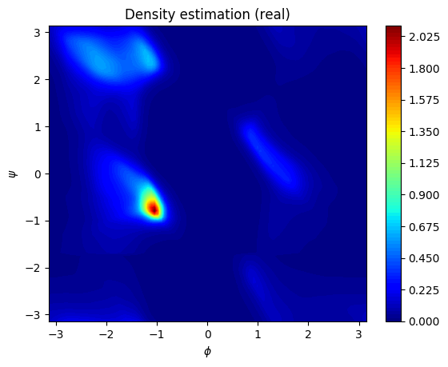}\par
    \includegraphics[width=0.47\linewidth]{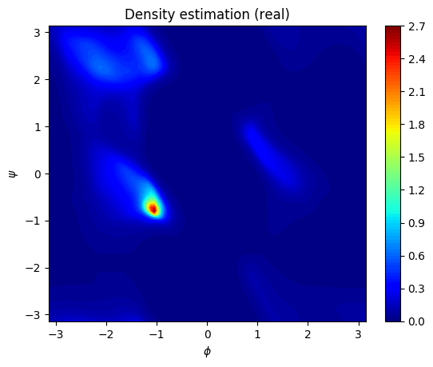}\hfil
    \includegraphics[width=0.47\linewidth]{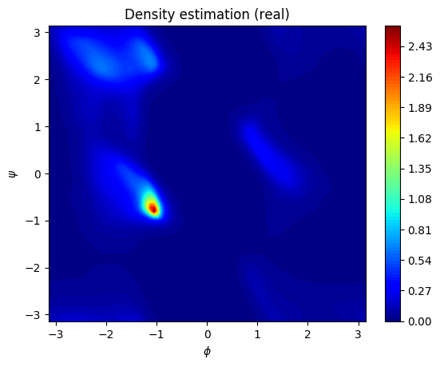}\par\medskip
    \includegraphics[width=0.47\linewidth]{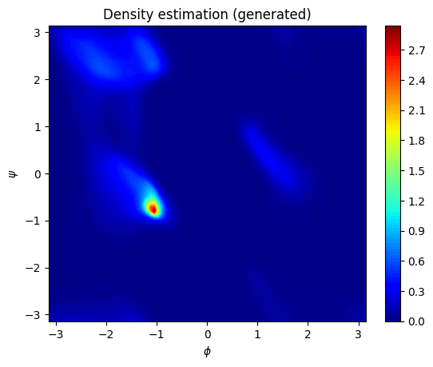}\hfil 
    \includegraphics[width=0.47\linewidth]{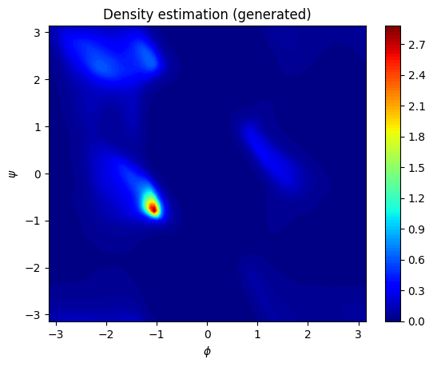}\par\medskip
\caption{Estimated density plots for the third experiment. Note that square root were applied on the estimated densities of angles for better visualization.\\ 
Top: $p_{\text{\tiny NCE}}(x|S=real)$, \\
middle left: $p_{\text{\tiny AC-GAN}}(x|S=real)$, 
middle right: $p_{\text{\tiny Semi-supervised GAN}}(x|S=real)$,\\ 
bottom left: $p_{\text{\tiny AC-GAN}}(x|S=fake)$, bottom right: $p_{\text{\tiny Semi-supervised GAN}}(x|S=fake)$}\label{fig:Density plot D without regression loss logit2}
\end{figure}
\begin{table}[H]
	\begin{tabular}{M{1.8cm}M{2.4cm}M{2.4cm}M{2.4cm}M{2.4cm}M{2.4cm}}
		{}&\footnotesize Aspartate (D)&\footnotesize Cysteine (C)&\footnotesize Glycine (G)&\footnotesize Histidine (H)&\footnotesize Serine (S)\\
		\footnotesize Real angles&\includegraphics[width=1.15\linewidth]{X_without_MSE/Ramachandran_plot__real__Aspartate__D_.png}&\includegraphics[width=1.15\linewidth]{X_without_MSE/Ramachandran_plot__real__Cysteine__C_.png}&\includegraphics[width=1.15\linewidth]{X_without_MSE/Ramachandran_plot__real__Glycine__G_.png}&\includegraphics[width=1.15\linewidth]{X_without_MSE/Ramachandran_plot__real__Histidine__H_.png}&\includegraphics[width=1.15\linewidth]{X_without_MSE/Ramachandran_plot__real__Serine__S_.png}\\ 
		\footnotesize Density estimation by NCE&\includegraphics[width=1.15\linewidth]{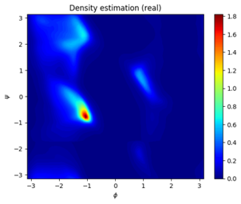}&\includegraphics[width=1.15\linewidth]{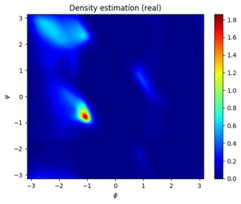}&\includegraphics[width=1.15\linewidth]{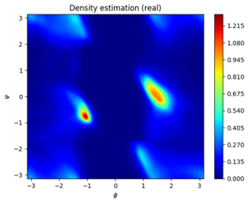}&\includegraphics[width=1.15\linewidth]{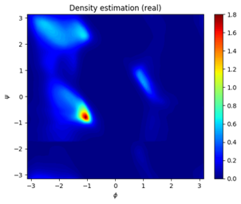}&\includegraphics[width=1.15\linewidth]{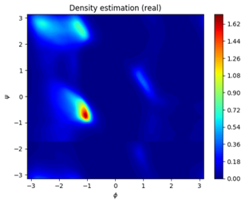}\\
		\footnotesize Density estimation by AC-GAN&\includegraphics[width=1.15\linewidth]{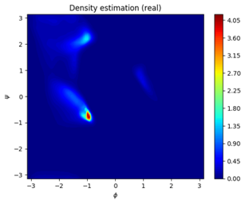}&\includegraphics[width=1.15\linewidth]{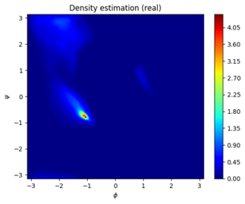}&\includegraphics[width=1.15\linewidth]{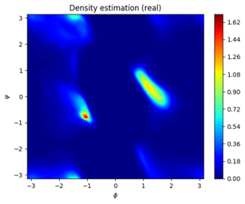}&\includegraphics[width=1.15\linewidth]{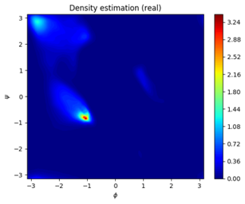}&\includegraphics[width=1.15\linewidth]{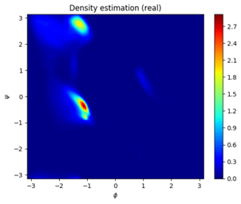}\\
		\footnotesize Density estimation by Semi-supervised GAN&\includegraphics[width=1.15\linewidth]{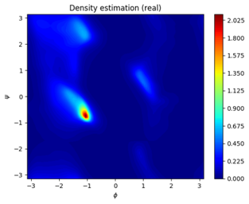}&\includegraphics[width=1.15\linewidth]{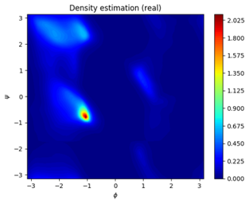}&\includegraphics[width=1.15\linewidth]{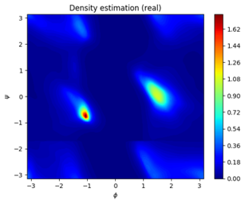}&\includegraphics[width=1.15\linewidth]{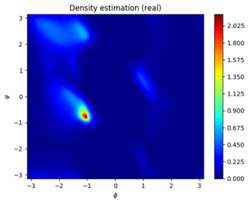}&\includegraphics[width=1.15\linewidth]{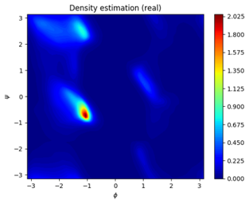}\\
		\footnotesize Predicted angles by regression model&\includegraphics[width=1.15\linewidth]{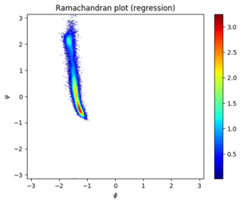}&\includegraphics[width=1.15\linewidth]{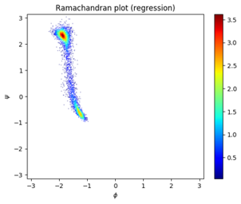}&\includegraphics[width=1.15\linewidth]{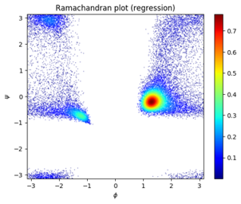}&\includegraphics[width=1.15\linewidth]{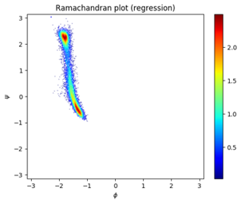}&\includegraphics[width=1.15\linewidth]{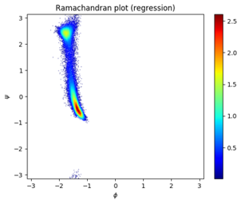}\\
		\footnotesize Predicted angles by C-GAN&\includegraphics[width=1.15\linewidth]{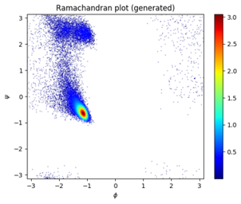}&\includegraphics[width=1.15\linewidth]{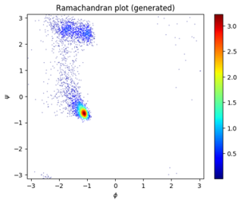}&\includegraphics[width=1.15\linewidth]{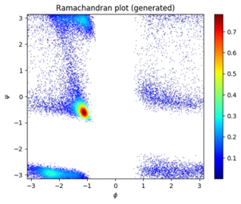}&\includegraphics[width=1.15\linewidth]{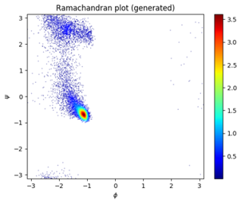}&\includegraphics[width=1.15\linewidth]{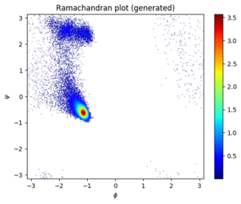}\\
		\footnotesize Predicted angles by AC-GAN&\includegraphics[width=1.15\linewidth]{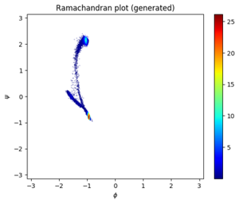}&\includegraphics[width=1.15\linewidth]{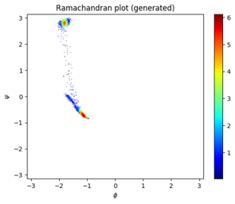}&\includegraphics[width=1.15\linewidth]{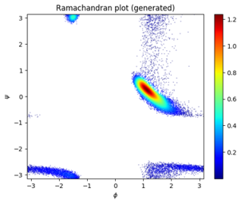}&\includegraphics[width=1.15\linewidth]{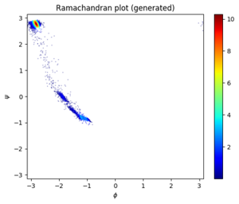}&\includegraphics[width=1.15\linewidth]{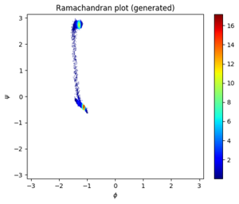}\\
		\footnotesize Predicted angles by Semi-supervised GAN&\includegraphics[width=1.15\linewidth]{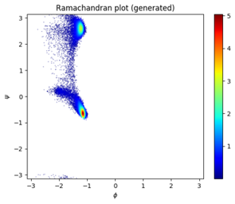}&\includegraphics[width=1.15\linewidth]{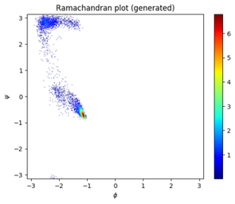}&\includegraphics[width=1.15\linewidth]{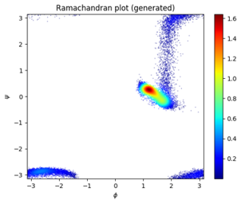}&\includegraphics[width=1.15\linewidth]{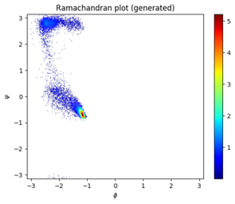}&\includegraphics[width=1.15\linewidth]{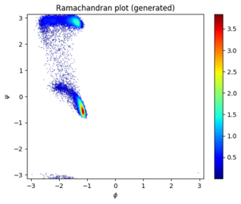}
	\end{tabular}
	\caption{Table of real angles, estimated densities $p_{model}(x|Center=c,S=real)$ and predicted angles plots for some central amino acids for the third experiment. Plotted real angles are from the test set and predicted angles are predicted by sequences in the test set. Same 5 chosen amino acids are depicted as in figure \ref{fig:Amino acids plots without regression loss}.}\label{fig:Amino acids plots D without regression loss minibatch}
\end{table}
\subsection{With regression loss, using predicted density by NCE and minibatch-wise loss in generators} \label{D With regression loss minibatch}
\begin{figure}[H]
	\centering
	\begin{subfigure}[t]{0.47\textwidth}
        \includegraphics[width=\textwidth]{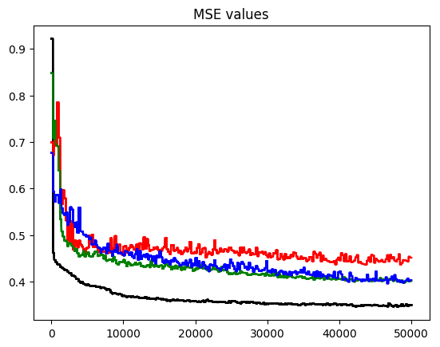}
        \caption{Test mean square errors of cosine and sine values of angles} 
    \end{subfigure}	%\hspace{0.05\textwidth}
    \begin{subfigure}[t]{0.47\textwidth}
        \includegraphics[width=\textwidth]{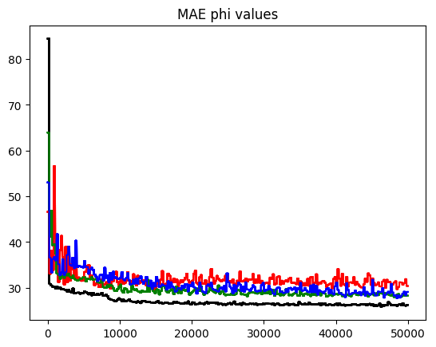}
        \caption{Test mean absolute errors of phi angle}
    \end{subfigure} 
	\begin{subfigure}[t]{0.47\textwidth}
        \includegraphics[width=\textwidth]{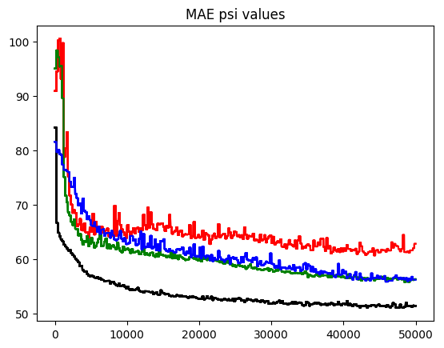}
        \caption{Test mean absolute errors of psi angle}
    \end{subfigure}     
    \caption{The regression errors of the fourth experiment.\\ 
Black line: Regression model, Red line: Conditional GANs, Green line: AC-GANs, Blue line: Semi-supervised GANs. x-axis indicates number of iterations. }\label{fig:errors D with regression loss logit2}
\end{figure}

\begin{figure}[H]
	\centering
	\begin{subfigure}[t]{0.47\textwidth}
        \includegraphics[width=\textwidth]{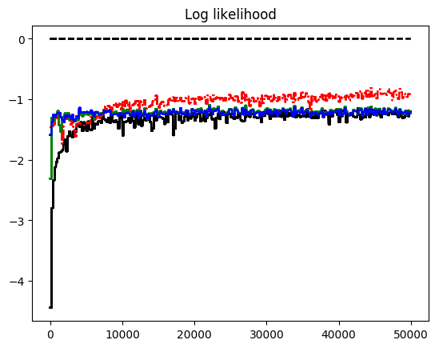}\hfil 
        \caption{Mean log-likelihood} 
    \end{subfigure}	%\hspace{0.05\textwidth}
    \begin{subfigure}[t]{0.47\textwidth}
        \includegraphics[width=\textwidth]{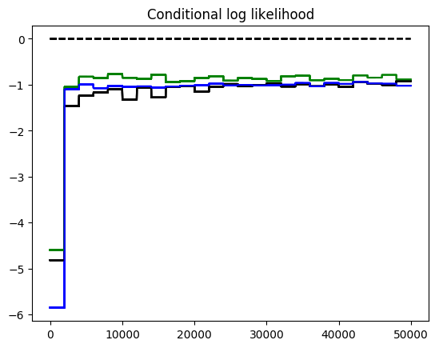}
        \caption{Weighted mean conditional log-likelihood along the middle amino acids}
    \end{subfigure}\par\medskip
	\begin{subfigure}[t]{0.47\textwidth}
        \includegraphics[width=\textwidth]{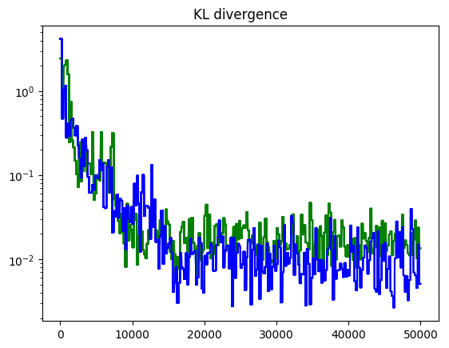}\par
        \caption{Kullback-Leibler divergence between\\ $p_{model}(x|S=real)$ and $p_{model}(x|S=fake)$}
    \end{subfigure}\par\medskip     
    \begin{subfigure}[t]{0.47\textwidth}
        \includegraphics[width=\textwidth]{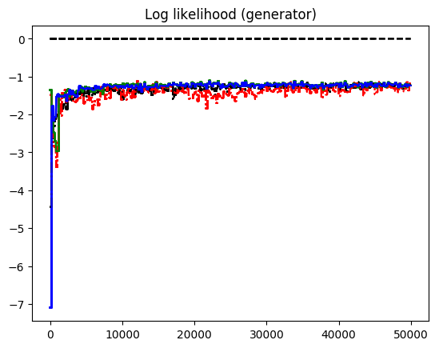}
        \caption{Mean log-likelihood for generated samples} 
    \end{subfigure}\medskip %\hspace{0.05\textwidth}
    \begin{subfigure}[t]{0.47\textwidth}
        \includegraphics[width=\textwidth]{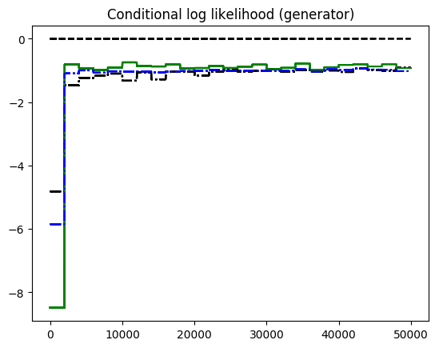}\par
        \caption{Weighted mean conditional log-likelihood along the middle amino acids for generated samples}
    \end{subfigure} 
    \caption{The generation metrics of the fourth experiment.\\
Black line: NCE model, Red line: Conditional GANs, Green line: AC-GANs, Blue line: Semi-supervised GANs. The x-axis indicates the number of iterations. \\
$\ast$ Red line in (a): $p_{model}(x,y|S=real)$, red line in (d): $p_{model}(x,y|S=fake)$, \\
black and blue line in (e): Weighted mean conditional log-likelihood for real data (not the generated data). They are depicted only for comparison.}\label{fig:estimations D with regression loss logit2} %\textcolor{red}{I'm planning to change the KL divergence plot so that y axis indicates log KL divergence with base 10.}
\end{figure}

\begin{figure}[H]
\centering
    \includegraphics[width=0.47\linewidth]{Ramachandran_plot__real_.jpg}\par
    \includegraphics[width=0.47\linewidth]{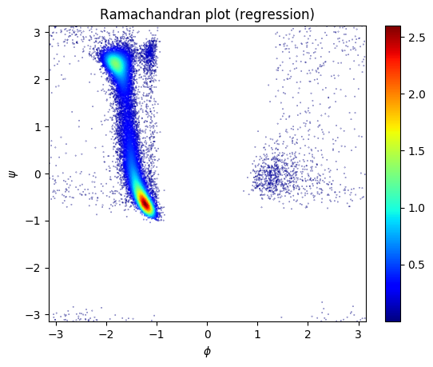}\hfil
    \includegraphics[width=0.47\linewidth]{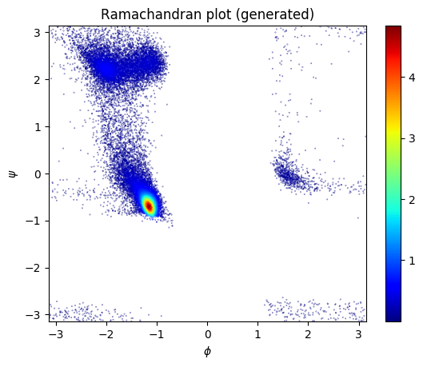}\par\medskip
    \includegraphics[width=0.47\linewidth]{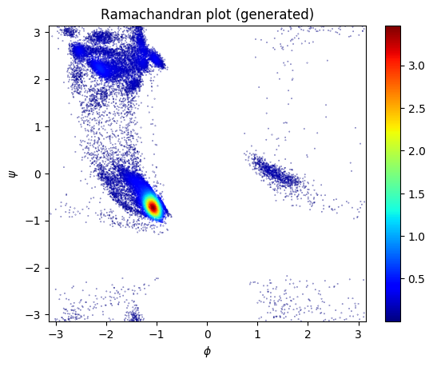}\hfil
    \includegraphics[width=0.47\linewidth]{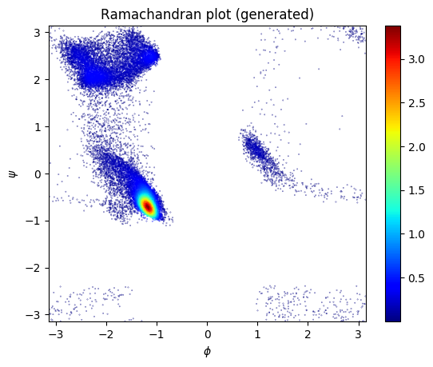}
\caption{Ramachandran plot of test angles and plots using predicted angles for the fourth experiment.\\
Top: real angles, \\
middle left: predicted angles by using regression model, middle right: predicted angles by using C-GAN, \\
bottom left: predicted angles by using AC-GAN, bottom right: predicted angles by using Semi-supervised GAN}\label{fig:Ramachandran plot D with regression loss logit2}
\end{figure}

We got high MSE and MAE values in the generators of previous experiments. As mentioned in \ref{Compare GAN}, in this experiment, we added regression loss \eqref{eq:Regression loss} in the loss of the generators to reduce these errors in the generators. \\\\
In table \ref{table:D with regression loss logit2}, MSE and MAE values of all generative models were reduced.
$p_{model}(x,y|S=fake)$ of C-GAN was improved and $p_{model}(x,y|S=real)$ was also increased probably due to improved stability of the generator as shown in \ref{fig:errors D with regression loss logit2}. Weighted mean conditional log-likelihood (CLL) of generated samples in AC-GAN was improved. This might be because of increased conditional generation performance due to added regression loss. However, mean log-likelihood (LL) of generated samples in AC-GAN was decreased. This might be because the generator put less effort in generating realistic samples due to added loss. \\\\
It appears that LL of both real and generated samples increased and CLL decreased slightly in Semi-supervised GAN. However, it was not clear whether this change was due to added loss or just random fluctuation. KL divergence of the Semi-supervised GAN seems to be reduced in the table \ref{table:D with regression loss logit2}, however when we considered figure \ref{fig:estimations D with regression loss logit2} (c), this was due to random fluctuation and it seems KL divergence got higher compared to previous experiment \ref{fig:estimations D without regression loss logit2} (c).\\\\
When distribution of predicted angles considered in figure  \ref{fig:Ramachandran plot D with regression loss logit2}, C-GAN could generate more samples in $\alpha_{L}$ region and distribution of predicted angles of AC-GAN is smoothed  probably because predicted points were more likely to be located in the region between clusters to reduce regression loss. The shape of $\beta_S$ and $\beta_P$ recovered slightly in Semi-supervised GAN.\\\\
When we compared Ramachandran plots based on predicted angles for each central amino acids in the table of figures \ref{fig:Amino acids plots D with regression loss minibatch}, all generative models generated samples in both $\alpha$ and $\alpha_L$ regions for central amino acid Glycine. In the result of AC-GAN, there were more number of predicted angles in the region between clusters.\\\\
\begin{table}[H]
	\begin{tabular}{ l | l | l | l | l | l }
    	\hline
    	\rowcolor{lightgray}
		 {}  &Regression &NCE &C-GAN &AC-GAN &Semi-supervised GAN\\ \hline
		MSE & $\boldsymbol{0.3496}$ & - &0.4513&0.4011&0.4032\\ \hline
		Phi MAE & $\boldsymbol{26.2809}$ & - &30.3906&28.2837&29.1065\\ \hline
		Psi MAE & $\boldsymbol{51.4477}$ & - &62.8351&56.3453&56.2262\\ \hline
		LL & - & -$\boldsymbol{1.2072}$ & -0.9146${}^\ast$ &-1.2174&-1.2307\\ \hline
		CLL &- & -0.9169 & - &-$\boldsymbol{0.8845}$&-1.0205\\ \hline
		KL divergence & -& - & -& 0.01348 &$\boldsymbol{0.005088}$\\ \hline
		LL (generated)&-& -&-1.1773${}^\ast$& -$\boldsymbol{1.2327}$ &-1.2351\\ \hline
		CLL (generated) &- & - & - &-0.9232&-\\ \hline
	\end{tabular} 
\caption{Comparision of MSEs, MAEs, mean log-likelihood, weighted mean conditional log-likelihood along the middle amino acids and Kullback-Leibler divergence between $p_{model}(x|S=real)$ and $p_{model}(x|S=fake)$ where $x$ indicates dihedral angles. The result of the fourth experiment after 50000 iterations. \\
$\ast$ Note that LL for C-GAN is not actually a mean log-likelihood, but $p_{model}(x,y|S=real)$. Likewise, LL (generated) for C-GAN indicates $p_{model}(x,y|S=fake)$. }\label{table:D with regression loss logit2}%\textcolor{red}{It might be better to use multi-class NCE as a base line model for density estimation to also estimate baseline CLL.} 
\end{table}
\newpage
\begin{figure}[H]
\centering
    \includegraphics[width=0.47\linewidth]{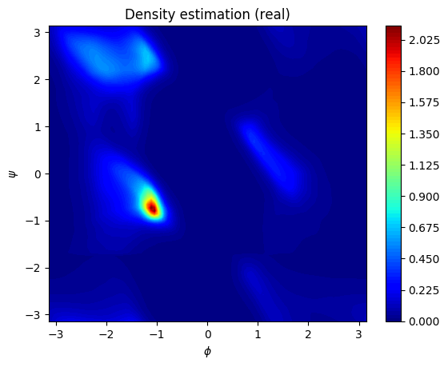}\par
    \includegraphics[width=0.47\linewidth]{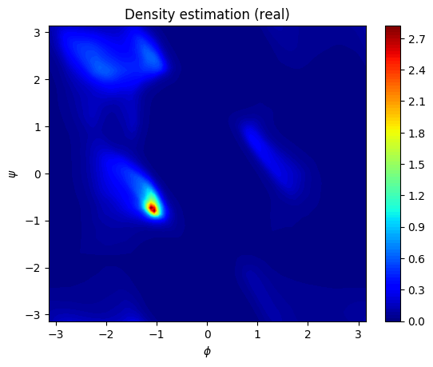}\hfil
    \includegraphics[width=0.47\linewidth]{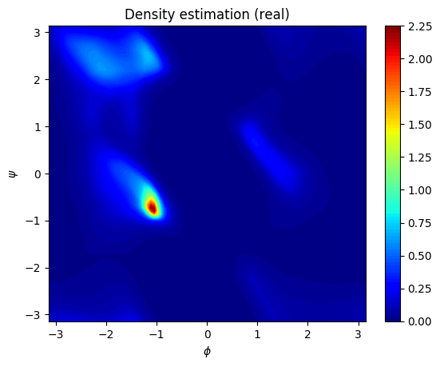}\par\medskip
    \includegraphics[width=0.47\linewidth]{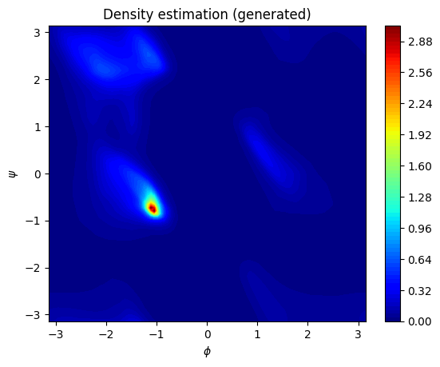}\hfil 
    \includegraphics[width=0.47\linewidth]{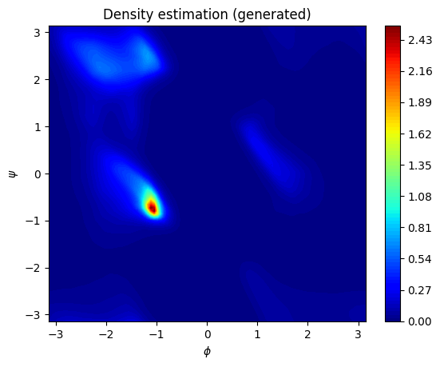}\par\medskip
\caption{Estimated density plots for the fourth experiment. Note that square root were applied on the estimated densities of angles for better visualization.\\ 
Top: $p_{\text{\tiny NCE}}(x|S=real)$, \\
middle left: $p_{\text{\tiny AC-GAN}}(x|S=real)$, 
middle right: $p_{\text{\tiny Semi-supervised GAN}}(x|S=real)$,\\ 
bottom left: $p_{\text{\tiny AC-GAN}}(x|S=fake)$, bottom right: $p_{\text{\tiny Semi-supervised GAN}}(x|S=fake)$}\label{fig:Density plot D with regression loss logit2}
\end{figure}
\begin{table}[H]
	\begin{tabular}{M{1.8cm}M{2.4cm}M{2.4cm}M{2.4cm}M{2.4cm}M{2.4cm}}
		{}&\footnotesize Aspartate (D)&\footnotesize Cysteine (C)&\footnotesize Glycine (G)&\footnotesize Histidine (H)&\footnotesize Serine (S)\\
		\footnotesize Real angles&\includegraphics[width=1.15\linewidth]{X_without_MSE/Ramachandran_plot__real__Aspartate__D_.png}&\includegraphics[width=1.15\linewidth]{X_without_MSE/Ramachandran_plot__real__Cysteine__C_.png}&\includegraphics[width=1.15\linewidth]{X_without_MSE/Ramachandran_plot__real__Glycine__G_.png}&\includegraphics[width=1.15\linewidth]{X_without_MSE/Ramachandran_plot__real__Histidine__H_.png}&\includegraphics[width=1.15\linewidth]{X_without_MSE/Ramachandran_plot__real__Serine__S_.png}\\ 
		\footnotesize Density estimation by NCE&\includegraphics[width=1.15\linewidth]{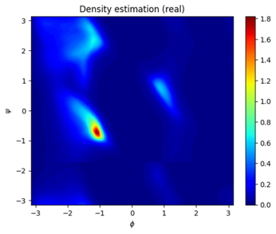}&\includegraphics[width=1.15\linewidth]{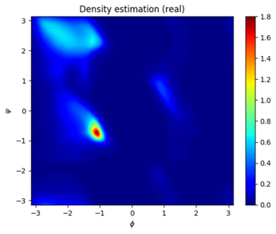}&\includegraphics[width=1.15\linewidth]{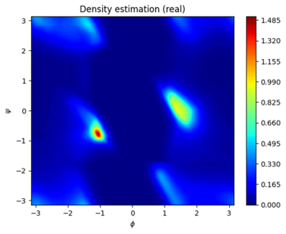}&\includegraphics[width=1.15\linewidth]{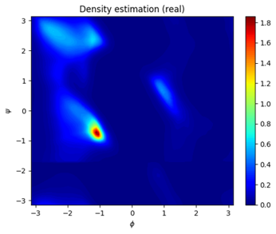}&\includegraphics[width=1.15\linewidth]{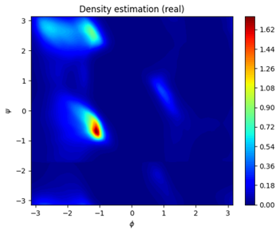}\\
		\footnotesize Density estimation by AC-GAN&\includegraphics[width=1.15\linewidth]{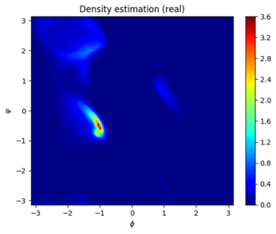}&\includegraphics[width=1.15\linewidth]{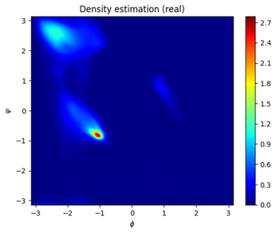}&\includegraphics[width=1.15\linewidth]{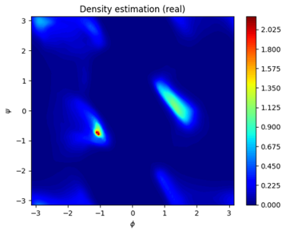}&\includegraphics[width=1.15\linewidth]{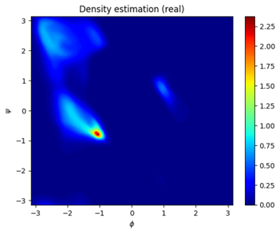}&\includegraphics[width=1.15\linewidth]{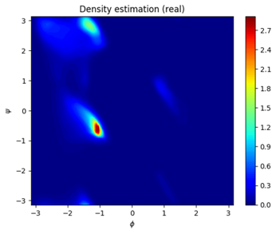}\\
		\footnotesize Density estimation by Semi-supervised GAN&\includegraphics[width=1.15\linewidth]{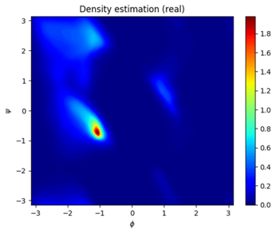}&\includegraphics[width=1.15\linewidth]{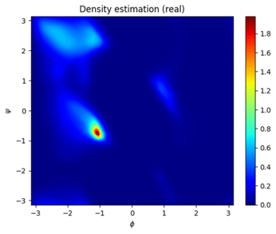}&\includegraphics[width=1.15\linewidth]{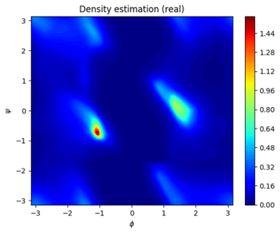}&\includegraphics[width=1.15\linewidth]{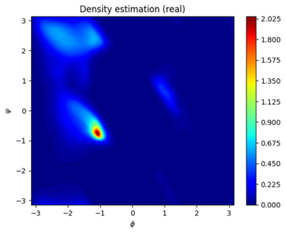}&\includegraphics[width=1.15\linewidth]{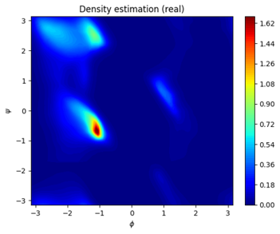}\\
		\footnotesize Predicted angles by regression model&\includegraphics[width=1.15\linewidth]{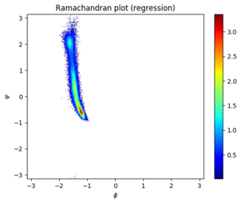}&\includegraphics[width=1.15\linewidth]{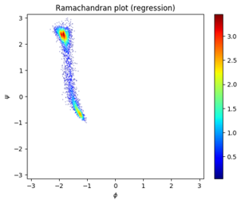}&\includegraphics[width=1.15\linewidth]{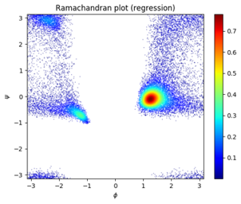}&\includegraphics[width=1.15\linewidth]{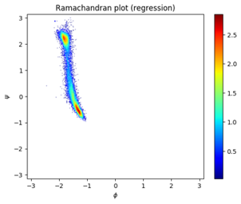}&\includegraphics[width=1.15\linewidth]{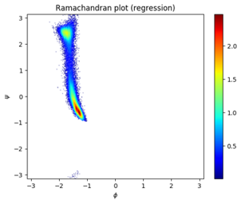}\\
		\footnotesize Predicted angles by C-GAN&\includegraphics[width=1.15\linewidth]{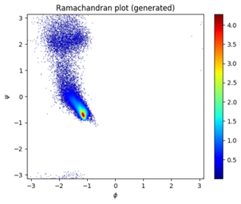}&\includegraphics[width=1.15\linewidth]{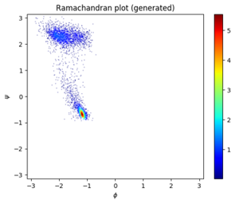}&\includegraphics[width=1.15\linewidth]{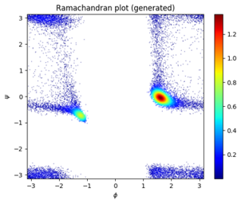}&\includegraphics[width=1.15\linewidth]{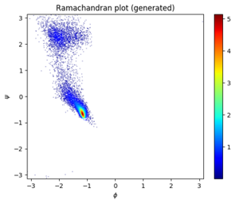}&\includegraphics[width=1.15\linewidth]{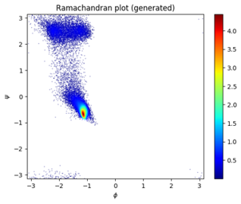}\\
		\footnotesize Predicted angles by AC-GAN&\includegraphics[width=1.15\linewidth]{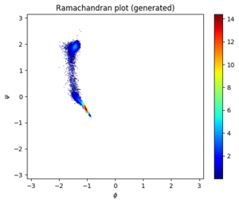}&\includegraphics[width=1.15\linewidth]{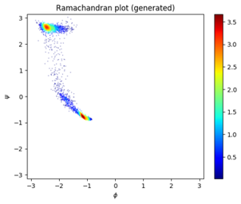}&\includegraphics[width=1.15\linewidth]{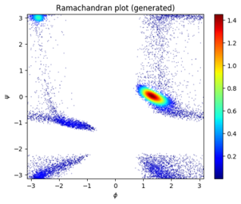}&\includegraphics[width=1.15\linewidth]{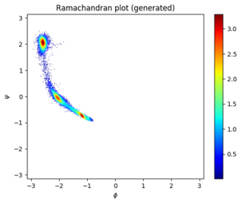}&\includegraphics[width=1.15\linewidth]{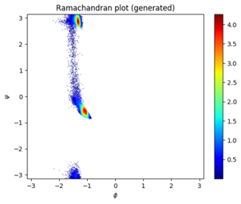}\\
		\footnotesize Predicted angles by Semi-supervised GAN&\includegraphics[width=1.15\linewidth]{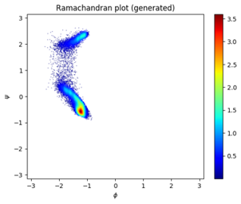}&\includegraphics[width=1.15\linewidth]{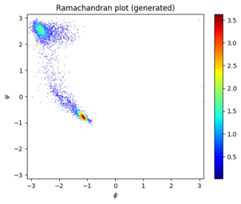}&\includegraphics[width=1.15\linewidth]{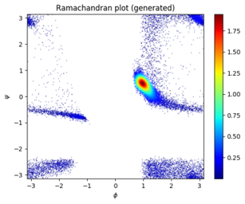}&\includegraphics[width=1.15\linewidth]{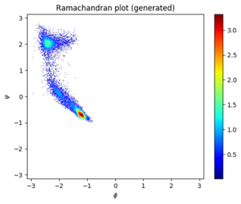}&\includegraphics[width=1.15\linewidth]{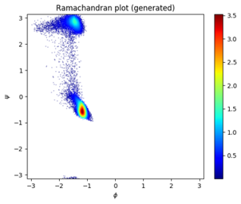}
	\end{tabular}
	\caption{Table of real angles, estimated densities $p_{model}(x|Center=c,S=real)$ and predicted angle plots for some central amino acids for the fourth experiment. Plotted real angles are from the test set and predicted angles are predicted by sequences in the test set. Same 5 chosen amino acids are depicted as in figure \ref{fig:Amino acids plots without regression loss}.
} \label{fig:Amino acids plots D with regression loss minibatch}
\end{table}
\subsection{With regression loss, using predicted density by NCE, minibatch-wise loss in generators and predicted angles by regression model} \label{Reg D With regression loss minibatch}
\begin{figure}[H]
\centering
    \includegraphics[width=0.47\linewidth]{Ramachandran_plot__real_.jpg}\par
    \includegraphics[width=0.47\linewidth]{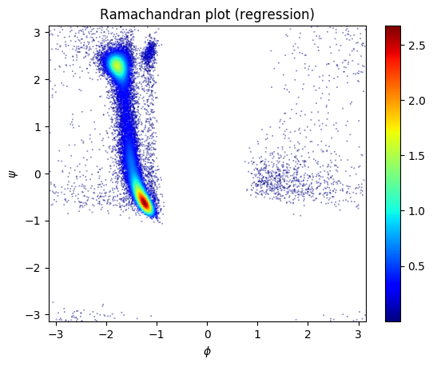}\hfil
    \includegraphics[width=0.47\linewidth]{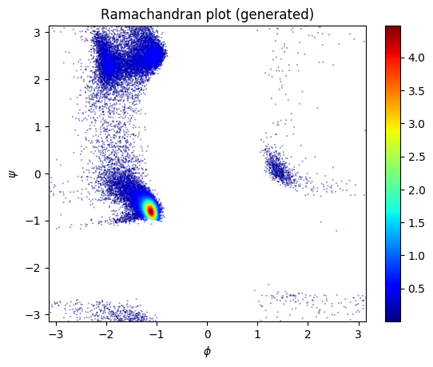}\par\medskip
    \includegraphics[width=0.47\linewidth]{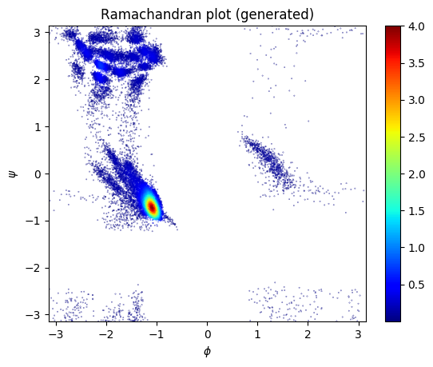}\hfil
    \includegraphics[width=0.47\linewidth]{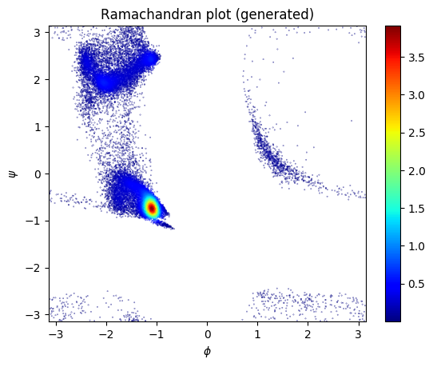}
\caption{Ramachandran plot of test angles and plots using predicted angles for the fifth experiment.\\
Top: real angles, \\
middle left: predicted angles by using regression model, middle right: predicted angles by using C-GAN, \\
bottom left: predicted angles by using AC-GAN, bottom right: predicted angles by using Semi-supervised GAN}\label{fig:Ramachandran plot Reg D with regression loss logit2}
\end{figure}

We still got high MSE and MAE values in the generators of the previous experiment. As mentioned in \ref{Compare GAN}, we used predicted results by regression model as an additional inputs of the generators in this experiment in order to further reduce these errors. \\\\
In table \ref{table:Reg D with regression loss logit2}, MSE and MAE values of all generative models were further improved.
$p_{model}(x,y|S=fake)$ of C-GAN was improved. %, however $p_{model}(x,y|S=real)$ is decreased. 
Weighted mean conditional log-likelihood (CLL) of generated samples in AC-GAN is also improved.\\\\ % and that of real samples in AC-GAN is decreased. 
In figure \ref{fig:Ramachandran plot Reg D with regression loss logit2},  it seems shape of $\beta_S$ regions of the C-GAN got worse.  Figure \ref{fig:Ramachandran plot Reg D with regression loss logit2} (bottom left) shows the smoothening effect of AC-GAN was lessened. This is probably because the generator is more focus on reducing generative loss as it can easily reduce the regression loss by using predicted angle information. The shape of the $\beta_S$ and $\beta_P$ region in Semi-supervised GAN appears to be worse.\\\\
When we compared Ramachandran plots based on predicted angles for each central amino acids in the table of figures \ref{fig:Amino acids plots Reg D with regression loss minibatch}, all generative models generated samples in both $\alpha$ and $\alpha_L$ regions for central amino acid Glycine.\\\\
\begin{table}[H]
	\begin{tabular}{ l | l | l | l | l | l }
    	\hline
    	\rowcolor{lightgray}
		 {}  &Regression &NCE &C-GAN &AC-GAN &Semi-supervised GAN\\ \hline
		MSE & $\boldsymbol{0.3504}$ & - &0.3823&0.3701&0.3678\\ \hline
		Phi MAE & $\boldsymbol{26.6377}$ & - &28.6309&27.7239&27.5119\\ \hline
		Psi MAE & $\boldsymbol{51.4816}$ & - &53.6731&51.5171&51.6804\\ \hline
		LL & - & -$\boldsymbol{1.2095}$ & -0.9789${}^\ast$ &-1.2233&-1.2301\\ \hline
		CLL &- & -$\boldsymbol{0.9107}$ & - &-0.9540&-1.0351\\ \hline
		KL divergence & -& - & -& 0.01684 &$\boldsymbol{0.01157}$\\ \hline
		LL (generated)&-& -&-1.0525${}^\ast$& -$\boldsymbol{1.2262}$ &-1.2448\\ \hline
		CLL (generated) &- & - & - &-0.9140&-\\ \hline
	\end{tabular} 
\caption{Comparision of MSEs, MAEs, mean log-likelihood, weighted mean conditional log-likelihood along the middle amino acids and Kullback-Leibler divergence between $p_{model}(x|S=real)$ and $p_{model}(x|S=fake)$ where $x$ indicates dihedral angles. The result of the fifth experiment after 50000 iterations. \\
$\ast$ Note that LL for C-GAN is not actually a mean log-likelihood, but $p_{model}(x,y|S=real)$. Likewise, LL (generated) for C-GAN indicates $p_{model}(x,y|S=fake)$. }\label{table:Reg D with regression loss logit2}%\textcolor{red}{It might be better to use multi-class NCE as a base line model for density estimation to also estimate baseline CLL.} 
\end{table}
\newpage
\begin{figure}[H]
	\centering
	\begin{subfigure}[t]{0.47\textwidth}
        \includegraphics[width=\textwidth]{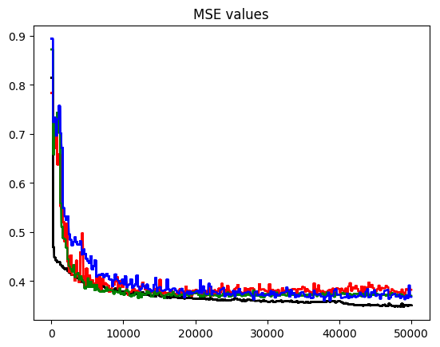}
        \caption{Test mean square errors of cosine and sine values of angles} 
    \end{subfigure}	%\hspace{0.05\textwidth}
    \begin{subfigure}[t]{0.47\textwidth}
        \includegraphics[width=\textwidth]{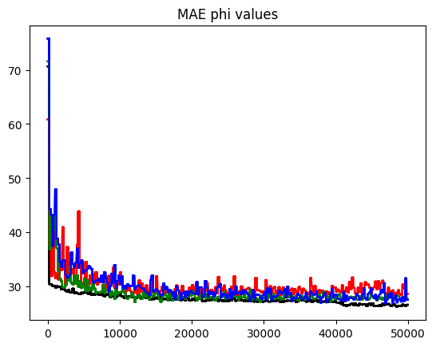}
        \caption{Test mean absolute errors of phi angle}
    \end{subfigure} 
	\begin{subfigure}[t]{0.47\textwidth}
        \includegraphics[width=\textwidth]{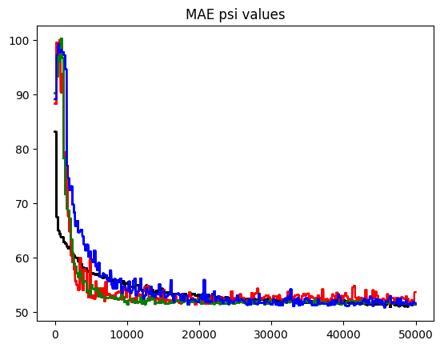}
        \caption{Test mean absolute errors of psi angle}
    \end{subfigure}     
    \caption{The regression errors of the fifth experiment.\\
Black line: Regression model, Red line: Conditional GANs, Green line: AC-GANs, Blue line: Semi-supervised GANs. The x-axis indicates the number of iterations. }\label{fig:errors Reg D with regression loss logit2}
\end{figure}
\begin{figure}[H]
	\centering
	\begin{subfigure}[t]{0.47\textwidth}
        \includegraphics[width=\textwidth]{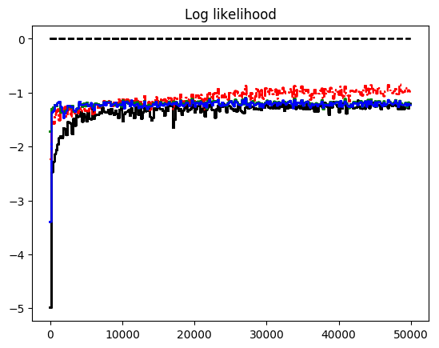}\hfil 
        \caption{Mean log-likelihood} 
    \end{subfigure}	%\hspace{0.05\textwidth}
    \begin{subfigure}[t]{0.47\textwidth}
        \includegraphics[width=\textwidth]{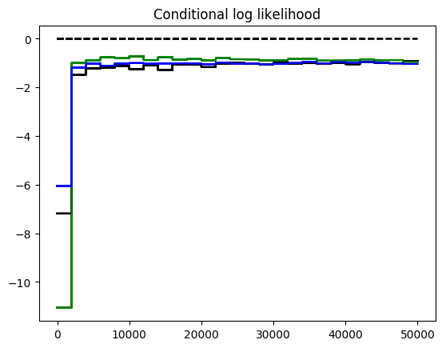}
        \caption{Weighted mean conditional log-likelihood along the middle amino acids}
    \end{subfigure}\par\medskip
	\begin{subfigure}[t]{0.47\textwidth}
        \includegraphics[width=\textwidth]{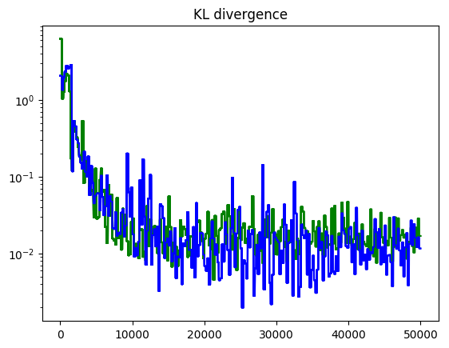}\par
        \caption{Kullback-Leibler divergence between\\ $p_{model}(x|S=real)$ and $p_{model}(x|S=fake)$}
    \end{subfigure}\par\medskip     
    \begin{subfigure}[t]{0.47\textwidth}
        \includegraphics[width=\textwidth]{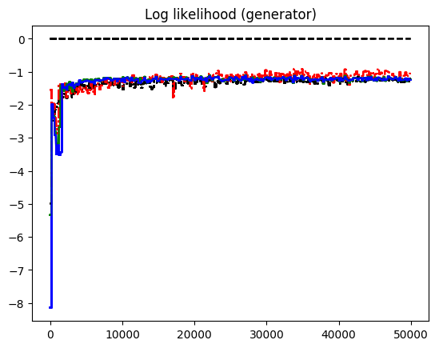}
        \caption{Mean log-likelihood for generated samples} 
    \end{subfigure}\medskip %\hspace{0.05\textwidth}
    \begin{subfigure}[t]{0.47\textwidth}
        \includegraphics[width=\textwidth]{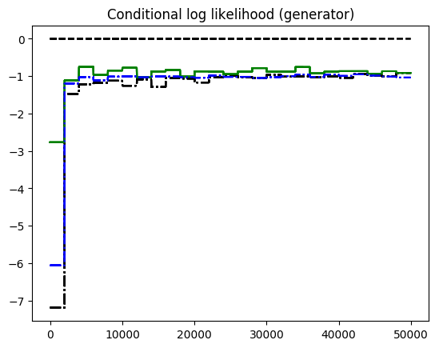}\par
        \caption{Weighted mean conditional log-likelihood along the middle amino acids for generated samples}
    \end{subfigure} 
    \caption{The generation metrics of the fifth experiment.\\
Black line: NCE model, Red line: Conditional GANs, Green line: AC-GANs, Blue line: Semi-supervised GANs. The x-axis indicates the number of iterations. \\
$\ast$ Red line in (a): $p_{model}(x,y|S=real)$, red line in (d): $p_{model}(x,y|S=fake)$, \\
black and blue line in (e): Weighted mean conditional log-likelihood for real data (not the generated data). They are depicted only for comparison.}\label{fig:estimations Reg D with regression loss logit2} %\textcolor{red}{I'm planning to change the KL divergence plot so that y axis indicates log KL divergence with base 10.}
\end{figure}
\begin{figure}[H]
\centering
    \includegraphics[width=0.47\linewidth]{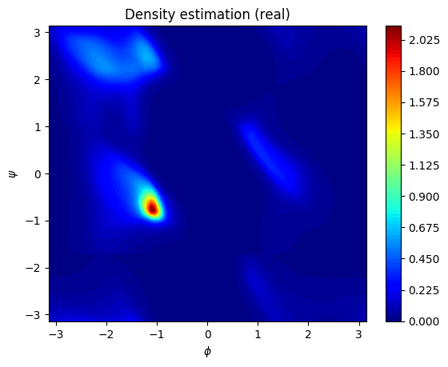}\par
    \includegraphics[width=0.47\linewidth]{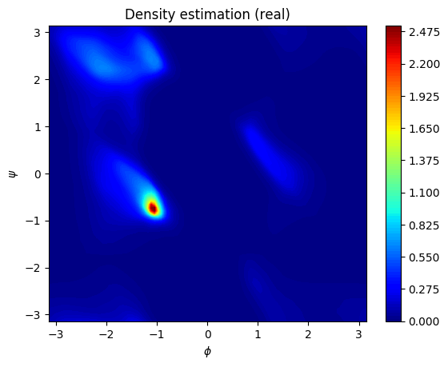}\hfil
    \includegraphics[width=0.47\linewidth]{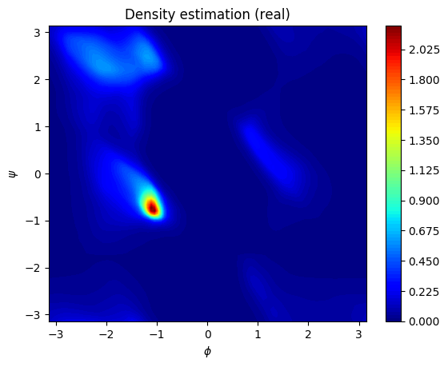}\par\medskip
    \includegraphics[width=0.47\linewidth]{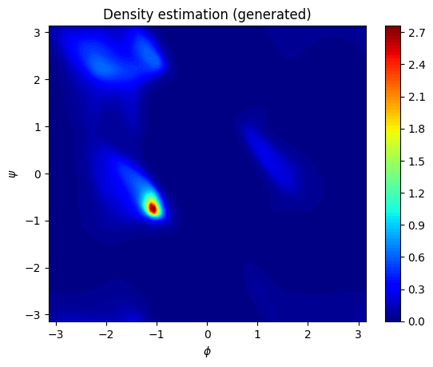}\hfil 
    \includegraphics[width=0.47\linewidth]{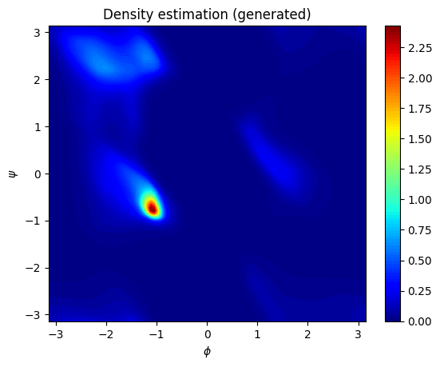}\par\medskip
\caption{Estimated density plots for the fifth experiment. Note that square root were applied on the estimated densities of angles for better visualization.\\ 
Top: $p_{\text{\tiny NCE}}(x|S=real)$, \\
middle left: $p_{\text{\tiny AC-GAN}}(x|S=real)$, 
middle right: $p_{\text{\tiny Semi-supervised GAN}}(x|S=real)$,\\ 
bottom left: $p_{\text{\tiny AC-GAN}}(x|S=fake)$, bottom right: $p_{\text{\tiny Semi-supervised GAN}}(x|S=fake)$}\label{fig:Density plot Reg D with regression loss logit2}
\end{figure}
\begin{table}[H]
	\begin{tabular}{M{1.8cm}M{2.4cm}M{2.4cm}M{2.4cm}M{2.4cm}M{2.4cm}}
		{}&\footnotesize Aspartate (D)&\footnotesize Cysteine (C)&\footnotesize Glycine (G)&\footnotesize Histidine (H)&\footnotesize Serine (S)\\
		\footnotesize Real angles&\includegraphics[width=1.15\linewidth]{X_without_MSE/Ramachandran_plot__real__Aspartate__D_.png}&\includegraphics[width=1.15\linewidth]{X_without_MSE/Ramachandran_plot__real__Cysteine__C_.png}&\includegraphics[width=1.15\linewidth]{X_without_MSE/Ramachandran_plot__real__Glycine__G_.png}&\includegraphics[width=1.15\linewidth]{X_without_MSE/Ramachandran_plot__real__Histidine__H_.png}&\includegraphics[width=1.15\linewidth]{X_without_MSE/Ramachandran_plot__real__Serine__S_.png}\\ 
		\footnotesize Density estimation by NCE&\includegraphics[width=1.15\linewidth]{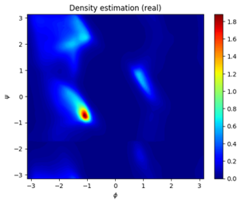}&\includegraphics[width=1.15\linewidth]{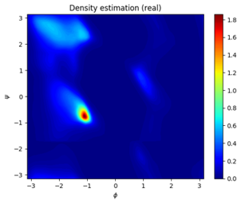}&\includegraphics[width=1.15\linewidth]{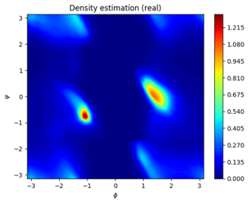}&\includegraphics[width=1.15\linewidth]{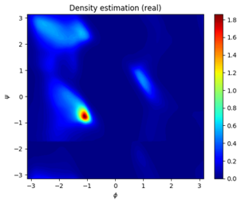}&\includegraphics[width=1.15\linewidth]{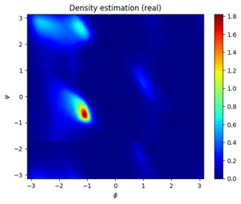}\\
		\footnotesize Density estimation by AC-GAN&\includegraphics[width=1.15\linewidth]{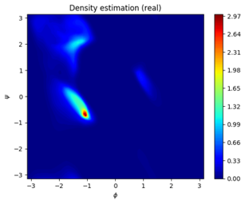}&\includegraphics[width=1.15\linewidth]{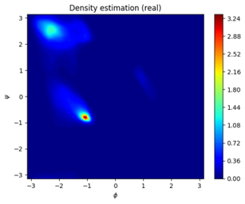}&\includegraphics[width=1.15\linewidth]{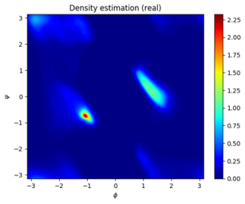}&\includegraphics[width=1.15\linewidth]{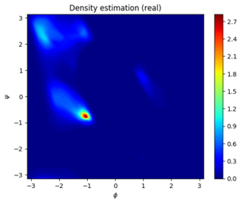}&\includegraphics[width=1.15\linewidth]{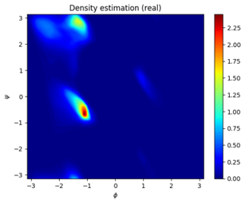}\\
		\footnotesize Density estimation by Semi-supervised GAN&\includegraphics[width=1.15\linewidth]{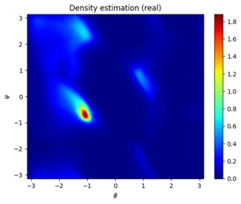}&\includegraphics[width=1.15\linewidth]{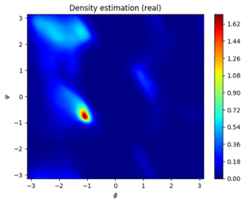}&\includegraphics[width=1.15\linewidth]{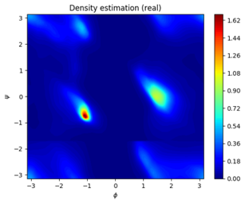}&\includegraphics[width=1.15\linewidth]{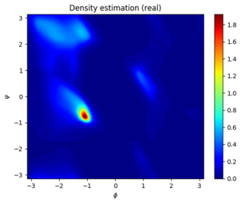}&\includegraphics[width=1.15\linewidth]{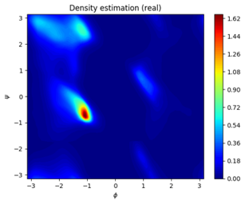}\\
		\footnotesize Predicted angles by regression model&\includegraphics[width=1.15\linewidth]{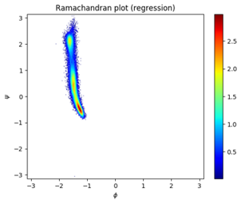}&\includegraphics[width=1.15\linewidth]{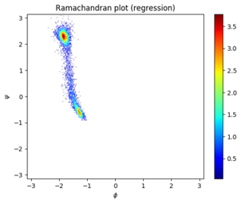}&\includegraphics[width=1.15\linewidth]{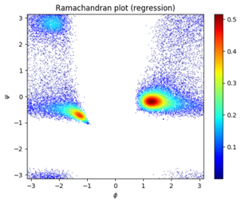}&\includegraphics[width=1.15\linewidth]{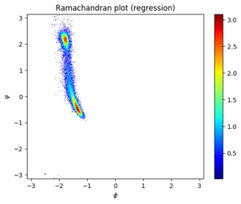}&\includegraphics[width=1.15\linewidth]{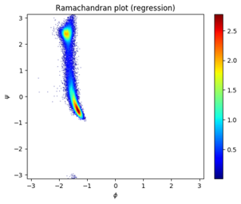}\\
		\footnotesize Predicted angles by C-GAN&\includegraphics[width=1.15\linewidth]{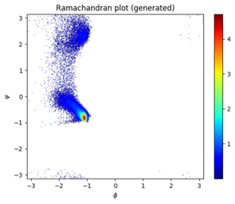}&\includegraphics[width=1.15\linewidth]{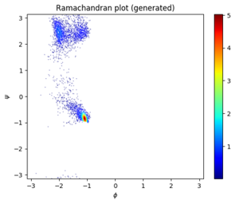}&\includegraphics[width=1.15\linewidth]{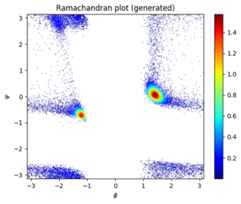}&\includegraphics[width=1.15\linewidth]{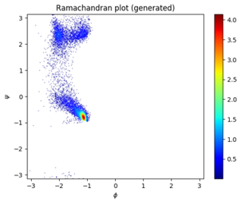}&\includegraphics[width=1.15\linewidth]{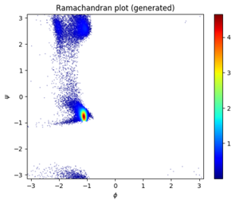}\\
		\footnotesize Predicted angles by AC-GAN&\includegraphics[width=1.15\linewidth]{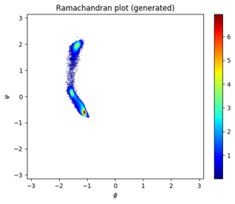}&\includegraphics[width=1.15\linewidth]{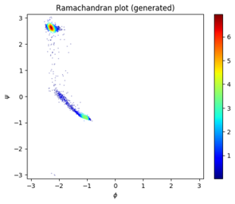}&\includegraphics[width=1.15\linewidth]{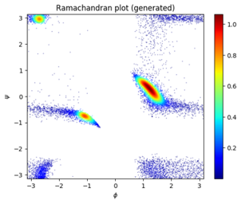}&\includegraphics[width=1.15\linewidth]{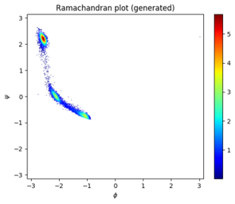}&\includegraphics[width=1.15\linewidth]{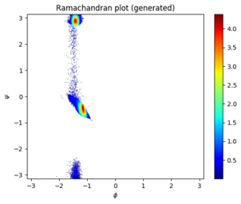}\\
		\footnotesize Predicted angles by Semi-supervised GAN&\includegraphics[width=1.15\linewidth]{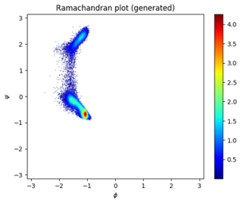}&\includegraphics[width=1.15\linewidth]{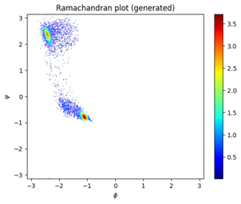}&\includegraphics[width=1.15\linewidth]{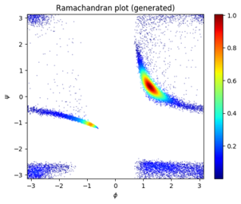}&\includegraphics[width=1.15\linewidth]{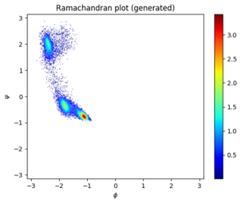}&\includegraphics[width=1.15\linewidth]{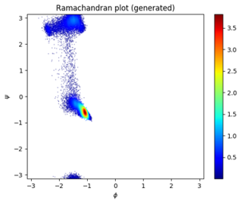}
	\end{tabular}
	\caption{Table of real angles, estimated densities $p_{model}(x|Center=c,S=real)$ and predicted angles plots for some central amino acids for the fifth experiment. Plotted real angles are from the test set and predicted angles are predicted by sequences in the test set. Same 5 chosen amino acids are depicted 
as in figure \ref{fig:Amino acids plots without regression loss}.
} \label{fig:Amino acids plots Reg D with regression loss minibatch}
\end{table}

%\subsection{With regression loss}
%\textcolor{red}{I'm planning to write that overall we get similar results when no regression loss was added. I will explain that we could improve in terms of conditional performances, like MSE, MAE values and conditional log-likelihoods, by adding regression loss. }
%\subsection{With regression loss and using predicted angles by regression}
%\textcolor{red}{I didn't get the updated result yet. I'm planning to write that we could further improve in terms of conditional performances by feeding predicted angle information (probably) without damaging other performances. The writing will be depend on the experiment results.}
%\subsection{With regression loss and using predicted angles by regression and using minibatch-wise loss in the generators}
%\textcolor{red}{I also didn't get the result yet. I'm planning to write that we could (probably) get more nicer Ramachandran plot and training was more stable. The writing will be depend on the experiment results.}\\\\
%\section{Comparison of methods for handling periodicity of the dihedral angle}
%\textcolor{red}{After finishing the explanation of the result of the previous section, I might choose a nicest model. And then, I will compare the performance of three periodicity handling methods by using chosen model.}

\chapter{Discussion}
In this thesis, we investigated dihedral angle prediction using three conditional generative adversarial models: C-GAN, AC-GAN and Semi-supervised GAN. We compared their characteristics and examined possible improvements. \\\\
When we compare the distributions of predicted angles, C-GAN was poor at capturing details of the Ramachandran plot, AC-GAN suffered from generating angles composed of unusual clusters and Semi-supervised GAN generated angles similar to the Ramachandran plot even though it could not generate angles in the $\zeta$ and $\gamma'$ regions separately in some experiments. So, we advise using Semi-supervised GAN for conditional generation of samples. We found that adding predicted density estimation by NCE model as an additional input of the discriminators can stabilize GAN training and help generating realistic angles. Minibatch-wise generation loss was only helpful in the training of C-GAN and AC-GAN, but not Semi-supervised GAN. Adding regression loss to the generators and adding predicted angles by regression model as an additional input of the generator was helpful improving conditional generation performances in C-GAN and AC-GAN addition to the MSE, MAE values.\\\\
%\textcolor{red}{I might need to add limitation of this work. For example, it was hard to distinguish whether the changes in the evaluation metric were due to change of the model or due to random fluctuation originate from the instability of the training. And we could not report conditional generation performance of the Semi-supervised GAN. To do this we can employ proposed Semi-supervised GAN structure \cite{2016arXiv160601583O}. Angles are directly fed into the discriminators. That is why density estimation plots are not continuous near boundary regions. However, it might be better to feed cosine and sine values of angles to the discriminators as we applied this approach in regression loss.}\\
This work is only focused on dihedral angle prediction using a window based method. Future works need to also predict secondary structure and solvent exposures. One can exploit bidirectional recurrent neural networks for this. The generator will output predicted angles and solvent exposures. Using predicted angles, secondary structure and solvent exposures by existing programs as an additional input of the generator would aid the generator to reduce the errors. Likewise, adding predicted secondary structure as an additional input of the discriminator would help the discriminator to stabilize training and one can make the discriminator to also output the predicted secondary structure.
%If the source discriminator also gets predicted classes of classification part of the discriminator as an intermediate input, then it can model the combination of classes.

%and ASA, HSE, CN by generator

%Semi-supervised generative adversarial networks might improve the secondary structure (SS) prediction of protein when generator tries to model the sequence profile and SS is the corresponding label of the sequence.

%\chapter*{Acknowledgements}
%\addcontentsline{toc}{chapter}{Acknowledgement}

\chapter*{Appendices}
\addcontentsline{toc}{chapter}{Appendices}
%\chapter{Appendix}

\markboth{\MakeUppercase{Appendices}}{\MakeUppercase{Appendices}}

\renewcommand*{\thesection}{A.\arabic{section}}
\renewcommand\thefigure{A.\arabic{figure}} 
\renewcommand\thetable{A.\arabic{table}} 
\section{Generating samples using NCE-GAN when target distribution is given} \label{Appendix generate samples}
\begin{figure}[H]
	\centering
	\begin{subfigure}[t]{0.4\textwidth}
        \includegraphics[width=\textwidth]{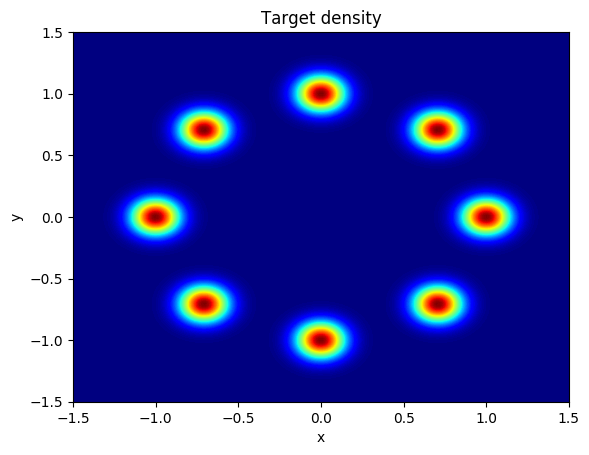}
        \caption{Target density} 
    \end{subfigure}	%\hspace{0.05\textwidth}
    \begin{subfigure}[t]{0.4\textwidth}
        \includegraphics[width=\textwidth]{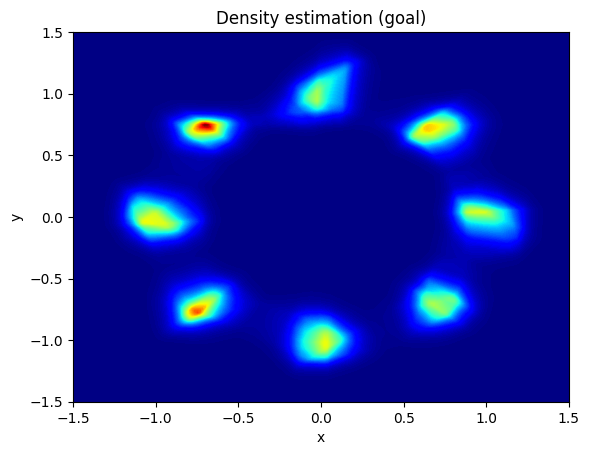}
        \caption{Estimated density of target distribution}
    \end{subfigure}    
    \begin{subfigure}[t]{0.4\textwidth}
        \includegraphics[width=\textwidth]{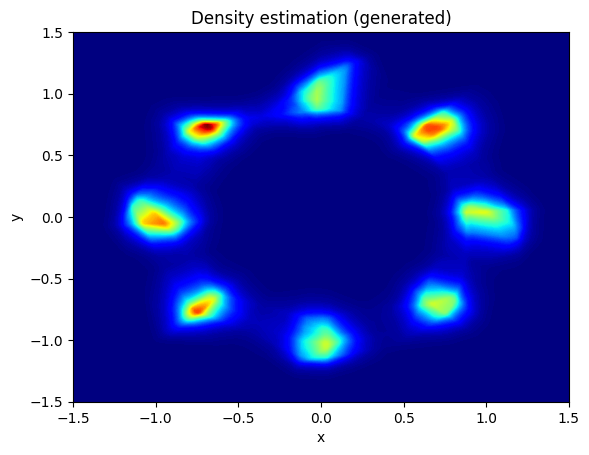}
        \caption{Estimated density of generated distribution}
    \end{subfigure} 
    \begin{subfigure}[t]{0.4\textwidth}
        \includegraphics[width=\textwidth]{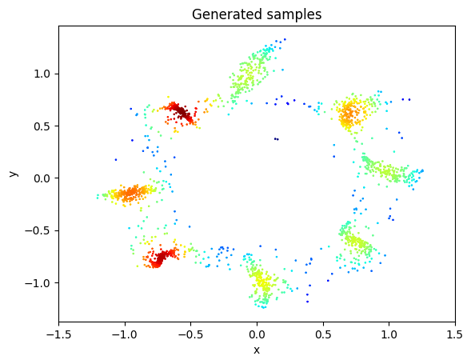}
        \caption{Generated samples}
    \end{subfigure}
    \begin{subfigure}[t]{0.4\textwidth}
        \includegraphics[width=\textwidth]{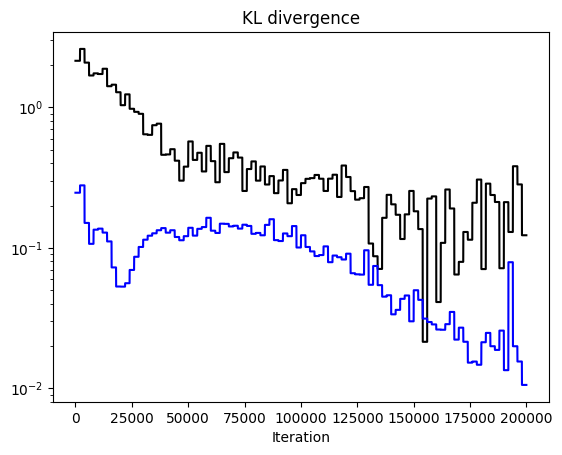}
        \caption{Estimated Kullback-Leibler divergence}
    \end{subfigure} 
    \caption{Generation of ring example \cite{metz2016unrolled} by described method. Note that black line in (e) indicates KL divergence between real target density and generated distribution. Blue line in (e) indicates estimated KL divergence between estimated target density and generated distribution. Minibatch-wise loss is used for training of the generator.}\label{Ring}
\end{figure}

\begin{figure}[H]
	\centering
	\begin{subfigure}[t]{0.4\textwidth}
        \includegraphics[width=\textwidth]{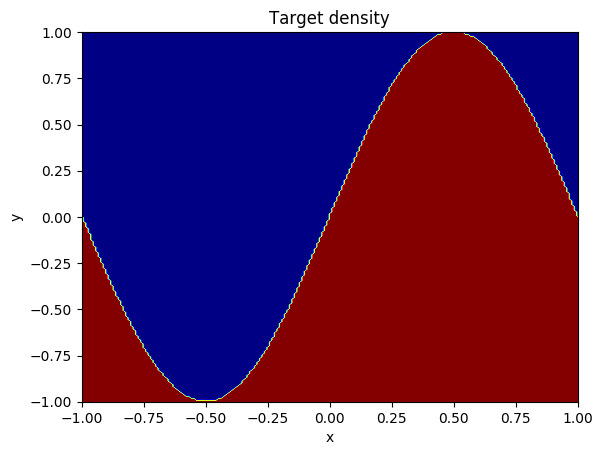}
        \caption{Target density} 
    \end{subfigure}	%\hspace{0.05\textwidth}
    \begin{subfigure}[t]{0.4\textwidth}
        \includegraphics[width=\textwidth]{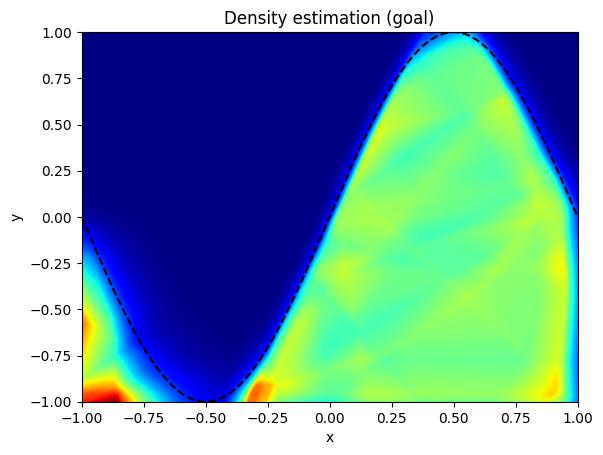}
        \caption{Estimated density of target distribution}
    \end{subfigure}    
    \begin{subfigure}[t]{0.4\textwidth}
        \includegraphics[width=\textwidth]{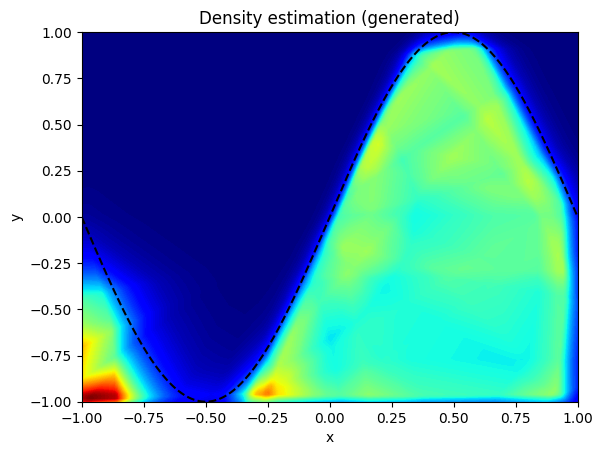}
        \caption{Estimated density of generated distribution}
    \end{subfigure} 
    \begin{subfigure}[t]{0.4\textwidth}
        \includegraphics[width=\textwidth]{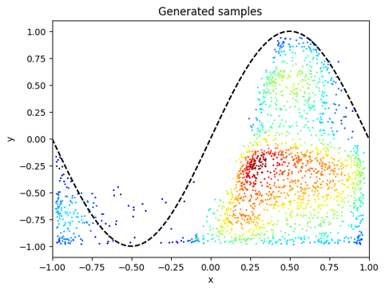}
        \caption{Generated samples}
    \end{subfigure}
    \begin{subfigure}[t]{0.4\textwidth}
        \includegraphics[width=\textwidth]{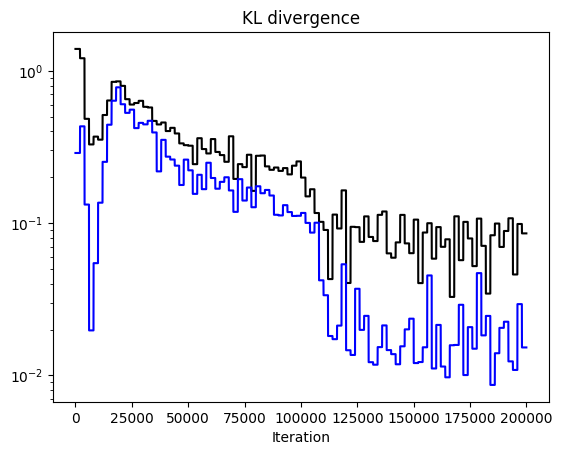}
        \caption{Estimated Kullback-Leibler divergence}
    \end{subfigure} 
    \caption{Generation of sine example by described method. Note that black line in (e) indicates KL divergence between real target density and generated distribution. Blue line in (e) indicates estimated KL divergence between estimated target density and generated distribution. Minibatch-wise loss is used for training of the generator.}\label{Sine}
\end{figure}

We can generate samples using NCE-GAN when target distribution is given by simply changing the training process. \\\\
The discriminator of the NCE-GAN will only get generated samples $x_{fake}$ and noise samples $x_{noise}$ as input because we do not have data samples $x_{real}$. Then it outputs corresponding predicted labels for both generated samples and noise samples. It also minimizes the estimated Kullback-Leibler (KL) divergence between estimated target density of the discriminator and target distribution. Remember that we can estimate $p_{model}(x|C=target)$ by equation \eqref{eq:density by NCE}. KL divergence will be estimated using numerical integration methods. \\\\ 
The generator tries to generate samples so that their estimated classes, which the discriminator predicts, match the target classes.\\\\
One problem with this method is that $p_{model}(C=target)$ will approach zero as no data samples are fed into the discriminator. Further research would be needed to handle this problem.\\\\
The following two examples \ref{Ring}, \ref{Sine} show NCE-GAN can generate target samples.\newpage
%\textcolor{red}{I'm planning to show examples of generated samples when some target density functions are given.}

\section{Experiment of minibatch discrimination for density estimation} \label{Appendix minibatch discrimination}

\begin{figure}[H]
	\centering
	\begin{subfigure}[t]{0.475\textwidth}
        \includegraphics[width=\textwidth]{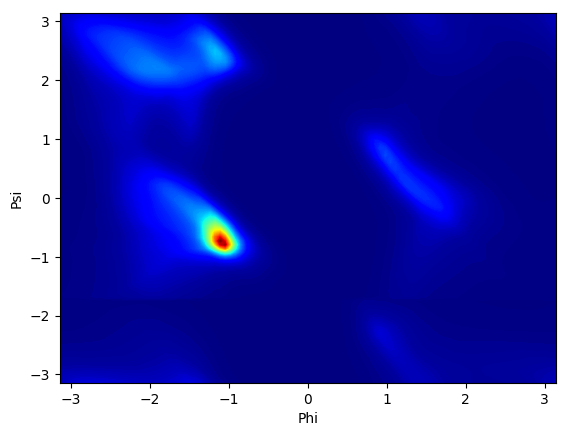}
        \caption{Estimated density using noise-contrastive estimation} 
    \end{subfigure}	%\hspace{0.05\textwidth}
    \begin{subfigure}[t]{0.475\textwidth}
        \includegraphics[width=\textwidth]{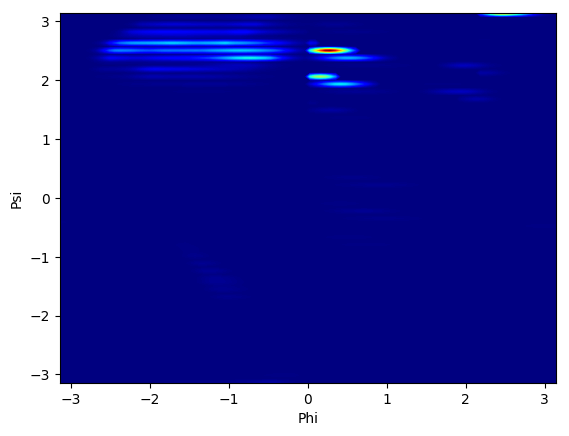}
        \caption{Estimated density using noise-contrastive estimation with minibatch discrimination}
    \end{subfigure}    
    \caption{Contour plots of estimated density of dihedral angles after 50000 iterations of mini-batch gradient updating. Noise samples are uniformly drawn from $(-\pi , \pi)^2$. Note that we also used batch size $64$ to calculate density estimation in (b). In other words, batch size $64$ is used for both training and test phase for (b). Estimated density for (b) is close to zero across all regions. Square root were applied on the estimated densities of angles for better visualization.}\label{Appendix MB figure}
\end{figure}

\begin{table}[H]
	\begin{tabular}{ | p{3.5cm} | p{3.5cm} | p{3.5cm} | p{3.5cm} |}
    	\hline
    	Discriminator & Estimated integration of estimated density function & LL for uniform (noise) samples & LL for test samples\\ \hline
    	Vanilla\newline discriminator & 0.9848 & -7.6532 & -1.2810\\ \hline
    	Suggested\newline minibatch\newline discriminator & 0.0001099 & -18.9252 & 9.3477\\ \hline
	\end{tabular} 
\caption{Monte Carlo integration method \cite{weinzierl2000introduction} was used to estimate the integration of the estimated density function. Remember that integration of probability density function should be 1. LL indicates mean log-likelihood.}\label{Appendix MB table}
\end{table}
From table \ref{Appendix MB table} we can notice that mean log-likelihood when minibatch discrimination is used (9.3477) is quite high compared to the result when the usual discriminator is used (-1.2810). However, if we consider other results like the small estimated integration of the estimated density function (0.0001099), small mean log-likelihood for noise samples (-18.9252) and poor estimated density plot in figure \ref{Appendix MB figure} (b), we can explain that high performance on log-likelihood was due to cheating, i.e., batch information is used to estimate the likelihood of the test samples.

\newpage\printbibliography[heading=bibintoc,
title={Bibliography}]

\end{document}